\documentclass[usenatbib, useAMS]{mn2e}
\usepackage{natbib, graphics, epsfig, color, times, amssymb, amsmath, verbatim,amsbsy}

\usepackage{times} 
\usepackage{amsmath}

\def\apj{ApJ}
\def\aap{A\&A}
\def\apjl{ApJL}
\def\apjs{ApJS}
\def\mnras{MNRAS}

\def\araa{ARA\&A}
\def\nat{Nature}
\def\prd{Phys. Rev. D}

\newcommand{\sm}{\small}

\newcommand{\msun}{{\rm M}_{\odot}} 
\newcommand{\msunh}{h^{-1}{\rm M}_{\odot}} 
\newcommand{\mhalo}{M_{\rm halo}}

\title[Galaxy and halo characteristics]
    {Moving mesh cosmology: characteristics of galaxies and haloes}

\author[D. Kere\v{s} et al.]{\parbox{18.5cm}{
Du\v{s}an Kere\v{s}$^{1,2}$\footnotemark[1], 
Mark Vogelsberger$^3$, 
Debora Sijacki$^3$\footnotemark[2], 
Volker Springel$^{4,5}$, 
Lars Hernquist$^3$}\vspace{0.2cm}\\
$^1$Department of Astronomy and Theoretical Astrophysics Center, University of California, Berkeley, CA 94720-3411, USA\\
$^2$Department of Physics, Center for Astrophysics and Space Sciences,
University of California at San Diego, 9500 Gilman Drive, La Jolla, CA
92093, USA\\
$^3$Harvard-Smithsonian Center for Astrophysics, 60 Garden Street, Cambridge, MA 02138, USA\\
$^4$Heidelberg Institute for Theoretical Studies, Schloss-Wolfsbrunnenweg 35, 69118 Heidelberg, Germany\\
$^5$Zentrum f\"ur Astronomie der Universit\"at Heidelberg, ARI, M\"onchhofstr. 12-14, 69120 Heidelberg, Germany\\
}
\begin{document}

\maketitle
\begin{abstract}
  We discuss cosmological hydrodynamic simulations of galaxy formation
  performed with the new moving-mesh code {\sm AREPO}, which promises
  higher accuracy compared with the traditional SPH technique that has
  been widely employed for this problem. In this exploratory study, we
  deliberately limit the complexity of the physical processes followed
  by the code for ease of comparison with previous calculations, and
  include only cooling of gas with a primordial composition, heating
  by a spatially uniform UV background, and a simple sub-resolution
  model for regulating star formation in the dense interstellar
  medium.  We use an identical set of physics in corresponding
  simulations carried out with the well-tested SPH code {\sm GADGET},
  adopting also the same high-resolution gravity solver. We are thus
  able to compare both simulation sets on an object-by-object basis,
  allowing us to cleanly isolate the impact of different
  hydrodynamical methods on galaxy and halo properties.  In
  accompanying papers, we focus on an analysis of the global baryonic
  statistics predicted by the simulation codes \citep{vogelsberger11},
  and complementary idealized simulations that highlight the
  differences between the hydrodynamical schemes
  \citep{sijacki11}. Here we investigate their influence on the
  baryonic properties of simulated galaxies and their surrounding
  haloes.  We find that {\sm AREPO} leads to significantly higher star
  formation rates for galaxies in massive haloes and to more extended
  gaseous disks in galaxies, which also feature a thinner and smoother
  morphology than their {\sm GADGET} counterparts. Consequently,
  galaxies formed in {\sm AREPO} have larger sizes and higher specific
  angular momentum than their SPH correspondents. Interestingly, the
  more efficient cooling flows in {\sm AREPO} yield higher densities
  and lower entropies in halo centers compared to {\sm GADGET},
  whereas the opposite trend is found in halo outskirts. The cooling
  differences leading to higher star formation rates of massive
  galaxies in {\sm AREPO} also slightly increase the baryon content
  within the virial radius of massive haloes.  We show that these
  differences persist as a function of numerical resolution.  While
  both codes agree to acceptable accuracy on a number of baryonic
  properties of cosmic structures, our results thus clearly
  demonstrate that galaxy formation simulations greatly benefit from
  the use of more accurate hydrodynamical techniques such as {\sm
    AREPO} and call into question the reliability of galaxy formation
  studies in a cosmological context using traditional standard
  formulations of SPH,  such as the one implemented in
   {\sm GADGET}.  Our new moving-mesh simulations demonstrate that a
  population of extended gaseous disks of galaxies in large volume
    cosmological simulations can be formed even without energetic
  feedback in the form of galactic winds, although such outflows
    appear required to obtain realistic stellar masses.
 \end{abstract}


\section{Introduction}
\renewcommand{\thefootnote}{\fnsymbol{footnote}}
\footnotetext[1]{Hubble Fellow. E-mail: dkeres@physics.ucsd.edu}
\footnotetext[2]{Hubble Fellow.}

\label{sec:intro}

Cosmological simulations of galaxy formation provide a powerful technique to
calculate the non-linear evolution of structure formation.  In principle, such
simulations can properly model the evolution and formation of galaxies from
first principles, but the enormous dynamic range and the many poorly
understood aspects of the baryonic physics of star formation make this task
extremely challenging.  At present, the limitations imposed by the coarse
numerical resolution achievable in practice and by the approximate treatment
of the physics introduce significant uncertainties in simulation predictions.
Regardless, previous simulations have already proven instrumental for
developing our current understanding of the formation and evolution of
galaxies \citep[e.g.][]{katz92b, weinberg97, pearce99, kravtsov02,
  springel03a, springel03b, borgani04, crain09}.

When only dark matter (DM) is considered, the newest generation of
cosmological simulation codes yields a consensus picture of important key
results about the matter distribution in the Universe, such as the large-scale
distribution and the detailed internal properties of dark matter haloes
\citep{springel05c, springel08, diemand08, boylankolchin09,navarro10,
  klypin10}.  The computational methods employed in the codes vary between
different implementations; however, the discretization of the equations needed
to follow the evolution of DM involves only gravity, and can only be done
efficiently in terms of the N-body technique.  Hence the requirements for
obtaining converged and consistent results with different simulation codes are
clear; all that is needed is an accurate gravity solver, a sufficiently large
number of simulation particles and an accurate time integration scheme.

Over the last decades, a number of hydrodynamic codes have been developed and
used in the field of computational cosmology \citep[e.g.][]{ryu93, bryan95,
  katz96, springel01a, teyssier02, wadsley04, springel05a}.  They all account
for the baryons in the Universe and try to follow their evolution
self-consistently in a $\Lambda$CDM dominated universe by solving the
equations of hydrodynamics coupled to gravity.  However, the discretization
schemes and numerical algorithms used for this task differ widely from code to
code, and unlike in pure DM simulations, large differences in the simulation
outcomes are often found.  Such simulations can be more directly related to
observations, unlike the pure dark matter simulations.  

In fact, hydrodynamic
simulations have already explained several features of the observed large scale
distribution of baryons, such as the existence and the properties of the
Ly-${\alpha}$ forest in the spectra of distant QSOs
\citep[e.g.][]{hernquist96, miralda-escude96}. In this, low density, low
temperature regime different hydrodynamic techniques show good agreement
between SPH and grid-based codes \citep{regan07}. Numerical simulations played
a crucial role in both our understanding of the properties of the absorbing
gas and in the quantitative measurements of the matter distribution in the
Universe based on Lyman-$\alpha$ forest data \citep[e.g.][]{croft98,
  seljak05}.  

Unfortunately, at higher densities characteristic of galaxies
and their haloes, qualitative differences exist already at the simplest
modeling level where radiative gas cooling and energetic feedback processes
due to star formation are ignored \citep[e.g.][]{frenk99, oshea05, agertz07,
  tasker08, vazza11}. Such differences are sometimes apparent even between
codes using quite similar discretization techniques \citep{springel02,
  mythesis}. This emphasizes the need to find numerical methods that
give the most accurate and reliable results. Understanding the origin and size
of errors in a given computational technique is clearly vital for fully 
exploiting
the predictive power of cosmological simulations of galaxy formation.

The most widely used methods to follow the evolution of baryons in
cosmological simulations are Lagrangian smoothed particle hydrodynamics (SPH)
and Eulerian mesh-based hydrodynamics with adaptive mesh refinement (AMR).
SPH has the great advantage of being naturally adaptive, making it suitable
for simulations with a large dynamic range in density, such as cosmological
simulations where SPH automatically increases its resolution in collapsing
regions. In addition, SPH can be easily combined with very accurate gravity
solvers, which are often based on gravitational tree algorithms
\citep[e.g.][]{hernquist89} or particle-mesh+tree
\citep[e.g.][]{bagla02,springel05a} methods. SPH has excellent conservation
properties. In particular, it still manifestly conserves total energy even
when coupled to self-gravity, something not available in hydrodynamical mesh
codes \citep{mueller95}.  On the other hand, AMR codes offer higher accuracy
in representing shocks, large density gradients and hydrodynamic fluid 
instabilities \citep[e.g.][]{agertz07}.

It has become clear over recent years that both SPH and AMR codes suffer from
weaknesses which are likely the primary cause for the disagreements found in
some cosmological results obtained with these methods
\citep[e.g.][]{mitchell09}. For example, it has been shown that SPH in its
standard implementation has limited ability to accurately follow fluid
instabilities \citep{agertz07}.  Furthermore, in SPH, shock capturing is done
through the addition of an artificial viscosity, which increases the
dissipativeness of the scheme and can have unwanted consequences such as
angular momentum transport in disks. Also, while the introduction of an
artificial viscosity results in the correct properties of post-shock flow, the
width of a shock in SPH can be significantly broadened
\citep[e.g.][]{hernquist89}, potentially affecting the evolution of gas in
complex cosmological environments \citep{hutchings00}, for example causing
in-shock cooling \citep{creasey11}.  On the other hand, Eulerian AMR codes are
not free of unwanted numerical effects either. They may suffer from advection
errors, mesh-alignment effects, and potentially over-mixing \citep{wadsley08}.
In addition, AMR codes are intrinsically not Galilean-invariant, i.e.~their
truncation error depends on the bulk velocity relative to the grid. This can
be a significant issue when the velocities of the flow are large and the
resolution is limited \citep[e.g.][]{springel10, robertson10}.

Recently, \citet{springel10} (S10 hereafter) proposed a new ``moving-mesh''
technique that promises accuracy advantages over both of these traditional
approaches.  This new method has been realized in the {\sm AREPO} code and can
be viewed as a hybrid of Lagrangian SPH and Eulerian AMR, combining positive
features of both methods while avoiding their most important weaknesses.
Instead of a stationary structured mesh as in AMR codes, {\sm AREPO} uses an
unstructured mesh given by the Voronoi tessellation of a set of
mesh-generating points distributed within the computational domain. These
points are allowed to move with the fluid, causing the mesh to be moved along
as well, so that a natural adaptivity is achieved, just as in SPH. On the
other hand, {\sm AREPO} employs the same concepts of spatial reconstruction
and Riemann-solver based flux computation as ordinary AMR codes, thereby
retaining their high accuracy for shocks and fluid instabilities, and their
low level of numerical dissipation. In addition, because the mesh can move
with the fluid, the numerical solutions of {\sm AREPO} are fully
Galilean-invariant and have lower advection errors compared to codes with a
stationary mesh.
 
The {\sm AREPO} code has already been successfully used in problems of first
star formation \citep{greif11a}, and extensions for radiative transfer
\citep{petkova11} and magnetohydrodynamics \citep{pakmor11} exist. Here we
apply this new method for the fist time to cosmological galaxy formation
simulations in representative parts of the Universe.  In what
follows, we are primarily
interested in clarifying to which extent the new hydrodynamic solver affects
the properties of galaxies when compared to the widely employed SPH code {\sm
  GADGET} \citep[last described in][]{springel05a}. {\sm GADGET} shares an
identical high-resolution Tree-PM gravity solver with {\sm AREPO}, allowing us
to study differences caused by the hydrodynamics alone, unaffected by possible
systematics in the gravitational dynamics of the collisionless components,
which were present in previous comparisons of mesh-based and SPH techniques
\citep[e.g.][]{oshea05,heitmann08}.  Furthermore, we include an identical
implementation of the basic physical processes that govern the evolution of
the gaseous component in galaxy formation, such as radiative cooling and
heating, star formation and feedback.  Finally, we let {\sm AREPO} carry out
refinement and de-refinement operations on cells (if needed despite the moving
mesh which normally keeps the mass per cell constant to good accuracy) in
order to guarantee that the masses of cells never deviate significantly
from a target given by the particle mass in the corresponding SPH
comparison run.  This ensures that we compare runs carried out with the two
codes at about the same mass resolution.

 In the last few years significant strides have been made in
  modeling disk galaxy formation with cosmological zoom-in simulations
  based on the SPH technique \citep[e.g.][]{governato10, guedes11}.
  These simulations often contain millions of resolution elements per
  halo and incorporate feedback processes that prevent an excessive
  collapse of baryons into galactic components. However, these
  successes in reproducing properties of observed galaxies have been
  reached for individual halos over a limited mass range. It is not
  clear yet if the same model would be able to reproduce the observed
  population mix of galaxies over a wide range of properties.  We
  stress, however, that our goal here is not to model galaxies with
  the highest possible resolution in a single system or to produce the
  most realistic looking galactic disks.  Our comparison strategy is
  designed to detect systematic differences (that are projections of
  inaccuracies in used simulations techniques) for a large population
  of galaxies.  While strong galactic feedback is an important
  component in galaxy formation and can induce large variations in the
  results \citep[e.g.][]{scannapieco12}, we here prefer an extremely
  simple subresolution treatment of feedback which can be implemented
  numerically in a well-posed and identical way, both in our
  moving-mesh code and in SPH. Our strategy attempts to ensure a clean
  comparison by avoiding a modeling of strong feedback processes,
  which would significantly complicate the interpretation of
  differences induced by the underlying hydrodynamic technique because
  parameterizing such feedback in a code agnostic fashion is quite
  difficult.

In this Paper II of a series, we focus on an analysis of the properties of
galaxies and their haloes throughout cosmic time, and discuss differences that
the two simulation techniques imprint on the resulting population of galaxies.
In \citet[][hereafter Paper I]{vogelsberger11}, we discuss global properties
of baryons in different phases, the cosmic star formation history and various
code characteristics, including technical details of our updated versions of
the moving-mesh code {\small AREPO} and our SPH code {\small GADGET}.
Finally, in a further companion paper \citep[][hereafter Paper III]{sijacki11}
we use idealized setups of situations relevant for galaxy formation, such as
gas cooling in haloes and stripping of infalling galaxies to demonstrate
where and how differences between the two codes occur, which helps us to
understand and interpret the differences we find in a fully cosmological
environment.

 We note that the conclusions reached in this paper directly apply
  only to standard formulations of SPH \citep[as described, for
  example in][]{springel02}. It is now well established that noisy and
  inaccurate pressure gradient estimates \citep[for example
  see][]{read10}, an absence of mixing at the particle level, and
  comparatively large artificial viscosity lead to a number of
  accuracy problems in SPH, in particular a limited ability to model
  fluid instabilities \citep{agertz07, abel11} and problems in
  representing subsonic turbulence \citep{bauer12}, which are likely
  the most important factors contributing to the differences we
  identify in this paper.  We emphasise that this `standard SPH' method or
  quite similar variants thereof have been used in the majority of the
  works on cosmological simulations of galaxy formation to date (both
  using large boxes and zoom-ins).  A number of recent efforts proposed
  extensions or modifications of the traditional SPH formulation that
  aim to improve the accuracy of SPH \citep[e.g.][]{wadsley08,
    price08, hess10, read10, read12, abel11, saitoh12}. Our results do
  not necessarily apply to these new modified methods. They have yet
  to be tested for cosmological simulations of galaxy formation, and
  it remains to be seen whether they resolve the differences we find
  here in our comparison of standard SPH as implemented in GADGET3
  with a more accurate hydrodynamic scheme such as the one implemented
  in AREPO.

The structure of this paper is as follows. In Section~\ref{sec:methods}, we
describe our simulation setup, the physical processes included, the initial
conditions, and the most important code modifications and methods used to
identify haloes and galaxies. In Section~\ref{sec:results}, we present
properties of galaxies and haloes in our moving-mesh and SPH simulations. In
particular, we discuss the baryon content of haloes, galaxy mass functions, gas
fraction and star formation rates of haloes, disk sizes and specific angular
momenta of galactic baryons.  We interpret and discuss our new results from
the moving-mesh cosmological hydrodynamics in Section~\ref{sec:discussion}.

\section{Methods}
\label{sec:methods}

In this paper, we describe only the essential setup information for our
simulations, in the interest of brevity. We refer readers to the original code
papers for a detailed discussion of {\sm GADGET} \citep{springel05a} and {\sm
  AREPO} (S10), as well as to Paper I for technical information about specific
parameters adopted for the codes and about their relative CPU-time
performance.

\subsection{Simulation suite and initial conditions}

We construct the initial conditions at redshift $z=99$ and evolve them to
$z=0$, adopting the following $\Lambda$CDM cosmology: $\Omega_m = 0.27$,
$\Omega_{\Lambda} =0.73$, $\Omega_b = 0.045$, $\sigma_8 = 0.8$, $n_s = 0.95$
and $H_0 =70\, \rm km s^{-1} Mpc^{-1}$ ($h=0.7$).  These parameters are
consistent with recent WMAP-7 measurements \citep{komatsu11} at the $1\sigma$
level, except for $n_s$ which is consistent at $2\,\sigma$ level.  The initial
power spectrum was approximated with the fit of \citet{eisenstein99}.

Our simulation suite follows a periodic box of size $20\,h^{-1}{\rm Mpc}$ at
three different resolution levels: $2\times 128^3$, $2\times 256^3$ and
$2\times 512^3$ particles/cells, respectively. Initially we start with an
equal number of gaseous fluid elements and DM particles in both codes. While
the number of DM particles stays fixed, the number of gaseous elements changes
over time. In {\small GADGET} this change owes to transformations of gas
particles into star particles during star formation events, keeping the total
number of baryonic resolution elements constant. Likewise, in {\small AREPO}
gaseous cells turn into star particles, but occasionally can also be refined
and de-refined if needed to keep the mass resolution always close to the
nominal resolution present in the initial conditions.  This leads to further
changes in the number of cells, but overall, the total number of baryonic
resolution elements stays approximately constant as well (see Paper I for more
details).

We note that we have used the same random phases and amplitudes for
overlapping modes on large scales for the different resolution levels, so that
our simulation set can be used for a resolution study of individual objects
formed in our $\Lambda$CDM realization.  In Table~\ref{table:simulations}, we
list the symbolic names of the simulation runs, together with their principal
numerical parameters, such as the initial particle/cell masses, the number of
computational elements for all of our simulations, and the gravitational
softening used for collisionless particles.

\begin{table*}
\begin{tabular}{cccccccc}
\hline 
Name & Code & Boxsize $[(h^{-1}{\rm Mpc})^3]$ & hydro elements & DM particles & $m_{\rm target/SPH} [h^{-1}{\rm M}_\odot]$ & $m_{\rm DM}[h^{-1}{\rm M}_\odot]$ & $\epsilon$ [$h^{-1}{\rm kpc}$]\\
\hline
\hline
A\_L20n512  & {\sm AREPO}  & $20^3$ & $512^3$ & $512^3$  &
$7.444\times 10^5$ & $3.722\times 10^6$ & $1$ \\
G\_L20n512   & {\sm GADGET} & $20^3$ & $512^3$ & $512^3$  &
$7.444\times 10^5$ & $3.722\times 10^6$ & $1$ \\
\hline
A\_L20n256  & {\sm AREPO}  & $20^3$ & $256^3$ & $256^3$  &
$5.955\times 10^6$ & $2.977\times 10^7$ & $2$ \\
G\_L20n256   & {\sm GADGET} & $20^3$ & $256^3$ & $256^3$  &
$5.955\times 10^6$ & $2.977\times 10^7$ & $2$ \\
\hline
A\_L20n128  & {\sm AREPO}  & $20^3$ & $128^3$ & $128^3$  & 
$4.764\times 10^7$ & $2.382\times 10^8$ & $4$ \\
G\_L20n128   & {\sm GADGET} & $20^3$ & $128^3$ & $128^3$  &
$4.764\times 10^7$ & $2.382\times 10^8$ & $4$ \\
\hline
\end{tabular}
\caption{Basic parameters of our simulation set. All calculations were
  performed in a periodic box with side length of $20\,h^{-1} {\rm
    Mpc}$. The number of hydrodynamical resolution elements (SPH
  particles or Voronoi cells, respectively) becomes smaller with time
  due to star formation. In the {\sm AREPO} simulations, additional
  re- and de-refinement operations are invoked to keep the cell masses
  close to a target gas mass $m_{\rm target}$, which we set equal to
  the SPH particle mass of the corresponding {\sm GADGET} run. The
  comoving Plummer-equivalent gravitational softening length
  $\epsilon$ is constant in {\sm GADGET}, but adaptive for gas cells
  in {\sm AREPO}.  }
\label{table:simulations}
\end{table*}

\subsection{Gravity, cooling and star formation}

The calculation of gravitational forces in all of our simulations is done with
the algorithms described in \citet{springel05a}. In short, we use a
combination of a particle-mesh approach \citep[e.g.][]{hockney81} to calculate
large-scale forces with the help of fast Fourier transforms, and a
hierarchical multipole approximation \citep[a tree algorithm,][]{barnes86,
  hernquist87} for short-range distances. This combination ensures a high
computational speed nearly independent of the clustering state, and a
uniformly high force resolution throughout the simulation volume.

For the long-range PM computation, we employ a mesh size twice as large as the
corresponding particle grid, meaning that for our L20n128, L20n256 and L20n512
simulations, the mesh has $256^3$, $512^3$ and $1024^3$ elements,
respectively.  It is important to emphasize that the gravitational force
calculation is done in a consistent way in  {\small AREPO} and {\small
  GADGET}, which enables us to cleanly identify differences caused by the
hydrodynamics alone.  The gravitational forces are softened at small distances
as described in \cite{hernquist89}, with the corresponding Plummer-equivalent
softening length fixed in comoving coordinates at $4\,h^{-1} \rm kpc$,
$2\,h^{-1} \rm kpc$ and $1\,h^{-1} \rm kpc$, respectively.  These values are
applied to all collisionless particles in {\small AREPO} and all particle
types in {\small GADGET}.  Gaseous cells in {\small AREPO} use an adaptive
softening calculated as $\epsilon_{\rm gas}=2.5 \times (3 V_{\rm
  cell}/4\pi)^{1/3}$  with an imposed minimum 
gravitational softening length equal to the fixed softening of our
corresponding GADGET simulations.
However, we have also run cosmological simulations where
the gas cells used a fixed softening length as well, finding that this choice
does not affect the properties of galaxies found in our runs in any
significant way (see also Appendix of Paper I). 

We include all radiative cooling processes important for a hydrogen and helium
gas of primordial composition; i.e.~line cooling, free-free emission, and
inverse Compton cooling off the cosmic microwave background.  The rate
equations are solved assuming collisional ionization equilibrium, as in
\citet{katz96}. We also include a uniform ionizing background radiation field
with the normalization and time dependence of \cite{faucher-giguere09} that
heats and ionizes the gas in an optically thin approximation \citep{katz96}.

Dense star-forming gas is treated via a simple two-phase sub-resolution model
as in \cite{springel03a}, which gives rise to an effective equation of state
for the dense gas. We use the same parameters for the sub-resolution model as
in \cite{springel03a}, which give a threshold density for star formation of
$n_h = 0.13 \, \rm cm^{-3}$. The star formation timescale scales with density
as $t_{\star}\propto \rho^{-0.5}$, and we set it to $2.1\,{\rm Gyr}$ at the
threshold density in order to match the normalization of the local relation
between star formation rates and gas surface density of galaxies
\citep{kennicutt98}. The conversion of gas mass to collisionless stellar
particles proceeds stochastically based on the estimated star formation
rate. We simplify the original procedure by \citet{springel03a} further so
that each gas particle spawns only one generation of stars. In addition, we
modify the original model by letting gas above the star formation density
threshold with a temperature higher than the effective two-phase temperature
to radiatively cool in the ordinary way onto the equation of state, instead of
applying the relaxation timescale of the multi-phase medium.

We note that the pressurization of the star forming medium does not drive any
outflow from the galaxy. For reasons of simplicity, we do not include
additional violent feedback mechanisms (e.g. galactic winds) in the present
study that could significantly lower the mass of a galaxy. Our model for
supernova feedback can hence be viewed as a `minimal feedback model' which
cannot be expected to yield a particularly realistic galaxy luminosity
function, especially at the faint-end. In order to compare galaxy masses and
luminosities of our predicted galaxy population with observations we will
consider simulations with galactic winds in {\sm AREPO} in forthcoming work.
We work here with this simplified model in order to allow us to identify
differences between the hydro solvers as cleanly as possible.  
 
\subsection{Cell regularization and mass homogenization in {\sm AREPO}}

An important aspect for the accuracy and performance of {\small AREPO} is the
use of a mesh-regularization scheme that ensures a reasonable quality of the
Voronoi mesh, in the sense that unfavourable cells with very high aspect
ratios are avoided.  As described in detail in Paper I, we have adopted a
new method where cells are only regularized if the maximum angle under which a
face is seen from the mesh-generating point of a cell becomes large.
Regularization here means that a small velocity component towards the
geometric center of a cell is added to the motion of the mesh-generating
point, which is otherwise just given by the gas velocity in the cell. 

In addition to the mesh regularization, we also employ on-the-fly mesh
refinement and de-refinement operations, where appropriate, in order to ensure
a constant mass resolution.  In principle, this is not strictly required in
{\small AREPO}, as the pseudo-Lagrangian nature of the scheme moves the mesh
along with the mass, bringing more resolution to regions where this is needed
most and maintaining roughly constant mass in cells.  However, the resulting
spread in mass tends to grow with time, and without explicit refinement and
de-refinement measures, a relatively wide spectrum of cell masses may
develop. An unwanted consequence would also be that star particles with
considerable variation in mass would be created, which might lead to
undesirable two-body heating effects.  Therefore, we enforce cells to have a
mass close to our desired target mass by splitting and merging individual
cells if needed. In practice, we allow cell masses to deviate by at most a
factor of $\sim 2$ from the desired target mass.  The latter is set equal to
the SPH gas particle mass in the run with the corresponding initial
resolution, as listed in Table~\ref{table:simulations}, so that we can ensure
an equivalent mass resolution in the {\sm AREPO} simulations.  We note
however that different refinement criteria other than mass could be chosen in
{\sm AREPO}, which has a similar flexibility as AMR codes in this respect.
Further details on the technical implementation of refinement and
de-refinement in {\sm AREPO} are provided in S10 and Paper~I.

\subsection{Galaxy and halo identification}

We identify haloes in our simulations with the standard FOF approach
\citep{davis85} with a linking length of 0.2 times the mean interparticle
separation.  The FOF algorithm is applied only to the DM particles, whereas
baryonic particles/cells are later assigned to their nearest DM particle and
included in the halo to which the corresponding DM particle belongs.
Self-bound concentrations of mass, termed `subhaloes', are then identified
within each FOF halo using the {\small SUBFIND} algorithm \citep{springel01b},
which has recently been modified to also account correctly for baryons in
subhaloes \citep{dolag09}.

The {\small SUBFIND} approach uses an adaptively smoothed density field
(estimated from adding up the individual contributions of all particle/cell
types) to find isolated overdense regions by means of an excursion set
technique.  Each substructure candidate identified in this way is then
subjected to a gravitational unbinding procedure, leaving only the self-bound
material in the subhalo. This effectively produces a list of subhaloes for each
main FOF-halo. The bound part of the smooth background FOF-halo (after all
embedded substructures have been removed) is defined as the `main
subhalo'. For the centre of each FOF halo (selected as the minimum of the
gravitational potential of the main subhalo) we also determine virial mass
estimates with the spherical overdensity algorithm.  Following a common
convention in the literature, we adopt $M_{\rm halo}$=$M_{200}$ as estimate of
the halo mass, where $M_{200}$ is the mass contained within a spherical region
of radius $R_{200}$ around the centre with overdensity of 200 with respect to
the critical density.  We use the term `central galaxies' for the galaxies in
the central parts of the main subhalo of a given FOF halo.

Galaxy identification can be a complex task, especially in {\sm GADGET}, where
a large number of cold gaseous clouds form in massive haloes at large radii
\citep{kaufmann09, keres09a, puchwein10}. {\small SUBFIND} tends to associate
those with the central galaxies, which can skew some of their properties, like
their angular momentum, half-mass radius and, in extreme cases, even their
total mass.  In order to protect against this effect and to make the galaxy
identification closer to what would be inferred observationally, we adopt a
few further criteria after subhalo identification in order to select the
galactic baryons and construct the galaxy catalogue. Specifically, we assume
that galaxies include only reasonably cold gas with temperature $T <
30,000\,\rm K$, apart from the star-forming gas at high-density, whose
effective temperature can be somewhat higher (due to the pressurization of the
two-phase medium). We also require gas to have baryonic overdensities higher
than 2000 which allows us to include all of the star forming gas but also
extended  gaseous disk outside of the star-forming region.  Furthermore, we
apply a radial cut of $r < 30 \rm{kpc/h}$ comoving  to both gas and stars to
avoid that gaseous clouds and filaments in the  outskirts of haloes, or
substantial numbers of intra-group/intra-cluster stars \citep{puchwein10} are
included in our definition of a galaxy. While these criteria are clearly
crude, they are motivated by observed properties of galaxies that contain
mostly cold dense gas and stars. To exclude poorly resolved galaxies we
concentrate on galaxies more massive than $256\times m_{\rm target}$, or when
selecting by halo mass, we use $M_{200} = 256\times m_{\rm target}\times
\Omega_m/\Omega_b$ as our adopted resolution limit.

\section{Results}
\label{sec:results}

\subsection{Baryon content of haloes}

\begin{figure*}
\centering
\includegraphics[width=0.99\textwidth]{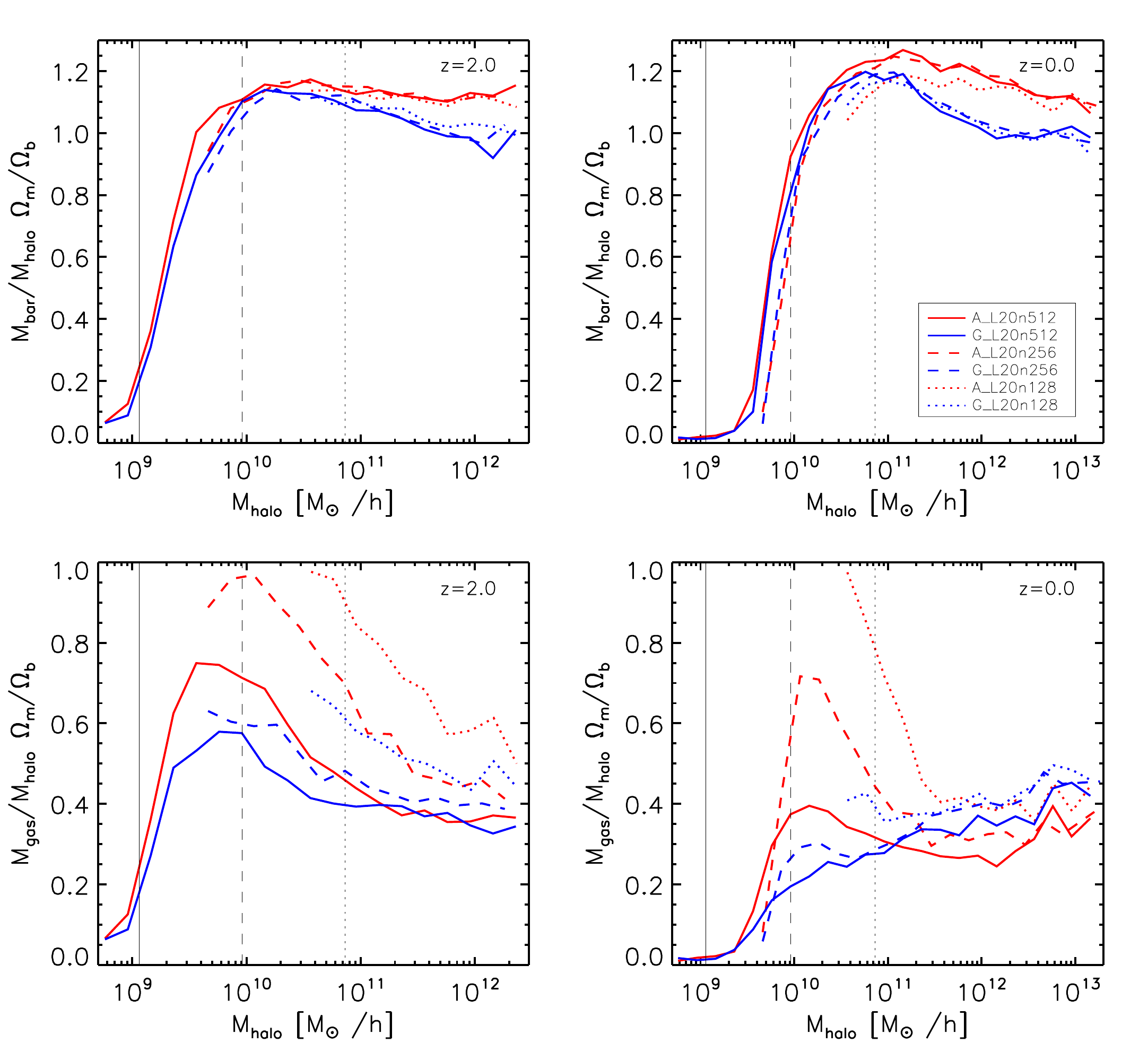}
\caption{Baryonic content of haloes for our simulation set at $z=2$
  (left panels) and $z=0$ (right panels). The lines show the total
  amount of baryons in haloes normalized by the baryon to total matter
  ratio in our adopted cosmology. The panels on top compare the total
  baryonic content in haloes of {\rm AREPO} simulations (shown in red)
  with those in corresponding {\sm GADGET} runs (blue
  lines). Different resolutions are displayed with different line
  styles, as labeled.  In the bottom panels, we show the same
  comparison for the baryons in gaseous form (i.e.~excluding stars),
  normalized the same way. The different vertical lines indicate a
  mass of $256 \times m_{\rm target} \times \Omega_m/\Omega_b$ for
  each resolution.}
\label{fig:hbaryons}
\end{figure*}

In Paper I, we have shown that the dark matter halo mass functions agree very
accurately between our different simulation techniques, and also converge well
as a function of numerical resolution. Here we concentrate on the baryonic
properties of these haloes.  In Figure~\ref{fig:hbaryons}, we show the mass
fraction of baryons and of gas in haloes as a function of parent halo virial
mass, comparing {\sm AREPO} with {\sm GADGET} at different resolutions.  Both
fractions are expressed in units of the value $\Omega_b/\Omega_m$ for the
universe as a whole, which would be the expected mass ratio if the baryons
would perfectly trace dark matter. Figure~\ref{fig:hbaryons} illustrates that
this is only approximately true, because the baryonic fraction is a function
of halo mass and often deviates significantly from unity.\\

In fact, in low mass haloes below $\sim 10^{10}\, \msunh$ at $z=0$, or $\sim
5\times 10^{9}\, \msunh$ at $z=2$, the fraction of collapsed baryons is
significantly lower than the universal value. This is a well known consequence
of the IGM photo-heating by the UV background, effectively weakly
``pre-heating'' the gas, modifying the peaks in the cooling 
curve \citep{katz96} and preventing collapse and accretion of gas into low
mass haloes \cite[e.g.][]{efstathiou92, thoul96, gnedin00}.  For the UV
background intensity used in this paper a rapid decline in baryonic fractions
is in fact expected below $\sim 5\times 10^{9}\, \msunh$ at $z=2$ and $\sim
10^{10}\, \msunh$ at $z=0$ \cite[see also][]{hoeft06,okamoto08,faucher-giguere11b}.
The mass scale at which
haloes transition into the baryon-poor regime is consistent between SPH and
{\sm AREPO} and slightly increases when the resolution is poor (see the
$2\times 128^3$ runs). 

More massive haloes have a baryonic content relatively close to the universal
value, but with interesting trends as a function of halo mass, redshift and
simulation technique.  At $z=2$, the baryonic fraction peaks  15\% above the
universal value at halo masses around $2\times 10^{10}\, \msunh$ in both
codes. However, while in {\sm GADGET} this drops again to about unity for
$M_{\rm halo} > 10^{12}\, \msunh$, in {\sm AREPO} there is still a baryon
excess of around 10\% in such massive haloes.  Similarly, at $z=0$ the peak in
the baryonic fraction has moved up a bit in halo mass scale to $\sim 10^{11}\,
\msunh$ and is even a bit higher.  At this epoch, well-resolved {\sm AREPO}
haloes can have a 25\% excess of baryons while in {\sm GADGET} the peak is
slightly lower. In more massive haloes, $M_{\rm halo}>10^{12}\, \msunh$, the
total baryon content drops again to around unity in SPH but remains around
10\% higher than the universal value in {\sm AREPO}.  This systematic
difference in massive haloes is very robust to changes in resolution.  While
this result may at first appear a bit puzzling, we note that such a systematic
offset between the baryon fractions in SPH and mesh-based hydro codes has in
fact been noticed earlier. Already the non-radiative simulations of the Santa
Barbara cluster comparison project \citep{frenk99} found that SPH runs had
only $\sim 90\%$ of the expected baryons, whereas grid codes found a ratio
close to unity \citep[see also][]{oshea05,crain07}. If radiative cooling is
included, simulations have been found to produce a higher baryon fraction at
any given radius, including out to the virial radius
\citep[e.g.][]{kravtsov05,ettori06}.  Our results are thus consistent with
these earlier studies, even though we find a somewhat larger difference
between SPH and mesh-codes compared to \citet{kravtsov05} when cooling is
included. Note however that our small simulation box does not probe the mass
scale of galaxy clusters examined in \citet{kravtsov05}, where the effect is
likely smaller.
 
We speculate that the origin of the difference between SPH and the mesh codes
ultimately lies in the higher dissipative heating rate we found for SPH haloes
in Paper~I, which effectively leads to a higher pressurization of the baryons
in the gaseous atmospheres, thereby pushing them out slightly in SPH relative
to the mesh-codes.  When cooling is included, the baryons removed in the halo
centres allow the outer gas to slide in, thereby increasing the baryon
fraction inside the virial radius. Since the large {\sm AREPO} haloes cool more
gas, this effect is stronger in the mesh code.  Also, the reduction in
strength of this effect towards higher masses can be understood as a
consequence of a declining cooling efficiency. In fact, when our results are
extrapolated to cluster scales, they appear entirely consistent with findings
in galaxy cluster simulations that include radiative cooling, and in
particular, with the weak trend of a falling baryon fraction with increasing
mass found there \citep{borgani09}.  Before moving on we remark that the
difference in baryon fraction also means that there is a slight but systematic
difference in the virial masses of matching large haloes in {\sm AREPO}
simulations compared to {\sm GADGET}. This difference is however $<$10\%
in most cases which means that the differences between the codes presented
as bin averaged results at a given halo mass are essentially the same as 
they would be for binned matched samples of halos.

\begin{figure*}
\includegraphics[width=0.49\textwidth]{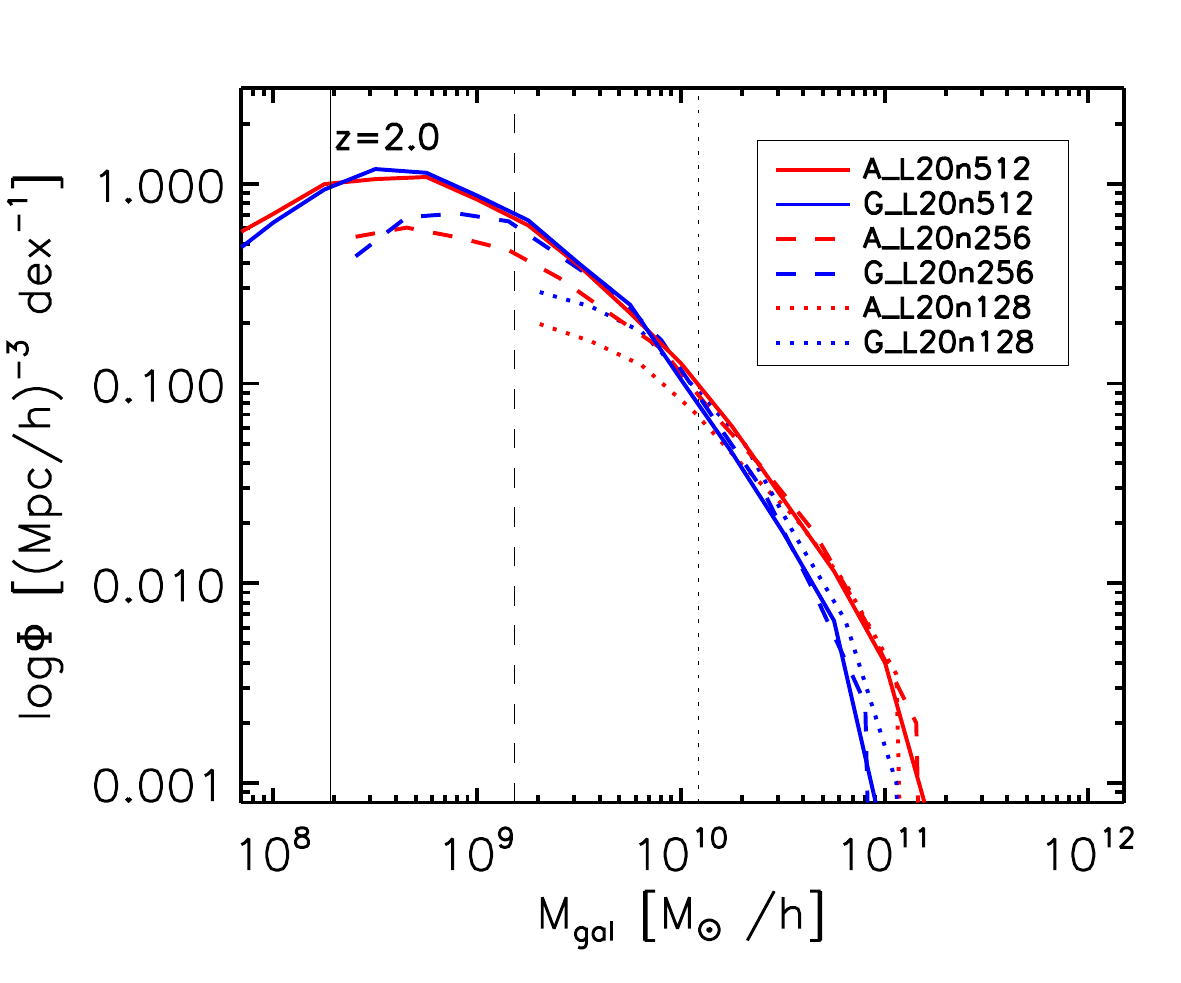}
\includegraphics[width=0.49\textwidth]{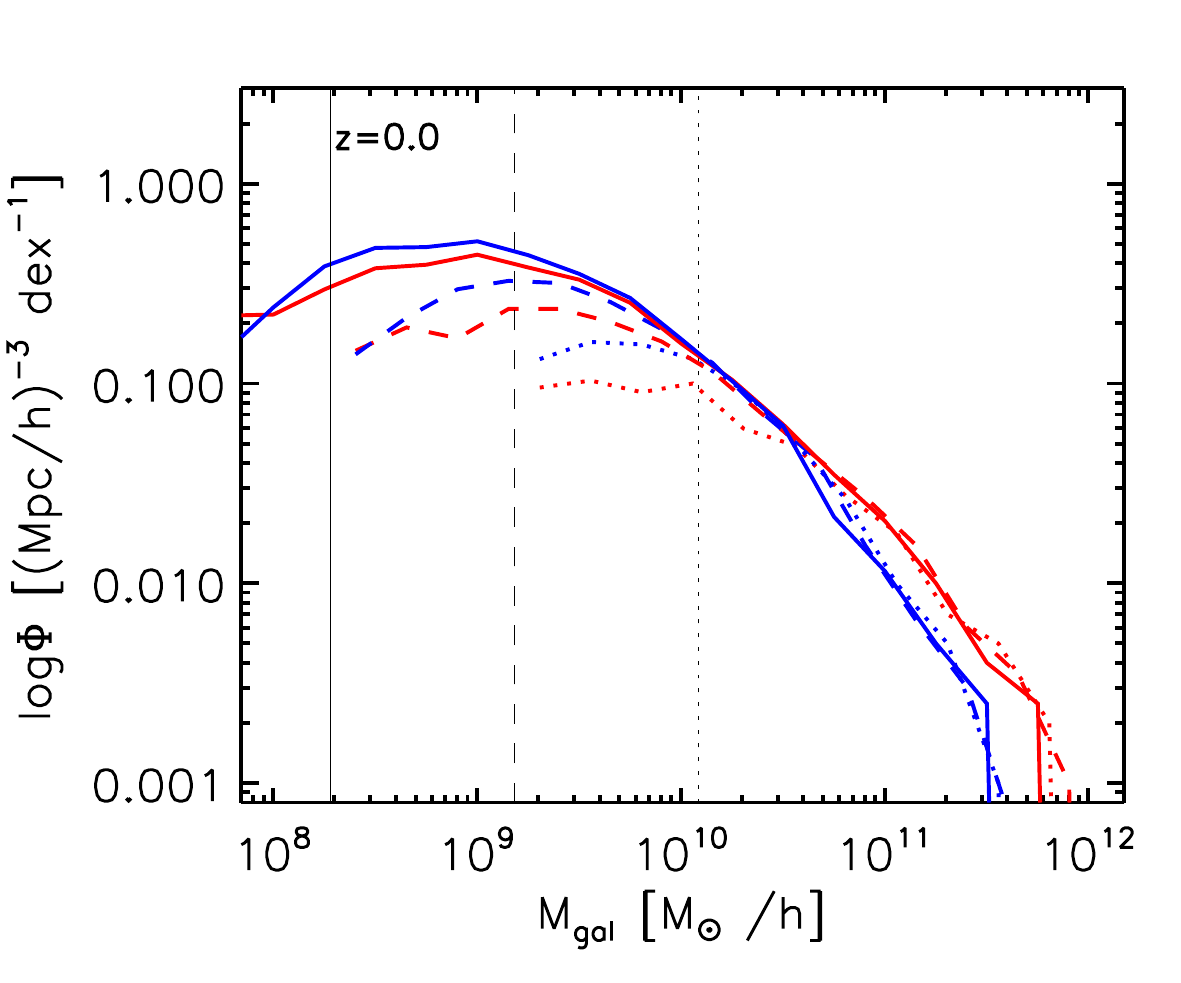}
\caption{Baryonic mass function (gas+stars) of galaxies identified in our
  simulation set at $z=2$ and $z=0$. Runs with {\sm AREPO} are shown as red
  lines, those with {\sm GADGET} as blue lines. The different resolution
  levels are indicated with different line styles, as labeled. The vertical
  lines indicate  the mass of 256 $\times$ target cell/particle mass.  At high
  redshift and at high resolution, the mass functions are similar between the
  codes.  At $z=0$, the differences are larger, as we discuss in the text. }
\label{fig:gmf}
\end{figure*}

\begin{figure*}
\includegraphics[width=0.49\textwidth]{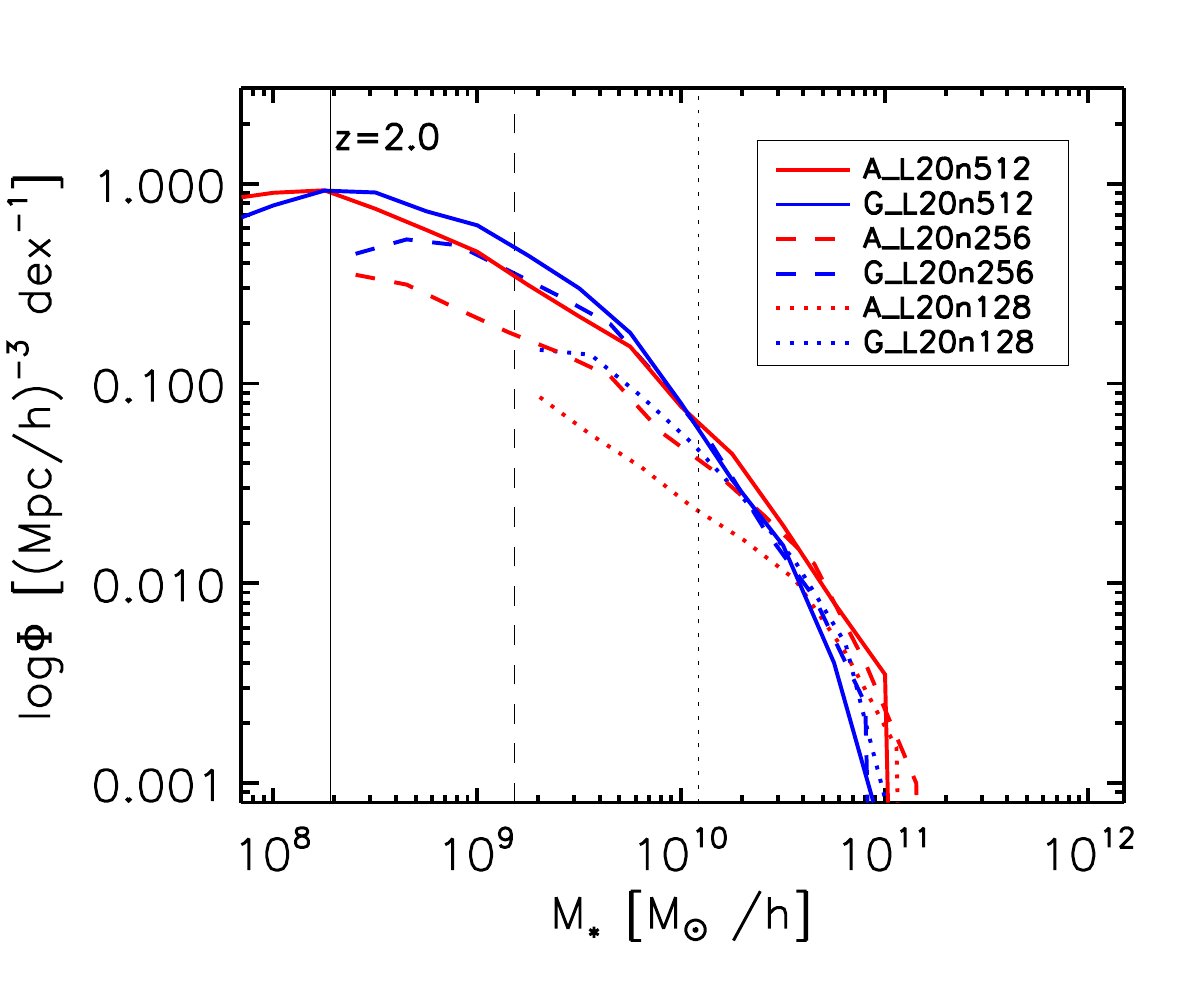}
\includegraphics[width=0.49\textwidth]{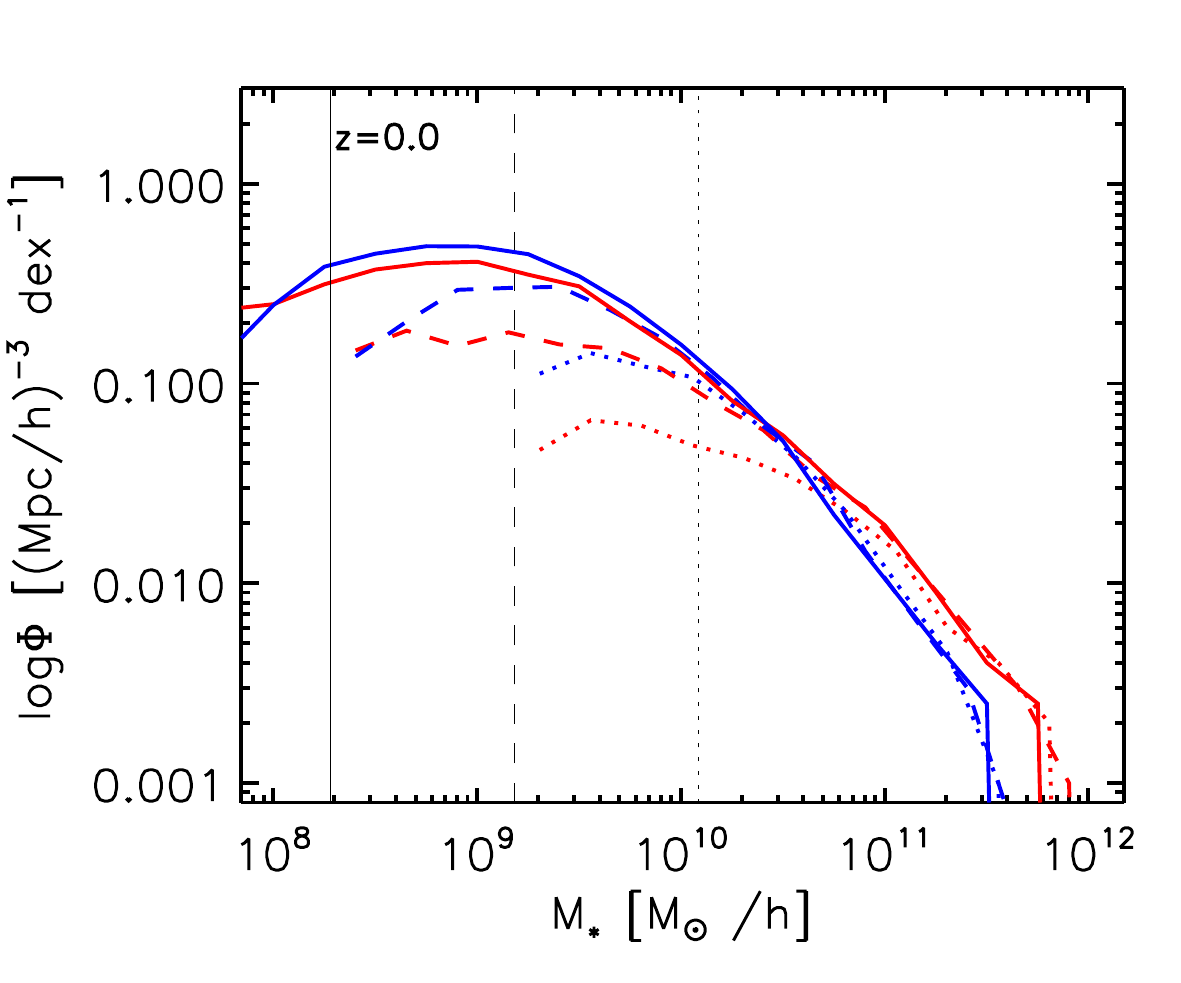}
\caption{Stellar mass functions of galaxies for our whole set of
  simulations at $z=2$ and $z=0$. As in Figure~\ref{fig:gmf}, we use red
  lines for {\sm AREPO} and blue lines for {\sm GADGET}, with
  different line styles corresponding to different resolutions, as
  labeled. The vertical lines indicate a mass of $256 \times m_{\rm target}$ for the corresponding
  resolution level.  We note that the differences in the stellar mass
  function are slightly larger than the differences in the galaxy mass
  function, caused by systematic offsets in the gas fractions of
  galaxies in {\small AREPO} and {\small GADGET} simulations.  }
\label{fig:smf}
\end{figure*}

In the bottom panels of Figure~\ref{fig:hbaryons}, we consider the gas fractions
of haloes, at redshifts $z=2$ and $z=0$.  The gas fraction of haloes is quite
low, amounting to $\sim 40\%$ of the universal baryon fraction for large
haloes, implying that most of the baryons in these haloes are actually locked
up in stars.  In fact, in our high resolution simulations at $z=0$ this is
true for all haloes above the UV background heating limit.  This extremely
efficient conversion of gas into stars is a consequence of the absence of very
strong feedback processes in our simulations that could eject material from
the bottom of the potential well of haloes in the form of winds or
superwinds. Instead, the amount of stars formed is primarily regulated by the
cooling efficiency of haloes, and the differences we find between the
simulation techniques reflect differences in this cooling
efficiency. In particular, there is less gas in large haloes at late times in
{\sm AREPO}, despite the fact that these haloes contain more baryons over-all
compared with {\sm GADGET}.  This is consistent with the stronger cooling
flows and higher star formation rates we found in these objects (see also
Paper I). On the other hand, in small mass systems just above the cooling
threshold, {\sm AREPO} retains a bit more gas, both at $z=0$ and $z=2$,
suggesting that here the mesh code is forming stars somewhat less
efficiently. In this regime, the moving-mesh code is also more sensitive to
numerical resolution than SPH, showing a stronger increase in the remaining
gas fraction if the resolution is degraded.  As we will discuss later on in
the paper, this is in part a consequence of differences in disk sizes of
galaxies and the inability of low resolution simulations to resolve high
density gas, both affecting the star formation efficiencies.

\begin{figure*}
\includegraphics[width=1.\textwidth]{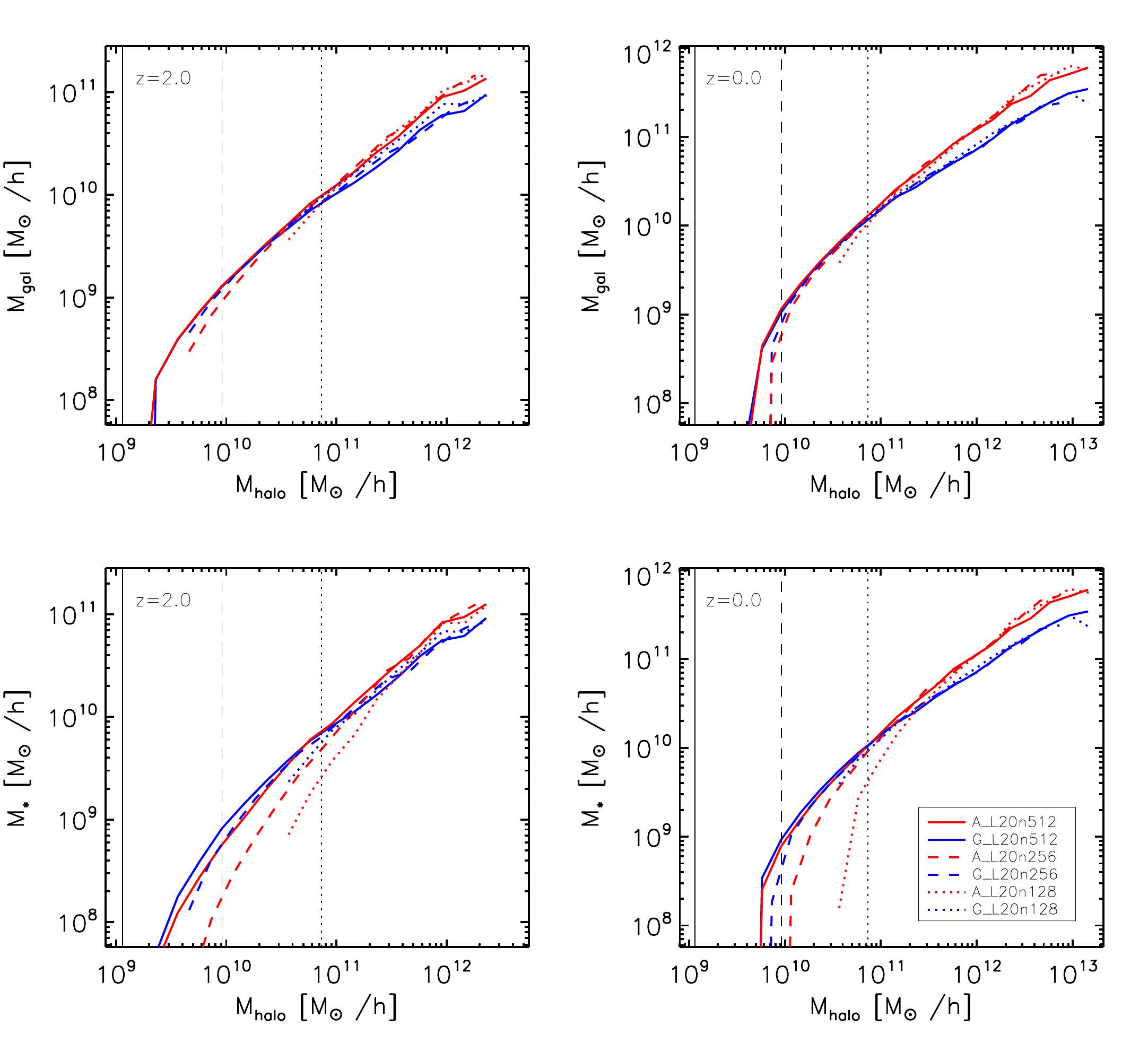}
\caption{Dependence of galaxy and stellar mass of central galaxies on
  halo mass at $z=2$ and $z=0$. Upper panels show the galaxy mass while
  the lower panels show the stellar mass. We compare results for {\sm
    AREPO} (red lines) and {\sm GADGET} (blue lines) at different
  resolutions, as labeled.  It is clear that at high resolution the
  galaxy masses of {\sm AREPO} and {\sm GADGET} galaxies are indeed
  very similar in haloes with $M_{\rm halo} \lesssim 10^{11}\, \msunh$,
  but in more massive haloes the galaxies in the moving-mesh
  simulations are systematically more massive.  This relative mass
  difference increases slowly with halo mass. Vertical lines show $256
  m_{\rm target} \Omega_m/\Omega_b$ at different resolutions.}
\label{fig:gal_to_halo}
\end{figure*}

\subsection{Galaxy mass functions}
\label{sec:mf}

Figures~\ref{fig:gmf} and \ref{fig:smf} show galactic baryonic mass functions
(stars plus gas associated with galaxies) and galactic stellar mass functions
at $z=2$ and $z=0$ for all of our simulations.  For our highest resolution
simulations, the galaxy mass functions are in relatively good agreement at
$z=2$ for both simulation techniques, up to masses of several $10^{10}\,
\msunh$. At the high-mass end, the masses of {\small AREPO} galaxies are
however systematically higher than those of {\small GADGET} galaxies, by a
factor of $\sim 1.5-2$.  

The agreement at the low mass end, is arguably to be expected.  Here the
galaxies accrete most of their gas rapidly with rates comparable to the gas
accretion rates onto haloes \citep{keres05, ocvirk08}. Furthermore, in this low
halo-mass regime, smooth accretion dominates over growth by mergers with other
galaxies \citep{murali02, keres05, guo08}, owing to rapid gas infall and a
limited mass range available for progenitors close to the UV-background
suppression mass.  Therefore, there is comparatively little opportunity in
these low mass haloes that differences in the treatment of hydrodynamic
interactions significantly affect the growth of galaxies. However, we note
that this is only approximately true as there are clearly some differences
when then resolution is low, as we discuss below.

\begin{figure*}
\includegraphics[width=1.\textwidth]{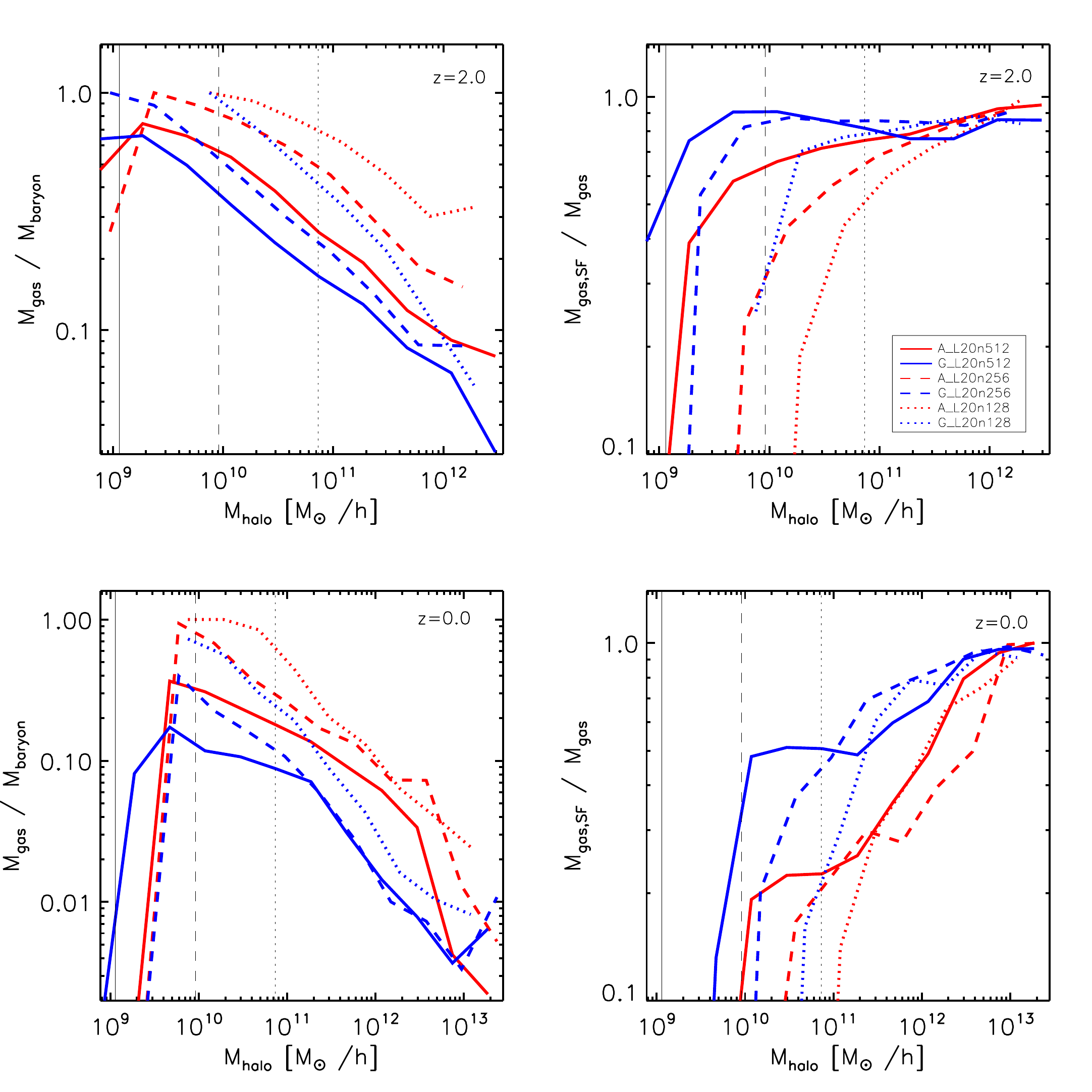}
\caption{Gas fractions of central galaxies, $M_{\rm gas}/M_{\rm gas+stars}$,
  as a function of halo mass for our simulation set (left panels).  We use red
  lines for {\sm AREPO} and blue lines for {\sm GADGET}, with different line
  style corresponding to different resolutions, as labeled. The gas fractions
  are systematically higher in {\sm AREPO} than in {\sm GADGET}. We also show
  the fractions of galactic gas that reach star-forming densities $M_{\rm
    SFgas}/M_{\rm gas}$ (right panels) to illustrate how this depends on halo
  mass, numerical technique and resolution.  Even though both quantities are
  resolution dependent, the overall systematic trends clearly show higher
  galactic gas fractions and a lower relative fraction of gas in star-forming
  regions in {\sm AREPO}.}
\label{fig:gas_fractions}
\end{figure*}

At $z=0$, some more prominent differences become apparent. While the highest
resolution simulations again show reasonably good agreement in the  galaxy
mass function below $2-3\times 10^{10}\, \msunh$, at higher masses, the
differences in the galaxy mass reach a factor of $\sim 2$.  While both
numerical methods converge to the {\em same} galaxy mass function at the low
mass end as the resolution is improved, at the high mass end, {\small AREPO}
and {\small GADGET} still converge individually, but they do so to a {\em
  different} converged result.  This systematic difference manifests itself in
higher galaxy masses in {\sm AREPO}, which we argue is linked to a stronger
cooling efficiency in massive haloes, an issue we will return to in
Section~\ref{sec:halo_profiles}.  

We find that {\small GADGET} simulations show convergence in the
galaxy mass function for galaxies more massive than $\sim 250$
baryonic particles (about $8\times N_{\rm ngb}$ for our runs).  To
compare this resolution limit to the convergence achieved with {\small
  AREPO}, we use vertical lines in Figures~\ref{fig:gmf} and
\ref{fig:smf} to mark the mass scale that corresponds to 256 times the
target cell mass. {\sm AREPO} appears to converge more slowly,
requiring up to $\sim 5$ times higher mass at $z=2$. At $z=0$, the
resolution dependence of both codes is even stronger and the
convergence of the high-resolution simulations is somewhat poorer than
at $z=2$, likely caused by differences in resolving the halo scale
affected by the UV background.  We caution that in the
  cosmological simulations without galactic outflows improved
  resolution always leads to a higher fraction of baryons locked up in
  stars, especially before reionization, as there is no efficient
  process to regulate formation of lower mass objects. Furthermore
  details of galactic structure will depend, in part, on resolution,
  especially since the hierarchical buildup proceeds from smaller
  objects that are always less well-resolved. Therefore, precise
  criteria for convergence depend on the galactic property one is
  interested in. Note also that faster convergence of SPH for some of
  the properties does not guarantee that it is a more accurate
  numerical scheme. In fact, in principle it is possible that one may
  even converge to a wrong result \citep[e.g.][]{springel10b}. This is 
  because in practice once
  typically only increases the total particle number and spatial
  resolution in an SPH convergence study. However, an increase in the
  number of neighbours would also be needed for convergence in order
  to progressively beat down the SPH noise \citep{rasio00}, as we discuss 
  in detail in Paper I.

The galaxy stellar mass functions shown in Figure~\ref{fig:smf} give a
qualitatively similar picture, if anything, the differences between the codes
are even more evident. Here the larger offsets between the codes at low masses
are caused by differences in galactic gas fractions at a given galaxy mass, as
we discuss more in Section~\ref{sec:gas_fractions}.  At the high mass end,
where stars form the bulk of galactic mass, a strong systematic offset in
stellar masses is again clearly evident.

To summarize, while some moderate differences exist in low mass galaxies,
these cannot be unambiguously attributed to systematic effects of the
hydrodynamical solvers, owing to numerical resolution and galaxy finder
uncertainties in this regime.  On the other hand, the differences we find
between the moving-mesh and SPH codes at the massive end are very 
significant and
robust. Here {\sm AREPO} produces more massive galaxies, which is one of the
key differences we will return to several times in the rest of the paper.

\subsection{Relation between galaxy and halo masses}
\label{sec:gal_to_halo}

Some of the trends we inferred from the galaxy and stellar mass functions are
more explicitly demonstrated in Figure~\ref{fig:gal_to_halo}, where we show
the  stellar mass and baryonic mass for central galaxies as a function of
their parent halo mass $M_{200}$.  The measurements make it clear that the
agreement of the baryonic masses of galaxies in our two simulation techniques
is indeed excellent in haloes of mass $M_{200} < 10^{11}\, \msunh$, whereas
galaxies in more massive host haloes are systematically more massive in {\sm
  AREPO} runs.

Because our galaxies have a relatively low gas fraction at $z=0$, the stellar
and baryonic masses are comparable at all halo masses at late times. In
contrast, at $z=2$, a lower amplitude of the stellar mass function when
compared with the galaxy mass function is apparent, caused by a much higher
gas content of galaxies at a given halo mass at this epoch.
The differences in the galaxy masses at low resolution are more dramatic 
which is caused by the sharply declining masses of poorly resolved galaxies 
in low mass host haloes. In this case,
simulations are not able to properly resolve galactic densities any more; in
fact, here the gravitational softening can become comparable to the galaxy
size, so that the collapse of baryons in low mass haloes is seriously affected.
We note that our findings also confirm previous results that the amount of
baryons accumulated in central parts of haloes substantially exceeds the
observationally inferred galaxy masses \cite[e.g.][]{guo10,behroozi10} if no
efficient feedback mechanism is invoked that either ejects gas from these
regions or somehow prevents the cooling in the first place
\citep{keres09b}. At the massive halo end, the {\sm AREPO} simulations make
this problem even more acute, strengthening the arguments for the need of very
strong feedback mechanisms.

\subsection{Galactic gas fractions}
\label{sec:gas_fractions}

In the left-hand panels of Figure~\ref{fig:gas_fractions}, we show the gas fractions $M_{\rm
  gas}/(M_{\rm gas}+M_{\rm stars})$ of central galaxies in our simulation
runs at $z=2$ and $z=0$. Lines give median gas fractions at a given parent
halo mass at the different resolutions.  At any given resolution, the gas
fractions are higher in {\sm AREPO} than in {\sm GADGET} runs, but the
magnitude of the difference depends on the resolution of the simulation. It is
interesting that even in our highest resolution simulations, the gas fraction of
{\sm AREPO} galaxies are systematically  higher than in {\sm GADGET}
indicating robust differences in galaxy  evolution between the codes.

\begin{figure*}
\begin{center}
\setlength{\unitlength}{1cm}
\includegraphics[width=8.\unitlength]{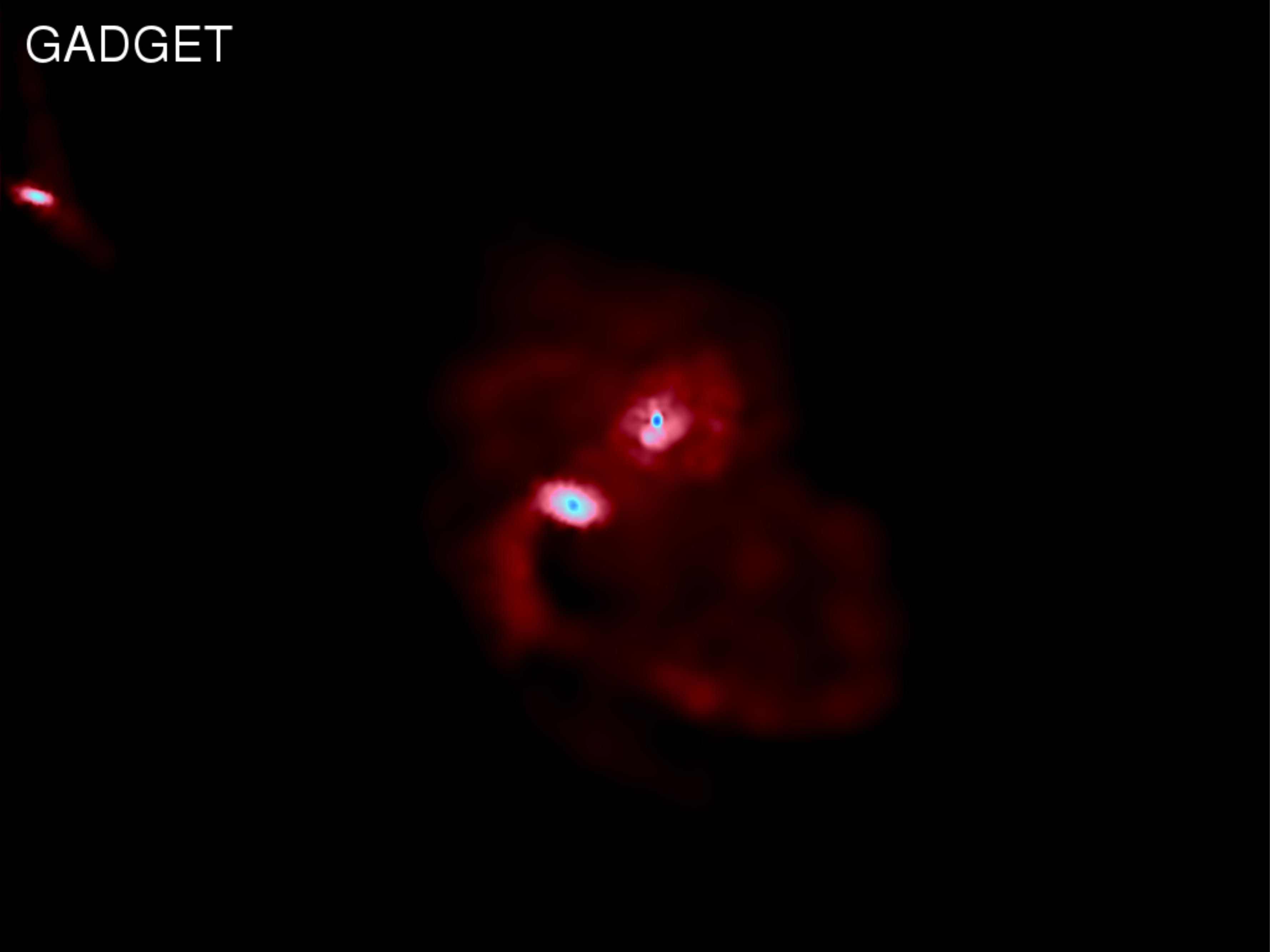}
\hskip 3pt
\includegraphics[width=8.\unitlength]{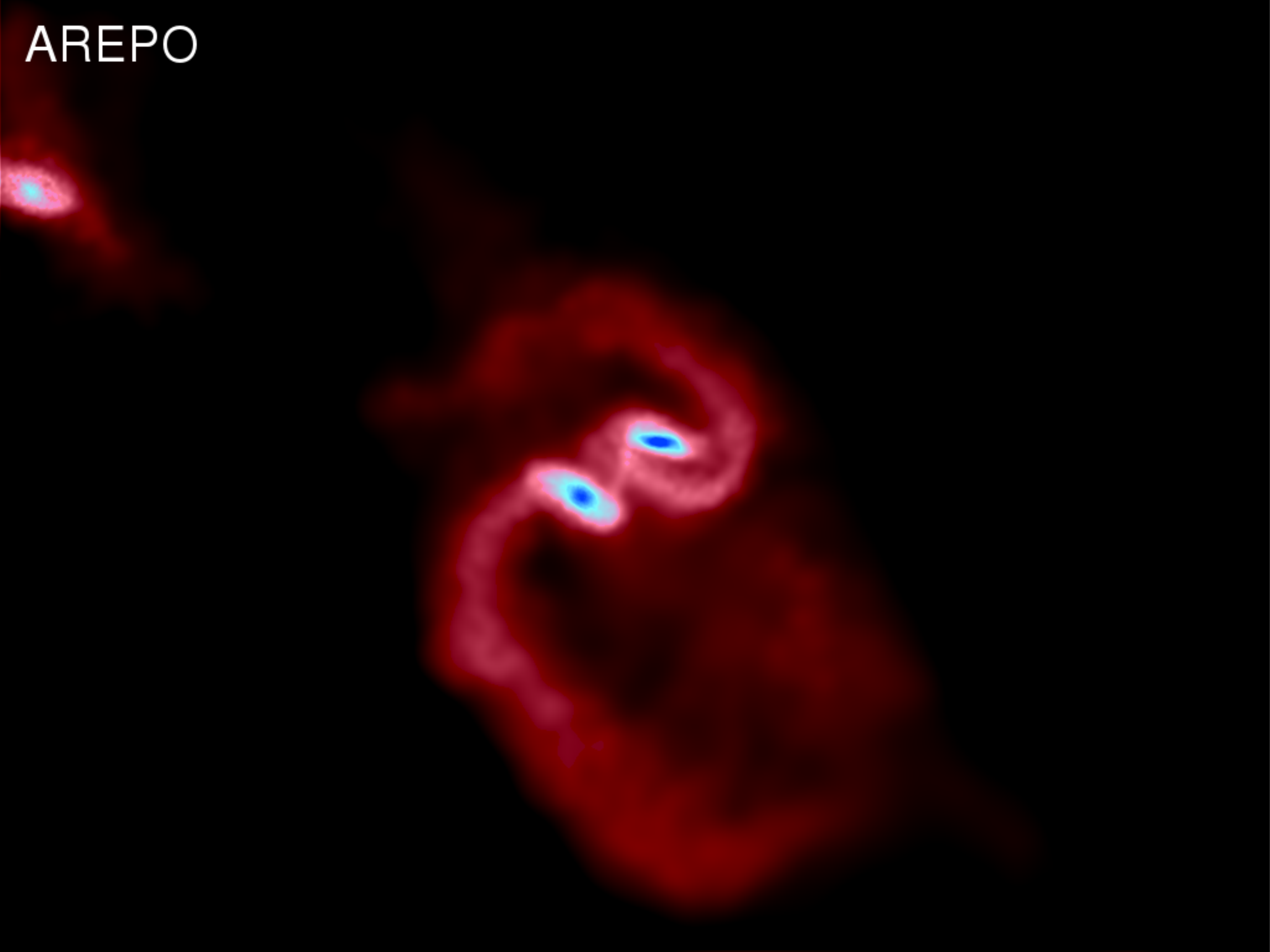}
\vskip 3pt
\includegraphics[width=8.\unitlength]{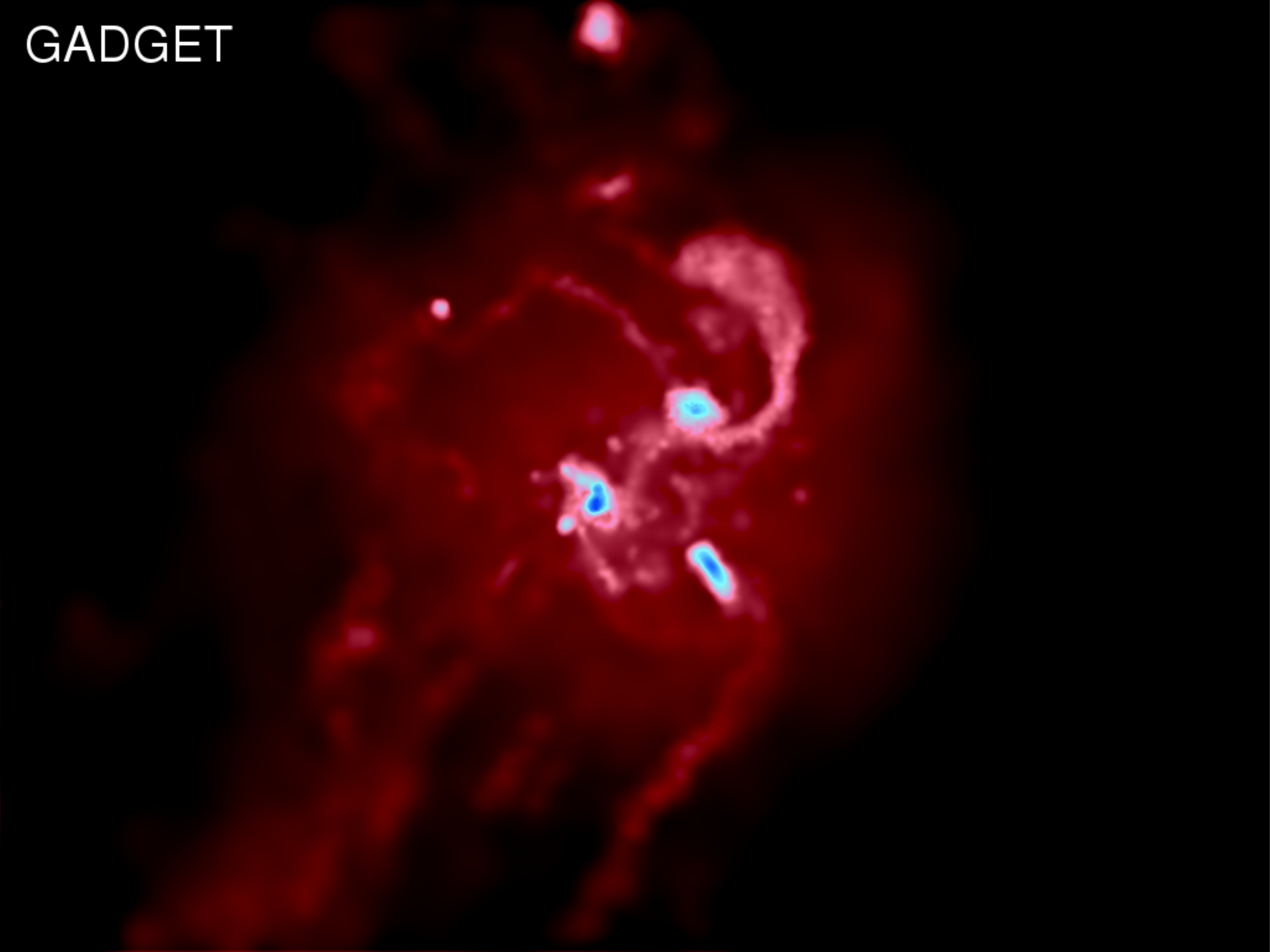}
\hskip 3pt
\includegraphics[width=8.\unitlength]{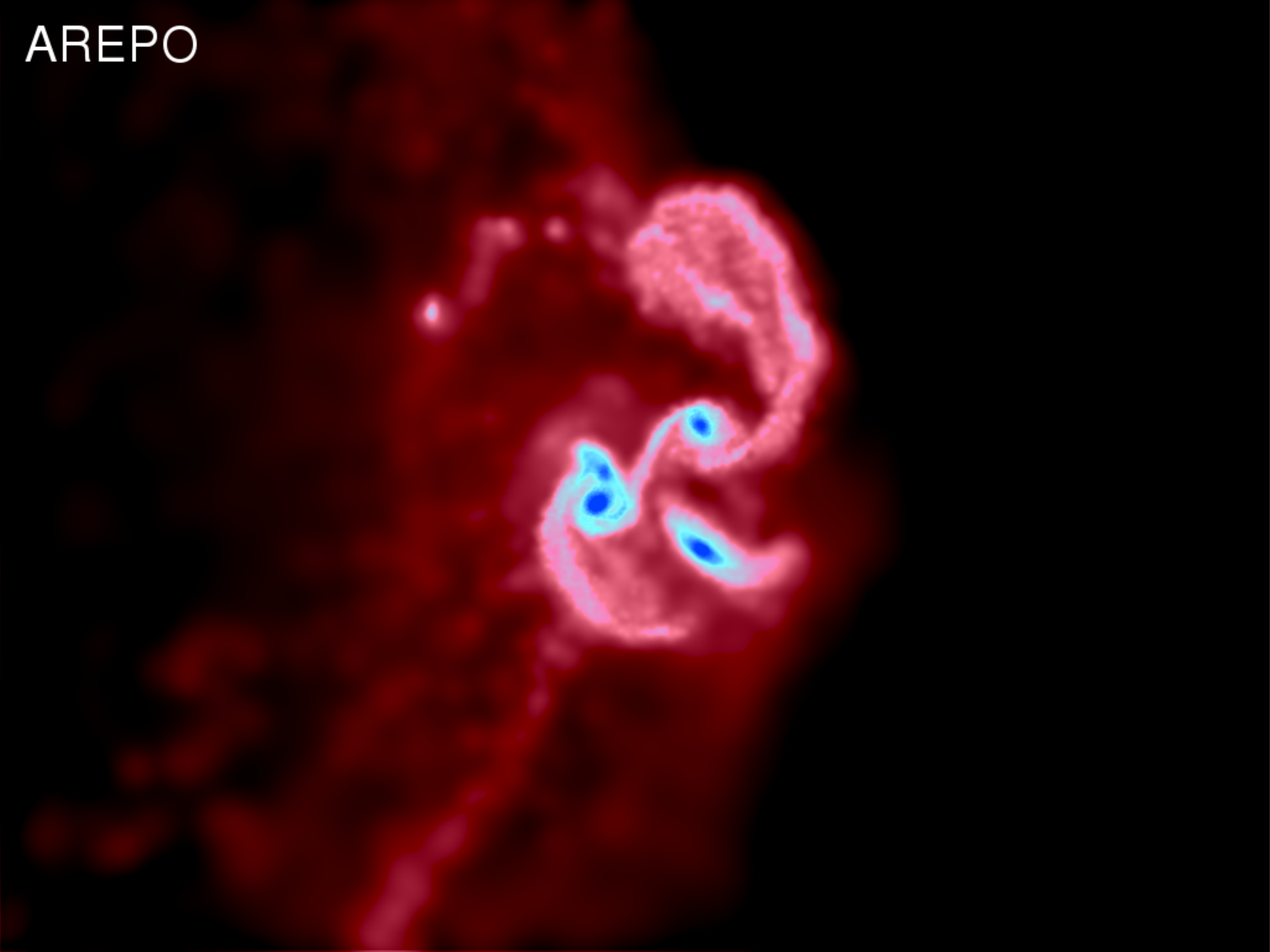}
\caption{Projected gas distributions of two interacting galaxy systems in {\sm
    GADGET} (left panels) and {\sm AREPO} (right panels) at redshifts
  $z=2$. The panels show regions of  100$\rm kpc/h$ comoving on a side. Upper
  panels show projections of gas in a major merger of two low mass galaxies
  while lower panels show a more complex merger involving three galaxies.  The
  tidal features, tails and bridges, seen in the gas are much more prominent
  in {\sm AREPO}, demonstrating that the galaxies simulated with the mesh code
  contain a larger amount of cold, rotationally supported gas.}
\label{fig:tidal} 
\end{center}
\end{figure*}

The fact that we find some resolution sensitivity is not too surprising as in
our lowest resolution runs the gravitational softening can be a sizable
fraction of the virial radius of low mass halo haloes and is comparable to the
sizes of the corresponding galaxies.  The higher resolution simulations are
expected to better resolve the dense gas in galaxies, leading to a higher star
formation rate, and consequently to a lower residual gas fraction at a given
halo mass.  Once the relevant densities are properly resolved, converged values
for the gas fractions should be reached, reflecting a balance between gas
supply and the depletion of gas by star formation.  However, the properties of
the galaxies and of their gas content will still be affected by their
hierarchical formation path.  The buildup of structure from quite poorly
resolved small sub-components can therefore impact the gas infall rates, and
also the density structure of much larger galaxies.  It is hence not
sufficient for convergence to achieve a sufficiently high resolution to
adequately represent a given galaxy, one also needs to represent well enough a
good part of the merger history leading to this object. This makes
hierarchical galaxy formation a particularly difficult numerical problem. 

It is clear that finite spatial resolution and the limited ability to follow
the full hierarchical buildup will affect both of the codes in some ways. One
might however expect numerical problems relevant to the density structure of
dense galactic gas to be more severe in SPH. This is because SPH relies on an
artificial viscosity in order to capture shocks and to produce the associated
entropy increase. It has been shown that the artificial viscosity can lead to
unwanted angular momentum transfer in the gas of a rotating galactic disk,
especially when the disk is sampled with a small number of particles
\citep{navarro97a,okamoto03}. The net effect is that the disk mass looses
angular momentum to material further out in the halo, making these systems
more compact in {\sm GADGET}, resulting in turn in a faster conversion of
baryons into stars and in lower residual gas fractions.

In
Figure~\ref{fig:tidal}, projected gas densities for two different interacting
galaxy systems in {\sm GADGET} and {\sm AREPO} are shown, supporting this
picture. The tidal features, such as bridges and tails seen in the gas are
prominent and well defined in {\sm AREPO}, demonstrating that the galaxies
simulated with the moving-mesh code contain a large amount of cold,
rotationally supported gas \citep{toomre72, donghia10}.
These features are much less prominent for the
same systems in {\sm GADGET}.  This is also consistent with our findings in
Section~\ref{sec:gal_radii} where we show that even in low mass haloes {\sm
  GADGET} galaxies are more compact than {\sm AREPO} galaxies.  We note that
since poorly resolved galaxies are also the progenitors of better resolved,
more massive galaxies, this effect can propagate to larger systems and induce
systematic differences in them.

The gas fractions in galaxies decrease with time in all simulations,
irrespective of the resolution and the code employed.  The bulk of this trend
is caused by a decreasing rate of gas supply into galaxies at a fixed halo
mass, as commonly predicted by semi-analytic models and cosmological
simulations \citep[][]{white91, murali02, hernquist03, keres05,
  dekel09}. Owing to the short (observationally motivated) star formation
timescale in our ISM sub-resolution model, the rate of conversion of dense gas
into stars is practically controlled by the rate of gas infall. However, this is
correct only to the extent that most of the galactic gas is at high enough
densities to form stars, which depends on the disk sizes.  Differences in disk
sizes will therefore influence the effective star formation timescale and the
gas fractions. The characteristic sizes of galactic disks at a fixed halo mass
increase with time (see below), enabling a larger amount of gas to reside in
a disk for a given star formation rate. This slows down the evolution of the
gas mass in the disk, resulting in a relatively modest drop in gas fractions
with time \cite[see also][]{dutton10}.

In the right-hand panels of Figure~\ref{fig:gas_fractions}, we also show the
fraction of galactic gas that resides in the star forming two-phase medium,
i.e. $M_{\rm gas,SF} / M_{\rm gas}$, as a function of halo
mass. At $z=2$, in central galaxies of low mass haloes, there is more gas
outside of the star forming region in {\sm AREPO} than in {\sm GADGET}. At
$z=0$, we find a similar difference, except for the most massive galaxies
where the fraction of star-forming gas is comparable between the codes.  Given
that {\sm AREPO} galaxies have also higher gas fractions, this implies that
their galactic gas is more extended, allowing a large quantity of gas to stay
below the star formation threshold. We confirm this in
Section~\ref{sec:gal_radii}, where we demonstrate that galaxies in the
moving-mesh code are more extended. It is interesting to note that in $\sim
10^{12}\, \msunh$ haloes at $z=0$ the gas fractions in {\sm AREPO} are a
factor of $\sim 2$ higher than in {\sm GADGET}, even though differences in the
fraction of star-forming gas are relatively minor. As it turns out these
galaxies are also accreting much more gas from their surrounding hot haloes
(see Section~\ref{sec:halo_profiles}), which contributes to their higher gas
fractions.

\subsection{Star formation rates of simulated galaxies}

\begin{figure*}
\includegraphics[width=1.\textwidth]{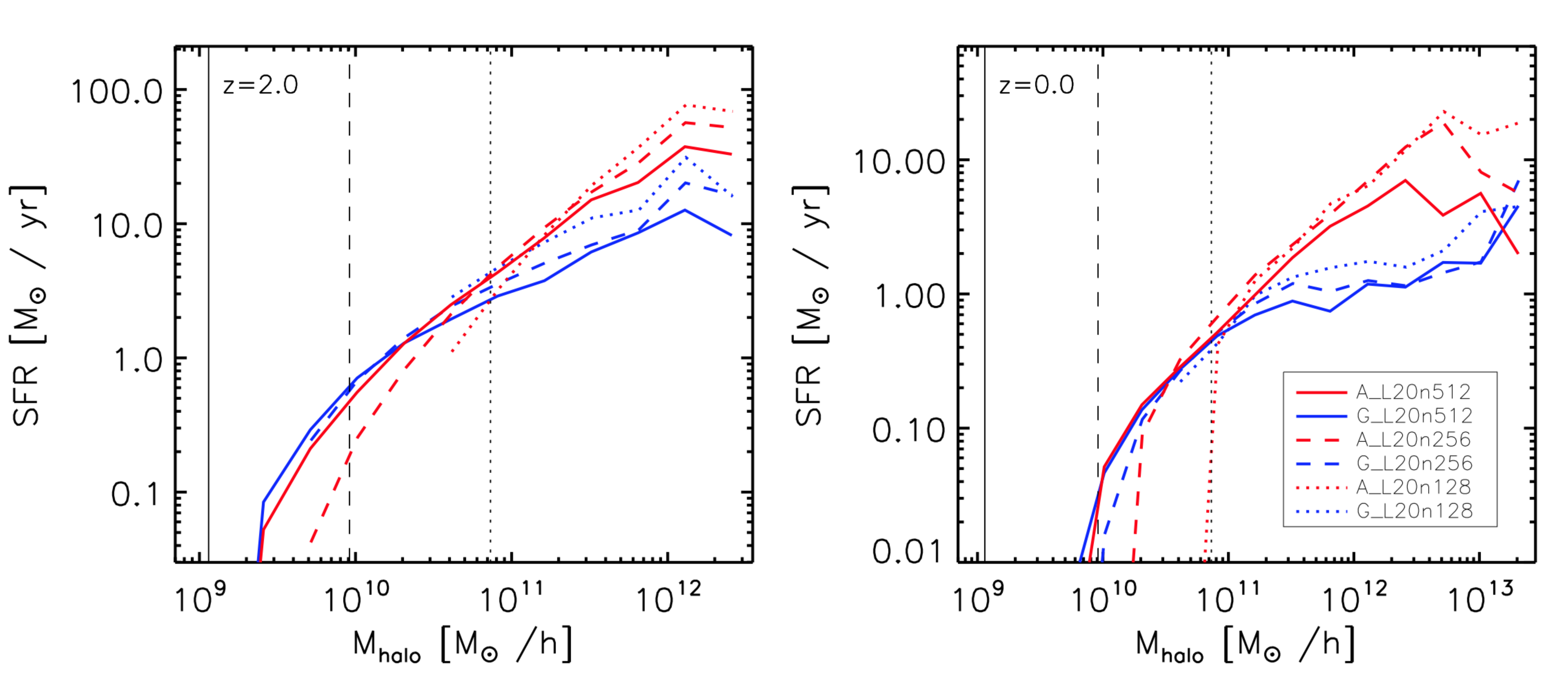}
\caption{Star formation rates of central galaxies as a function of their
  parent halo mass for all of our simulations. Blue lines show {\sm GADGET}
  galaxies while red lines show {\sm AREPO} galaxies at $z=2$ and $z=0$. In
  low mass haloes in our highest resolution  simulations galaxies form stars
  with similar rates. However in haloes more massive than $2\times 10^{11}\,
  \msunh$ {\sm AREPO}'s star formation rates are higher at a given mass. These
  differences reach a factor of $\sim 5$ in Milky Way-mass haloes at
  $z=0$. Lines show medium star formation rates at a given mass with different
  line types corresponding to different resolutions as labeled.}
\label{fig:sfr_hmass}
\end{figure*}

In Figure~\ref{fig:sfr_hmass}, we show the dependence of the star formation
rate of central galaxies on their parent halo mass at redshifts $z=2$ and
$z=0$ for both codes and at three different resolutions\footnote{Note that the
information displayed in this figure differs from Figure~10 in Paper~I by
restricting the star formation measurement to the central galaxy in a halo
instead of showing it summed for all galaxies within the virial radius of a
halo.}.

The most obvious trend in Figure~\ref{fig:sfr_hmass} is that the SFR is a strong
function of the parent halo mass. This dependence is stronger in the case of
{\sm AREPO} with almost linear dependence on halo mass for the $10^{10}\,
\msunh < M_h < 3\times 10^{12}\, \msunh$ range. It is clear that for {\sm
  GADGET} this relation is shallower, especially  in haloes above $\sim
3\times10^{11}\, \msunh$ where at $z=0$ the SFR is only weakly depended on
halo mass. Given the close to linear relation of galaxy mass and halo mass in
Figure~\ref{fig:gal_to_halo}, this also implies a steep relation between SFR
and galaxy mass over an extended mass range, similar to observed trends
\citep[e.g.][]{noeske07, daddi07, salim07}.  We find that in {\sm AREPO} this
relation has a higher normalization and stronger dependence on halo mass at
the massive end than in SPH simulations. This might help to explain the high
observed star formation rates of galaxies at $z=2-3$
\citep{forster-schreiber09}, which appear to form stars at rates higher than
in previous models \citep{dave08}. SFRs of simulated galaxies at a given mass
show a rapid decrease with time in both codes, qualitatively consistent with
observed trends and previous findings from cosmological simulations
\citep[e.g.][]{keres09a, wuyts11}.

At high redshift, {\sm GADGET} galaxies show good convergence in the SFRs of
large haloes, while {\sm AREPO} exhibits a slightly stronger residual drift
with resolution. However, for both methods, one typically finds that low
resolution causes a reduction of the SFR in low-mass objects, but leads to an
overestimate of the SFRs in high-mass objects. This can be understood as a
generic consequence of hierarchical galaxy formation. If limited resolution
suppresses some of the star formation in low-mass progenitor objects, more gas
is left over to support a higher SFR in large objects once they have formed. 
 
The SFRs in {\sm GADGET} and {\sm AREPO} reach similar values for low-mass
parent haloes in our highest resolution simulations, which is a consequence of
the relatively simple gas infall physics in this regime.  However, it is clear
that there is a substantial systematic difference in massive haloes where {\sm
  AREPO} shows a much higher star formation rate in the central galaxies. The
difference is especially acute for the mass range $\mhalo \sim 5\times10^{11}
- 3\times 10^{12}\, \msunh$ at $z=0$, where the SFRs of central galaxies in
{\sm AREPO} are a factor of 3-5 times larger than for the same galaxies in
{\sm GADGET}. This is also the reason why the global star formation rate in
{\sm AREPO} is significantly higher at low redshift than in {\sm GADGET}, as
discussed in Paper~I.  Since the SFR in these massive haloes is largely
regulated by the gas supply, and late-time mergers are relatively gas-poor,
the main cause of these high SFRs at $z=0$ in {\sm AREPO} must lie in a more
efficient cooling in massive haloes. Similar differences are apparent for
masses $M_{\rm halo} \sim 10^{12}\,\msun$ at $z=2$, where {\sm AREPO}'s SFRs
are a factor of 2-3 higher than in {\sm GADGET} simulations.  This is a clear
example of dramatic changes in the properties of galaxies when a more accurate
hydrodynamical method is used to model galaxy formation.

Interestingly, at high resolution, the SFRs of central galaxies in our most
massive haloes, $\mhalo \sim 10^{13}\, \msunh$ are quite comparable between the
codes and are relatively low relative to their halo mass.  In the next
subsection we explore the hot gas profiles in massive haloes to examine these
differences more explicitly.
  
\subsection{Gas profiles of massive haloes}
\label{sec:halo_profiles}

\begin{figure*}
\includegraphics[width=1.\textwidth]{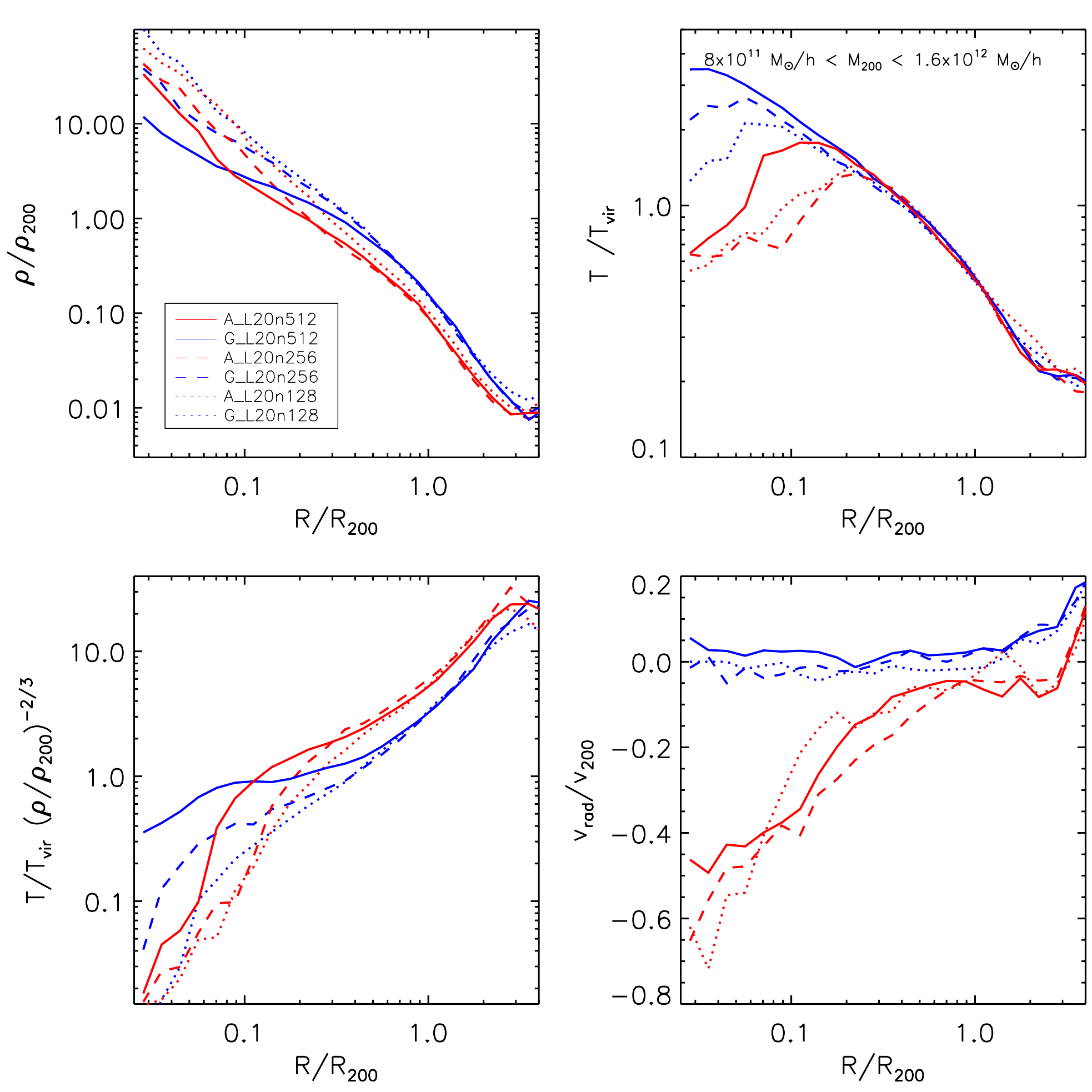}
\caption{Density, temperature, entropy and radial velocity profiles of $\sim
  10^{12}\, \msunh$ haloes at $z=0$.  We only include gas that is outside the
  two-phase star-forming medium and has temperature $T > 100, 000\,{\rm K}$,
  in order to concentrate on properties of hot halo gas. The radius is
  expressed in units of the virial radius ($R_{200}$), temperature is scaled
  to the virial temperature, density is expressed in units of the mean
  enclosed baryonic density of our haloes, $\rho_{200}$, and velocity is scaled
  to the circular velocity at $R_{200}$.  Red and blue lines are for haloes
  simulated with {\rm AREPO} and {\sm GADGET}, respectively, while different
  line styles indicate different numerical resolutions, as labeled in the
  density panel. The halo mass range stacked for the measurements is given in
  the temperature panel and includes between 14 and 20 haloes.}
\label{fig:halo_profiles12}
\end{figure*}

\begin{figure*}

\includegraphics[width=1.\textwidth]{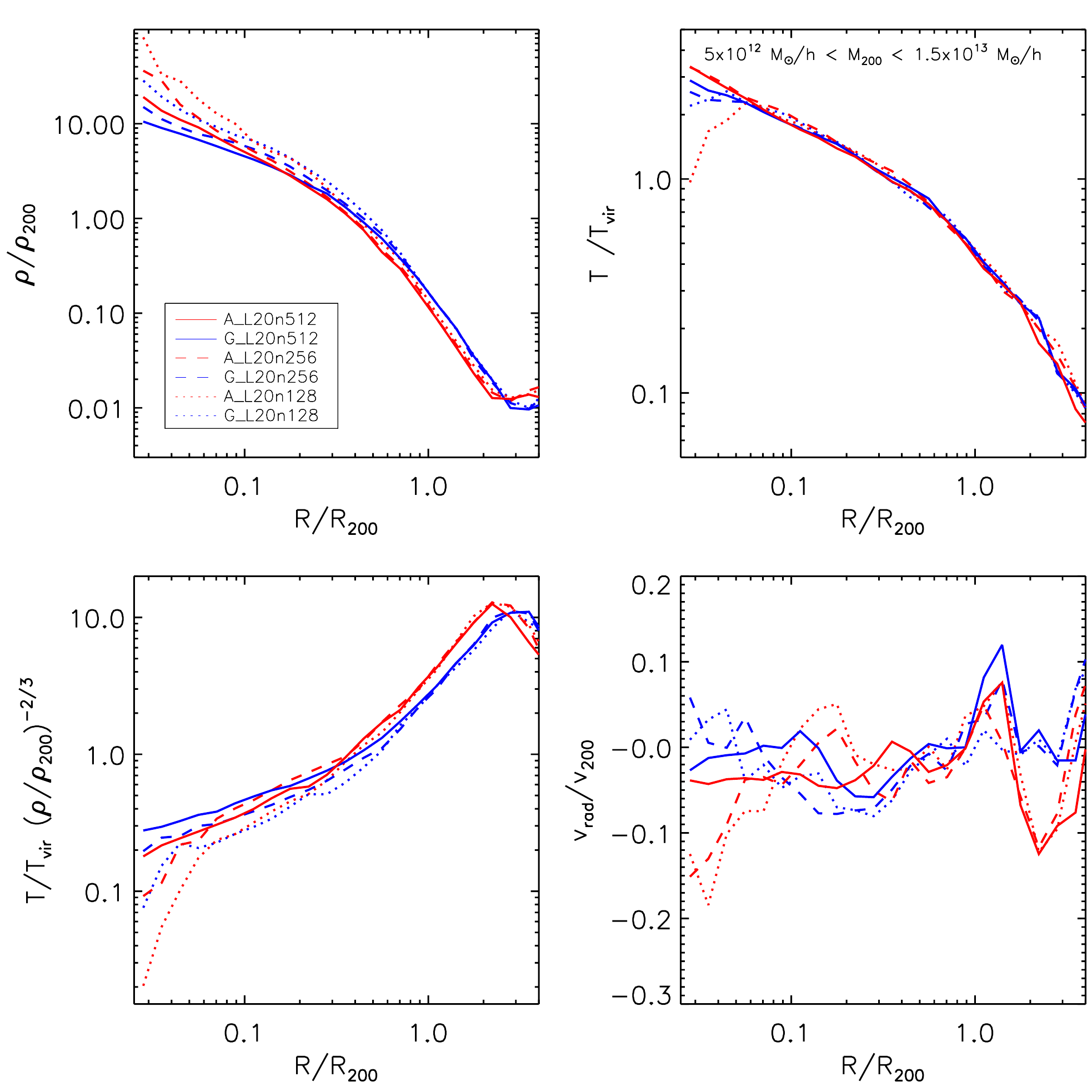}

\caption{Density, temperature, entropy and radial velocity profiles of $\sim
  10^{13}\, \msunh$ haloes at $z=0$. The selection criteria for the gas and the
  units chosen are as in Figure~\ref{fig:halo_profiles12}. The halo mass range
  used for the stacking is stated in the temperature panel and contains 5-6
  haloes.}
\label{fig:halo_profiles13}
\end{figure*}

In Figures~\ref{fig:halo_profiles12} and \ref{fig:halo_profiles13}, we show
average density, temperature, entropy and radial velocity profiles of $\sim
10^{12}\, \msunh$ and $\sim 10^{13}\, \msunh$ haloes at $z=0$, respectively. To
measure these profiles, we first select all haloes in a narrow mass range as
labeled in the temperature panels. For each halo, we find its spherically
averaged profiles for non star-forming gas with $T > 10^5\, \rm K$. We then
stack the haloes by finding the median value over all systems in the given mass
range. This is to avoid local distortions owing to infalling substructures
that often have slightly different locations in {\sm AREPO} and {\sm GADGET},
and hence helps to focus attention on the smooth hot halo gas only.  We note that
all the haloes in this mass range are dominated by the hot halo gas. To allow a
more faithful stacking, we express the radius in units of $R_{200}$, density
in terms of enclosed baryonic density within $R_{200}$, velocity in units of
$v_{200}=\sqrt{GM_{200}/R_{200}}$ and temperature as a virial temperature,
$T_{vir}=0.5 \mu m_{\rm H} v_{200}^2/k_{\rm B}$, where $m_{\rm H}$ is the proton mass, $\mu$ is
mean molecular weight of the gas, which we here take to be fully ionized, and
$k_{\rm B}$ is Boltzmann constant.

There are interesting systematic differences in the hot halo structure between
the different simulation codes, that are especially prominent for our lower
mass sample. The density profiles are steeper in {\sm AREPO} compared with
{\sm GADGET}, showing at large radii lower gas densities, and in the central
parts higher density. The temperature profiles are quite similar between the
two simulation techniques for $R > 0.2 R_{200}$. However, the inner region
clearly shows a decrease in gas temperature in {\sm AREPO} simulations
indicative of a cooling flow, while the temperature continues to rise at small
radii in {\sm GADGET}. At lower resolution, there is an even stronger decrease
in temperature and an increase in density in the innermost halo region in {\sm
  AREPO}.  Even {\sm GADGET} shows some mild decline of the central
temperature and a density increase, albeit with a much lower magnitude and
confined to smaller radii.  These density and temperature differences are
reflected in corresponding trends for the entropy of the halo gas. The entropy
in moving-mesh simulations is significantly higher in the halo outskirts but
is lower in the central $\sim 0.1\,R_{200}$ of the halo.  Finally, the mean
radial streaming velocity of the gas in Figure~\ref{fig:halo_profiles12} shows a
larger negative amplitude for {\sm AREPO} than for {\sm GADGET}. This can be
viewed as a tell-tale sign of a larger cooling rate out of the {\sm AREPO}
haloes.

These trends provide a clear explanation for the large differences in SFRs
between the two simulation techniques visible in Figure~\ref{fig:sfr_hmass},
which are induced by a much more efficient cooling in haloes with mass $\sim
10^{12}\, \msunh$ in {\sm AREPO}.  In the more massive, group-sized haloes
shown in Figure~\ref{fig:halo_profiles13}, the differences between the gas
profiles are however much smaller. There are still similar trends of a lower
gas density in the outskirts and higher densities and lower entropies in the
central parts of {\sm AREPO} haloes that indicate some residual cooling
difference. However, the temperature profiles are very similar between the
codes, except for a slightly steeper rise of the central temperatures in the
moving-mesh approach, which is likely caused by a deeper potential well owing
to a much higher accumulation of baryons in the centre.  Also, in both codes
the central entropy remains relatively high and the central temperature does
not show signs of strong cooling. This explains the relative low SFRs over
this halo mass range and the relative consistency between {\sm AREPO} and {\sm
  GADGET}. The long cooling times are also reflected in the radial infall
profile, which shows that the hot halo gas is essentially in hydrostatic
support and has very little average motion in the radial direction.  We note
that the high entropies in massive haloes in {\sm GADGET} are consistent with
previous results from cosmological simulations with cooling and star formation
\citep{keres09a}, but interestingly, the trend we find here between {\sm
  AREPO} and {\sm GADGET} is opposite to what is typically found in
non-radiative simulations of cluster formation where the central entropies of
massive haloes are found to be higher in mesh codes \citep{frenk99}.

Interestingly, we also find that the acceleration profiles due to
hydrodynamical pressure forces ($- \nabla P / \rho$) are very similar between
the two codes, as expected for the same haloes in quasi hydrostatic
equilibrium, because the pressure forces need to balance gravity, which is
dominated by the same dark matter distribution in both codes. Is is then at
first a bit puzzling to understand why the {\sm AREPO} haloes are cooling more
gas than {\sm GADGET}, as is clearly evident by the significantly different
radial streaming velocities. While the lower central temperatures and higher
central densities in {\sm AREPO} may suggest a higher cooling rate overall,
note that this is really only the case in the innermost regions of the haloes,
but much of the volume and gas mass of the haloes sit at larger radii. There,
the gas density in {\sm AREPO} is actually comparable or slightly lower than
in {\sm GADGET}, as is also expected due to the reduced remaining gas content
we have measured for these haloes in {\sm AREPO} at late times (because more
stars have been formed). This means that the total energy losses from emitted
cooling radiation are actually quite comparable for the two codes, especially
in the relevant region around the cooling radius in the outer parts of the
haloes. Given these cooling losses, the surprising thing is then actually not
that {\sm AREPO} cools out so much gas, it is that {\sm GADGET} cools out so
little gas, especially in $\sim 10^{12}\, \msunh$ haloes where the cooling
times are short. 

We argue that this is due to the difference in dissipative heating between the
two codes that we measure for such quasi-stationary haloes in Paper I. In SPH,
a combination of viscous dissipation of sonic velocity noise of the gas and
efficient damping of turbulent gas motions injected in the infall regions
leads to a significant heating of the outer parts of haloes, partially
offsetting the cooling losses there.  In contrast, {\sm AREPO} produces less
dissipative heating in this region, so that a stronger cooling flow and higher
radial infall velocities result. The adiabatic heating of the gas on its way
inward is then maintaining roughly constant or even rising temperatures for
the gas, despite the cooling, until it eventually drops catastrophically in
temperature at very small radius, and is then rapidly flowing onto the
star-forming phase.  In addition, stripping of more gas-rich disks in {\sm
  AREPO} and subsequent mixing of this low-entropy material (which is
suppressed in {\sm GADGET}) can further strengthen cooling in the central
region of {\sm AREPO} haloes.

\subsection{Characteristic size and specific angular momentum of galaxies}
\label{sec:gal_radii}
 
\begin{figure*}
\includegraphics[width=0.325\textwidth]{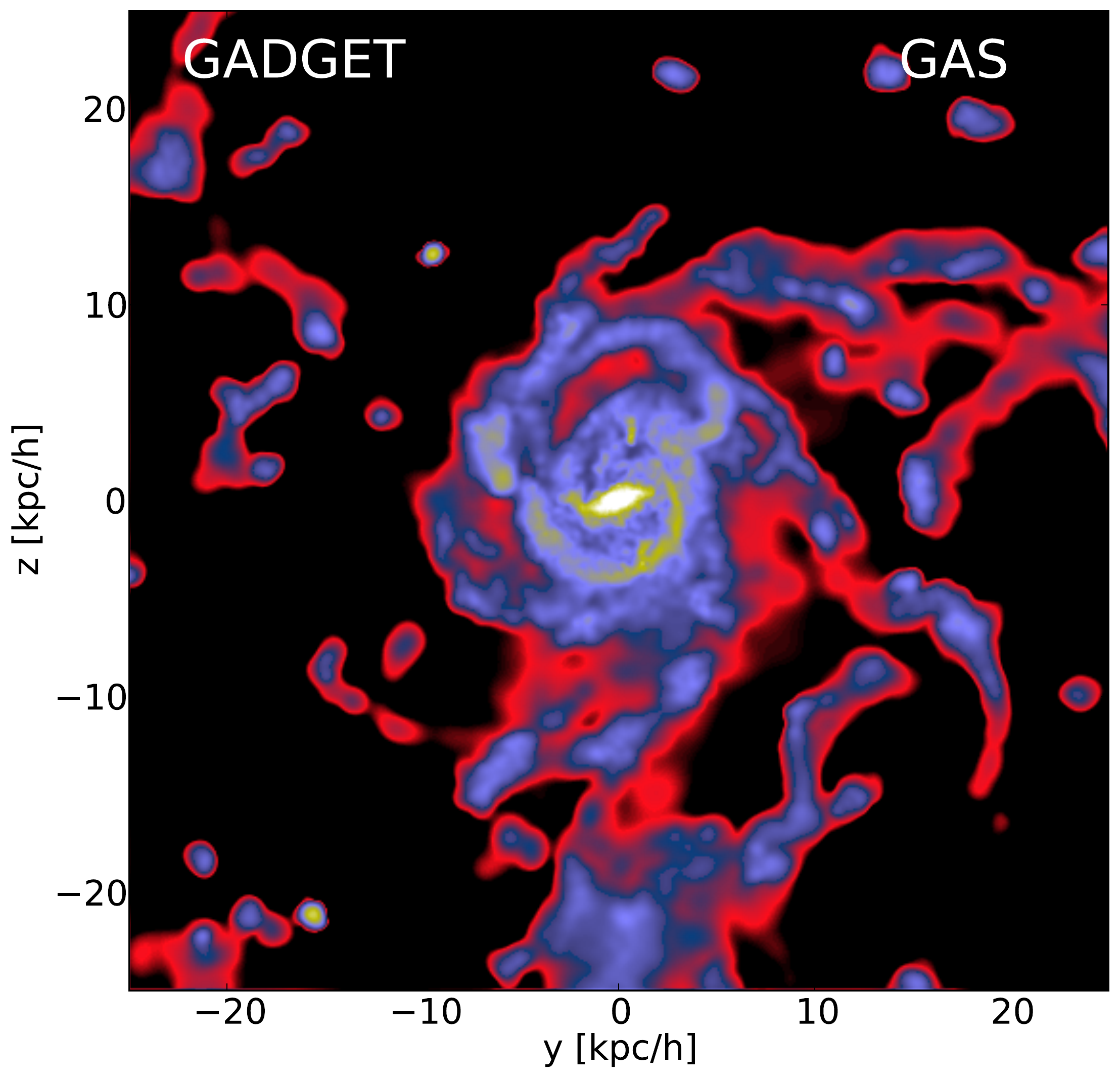}
\includegraphics[width=0.325\textwidth]{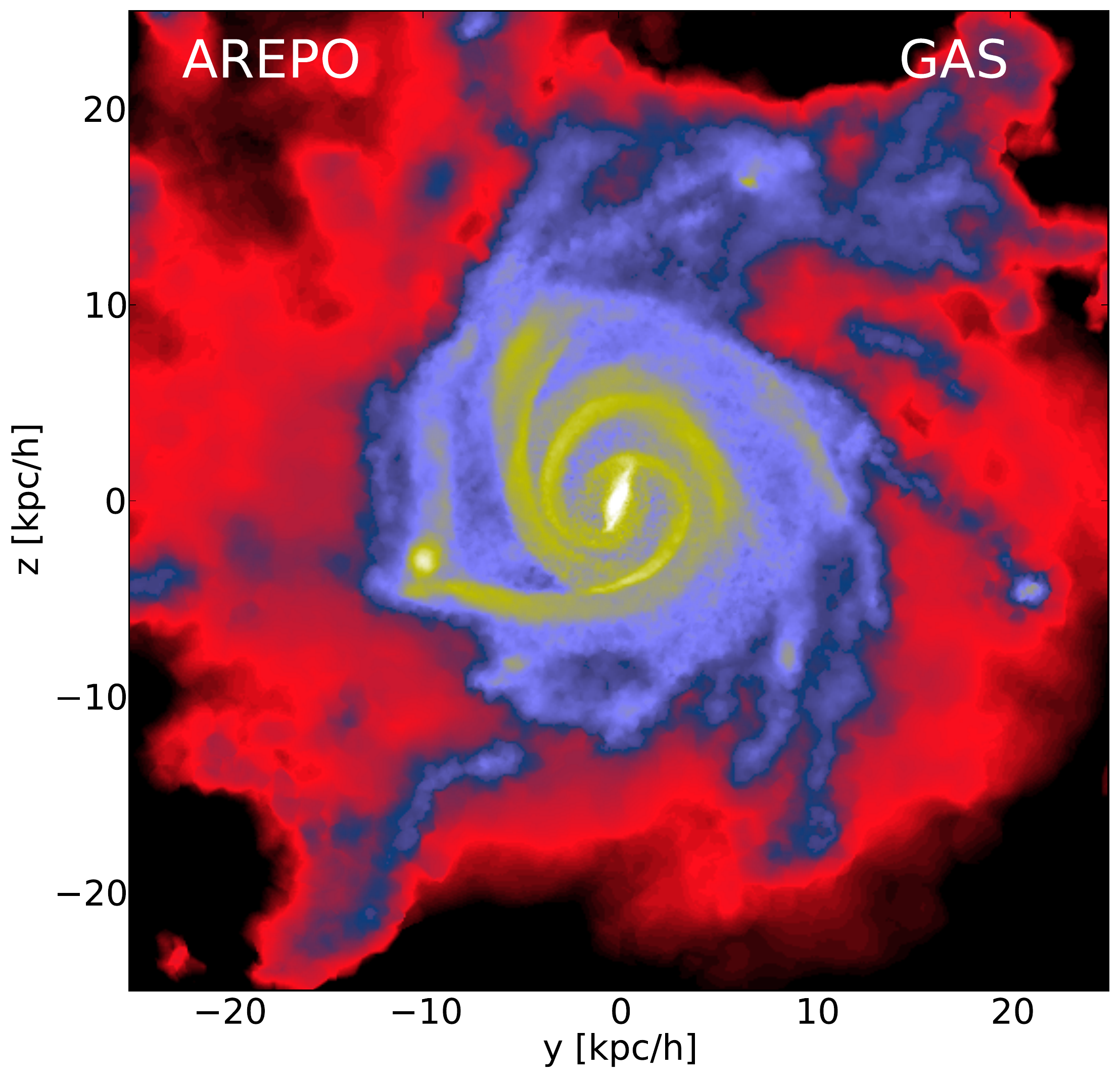}\\
\includegraphics[width=0.325\textwidth]{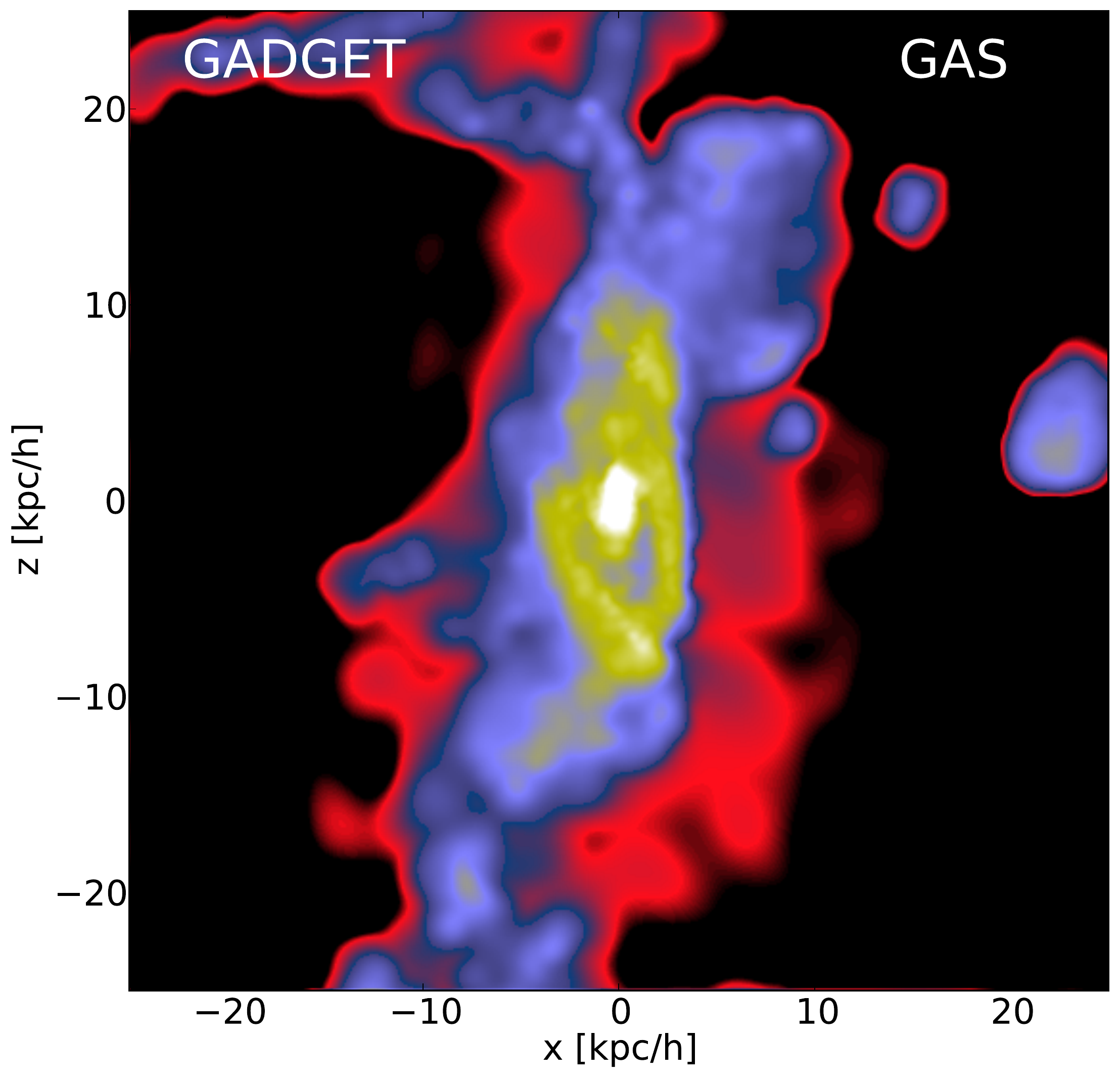}
\includegraphics[width=0.325\textwidth]{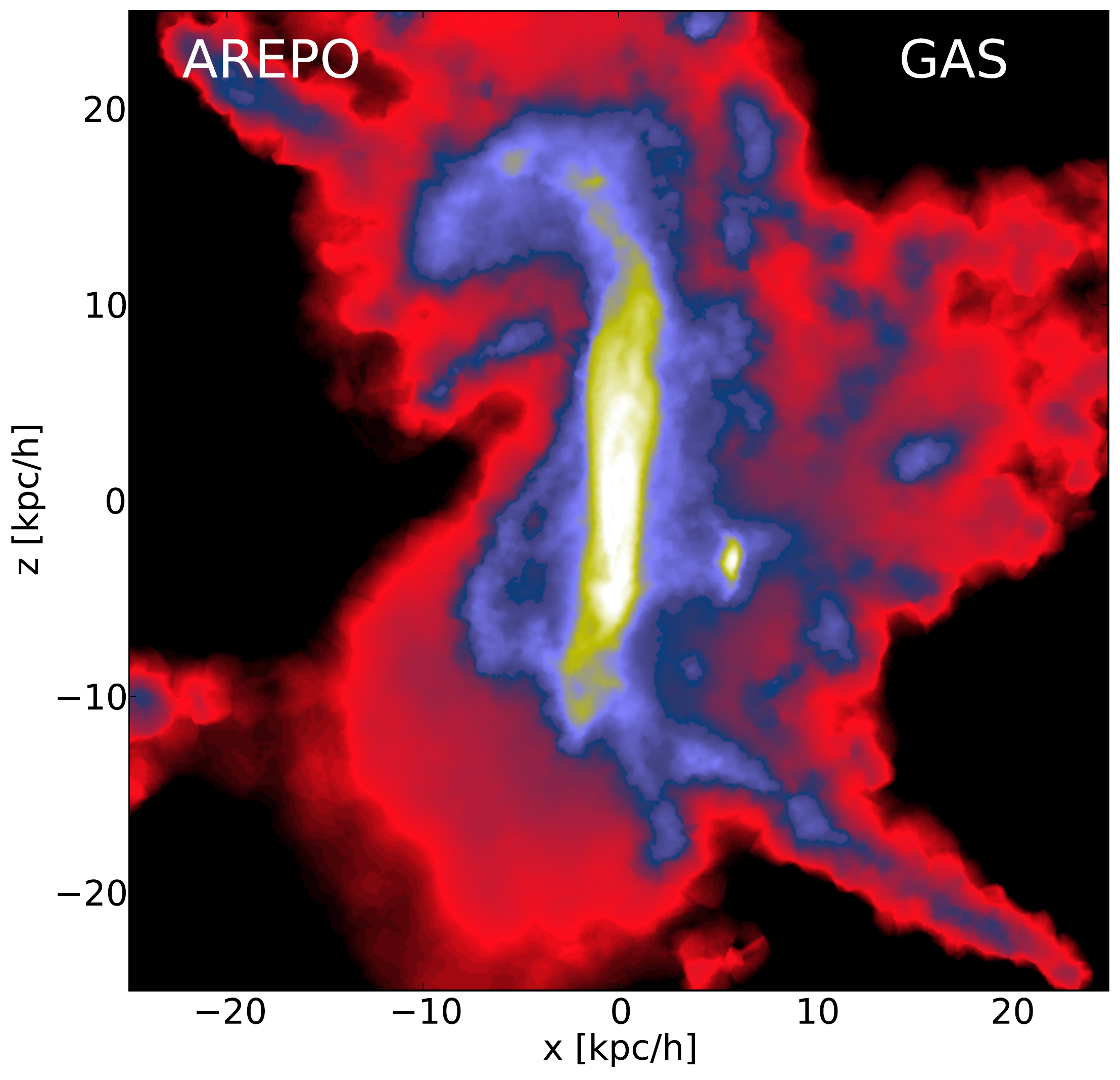}\\
\includegraphics[width=0.325\textwidth]{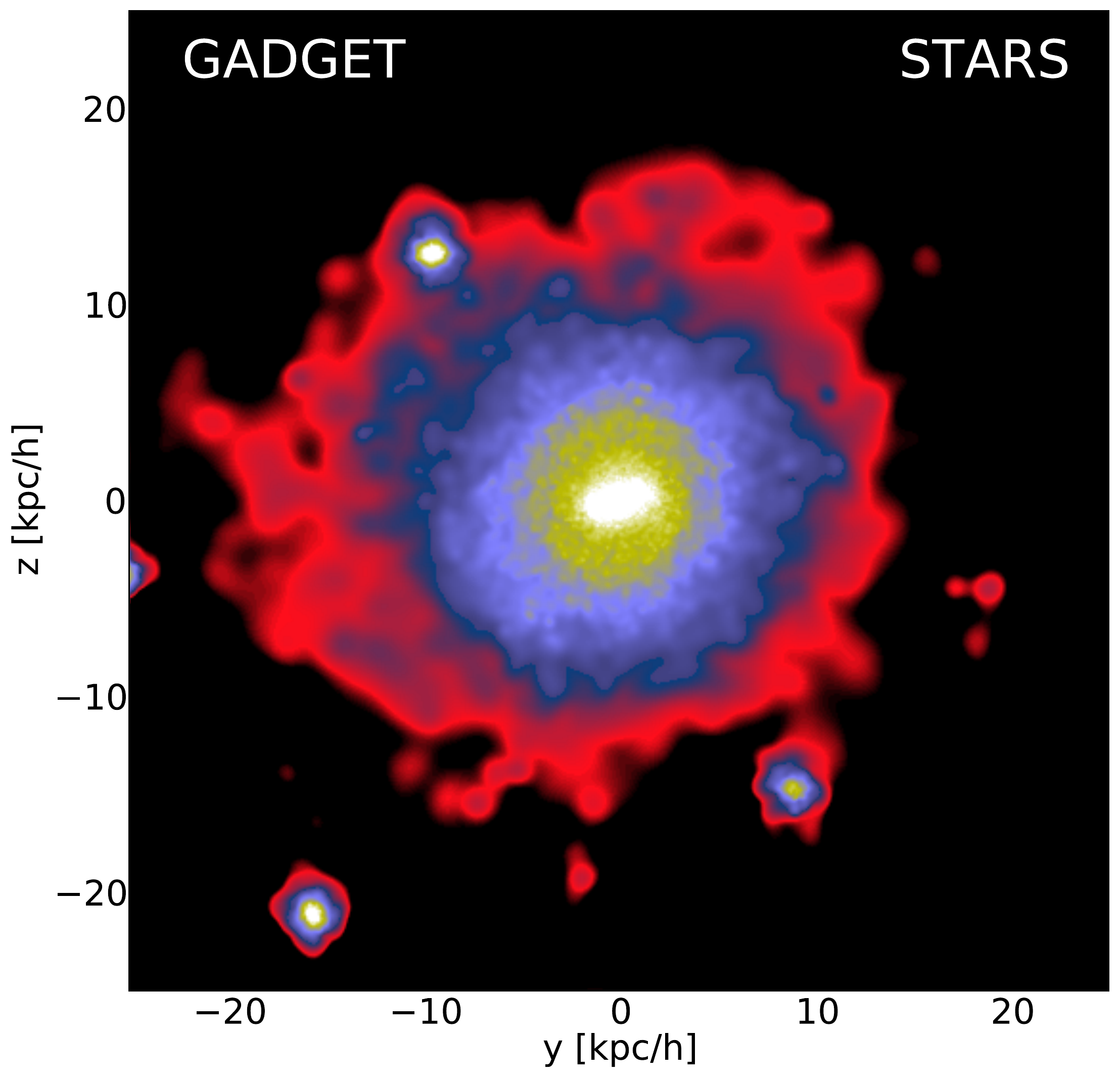}
\includegraphics[width=0.325\textwidth]{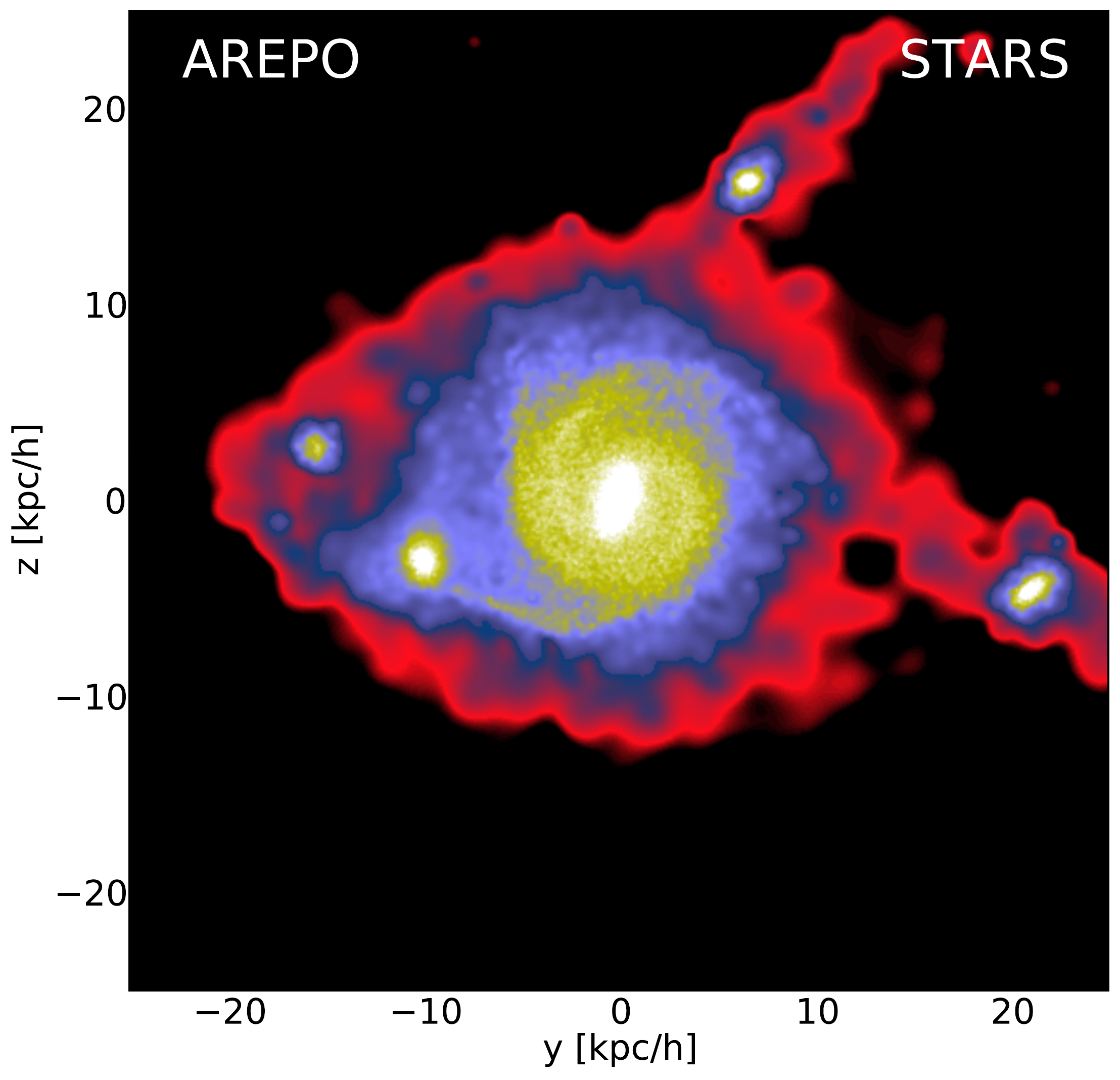}\\
\includegraphics[width=0.325\textwidth]{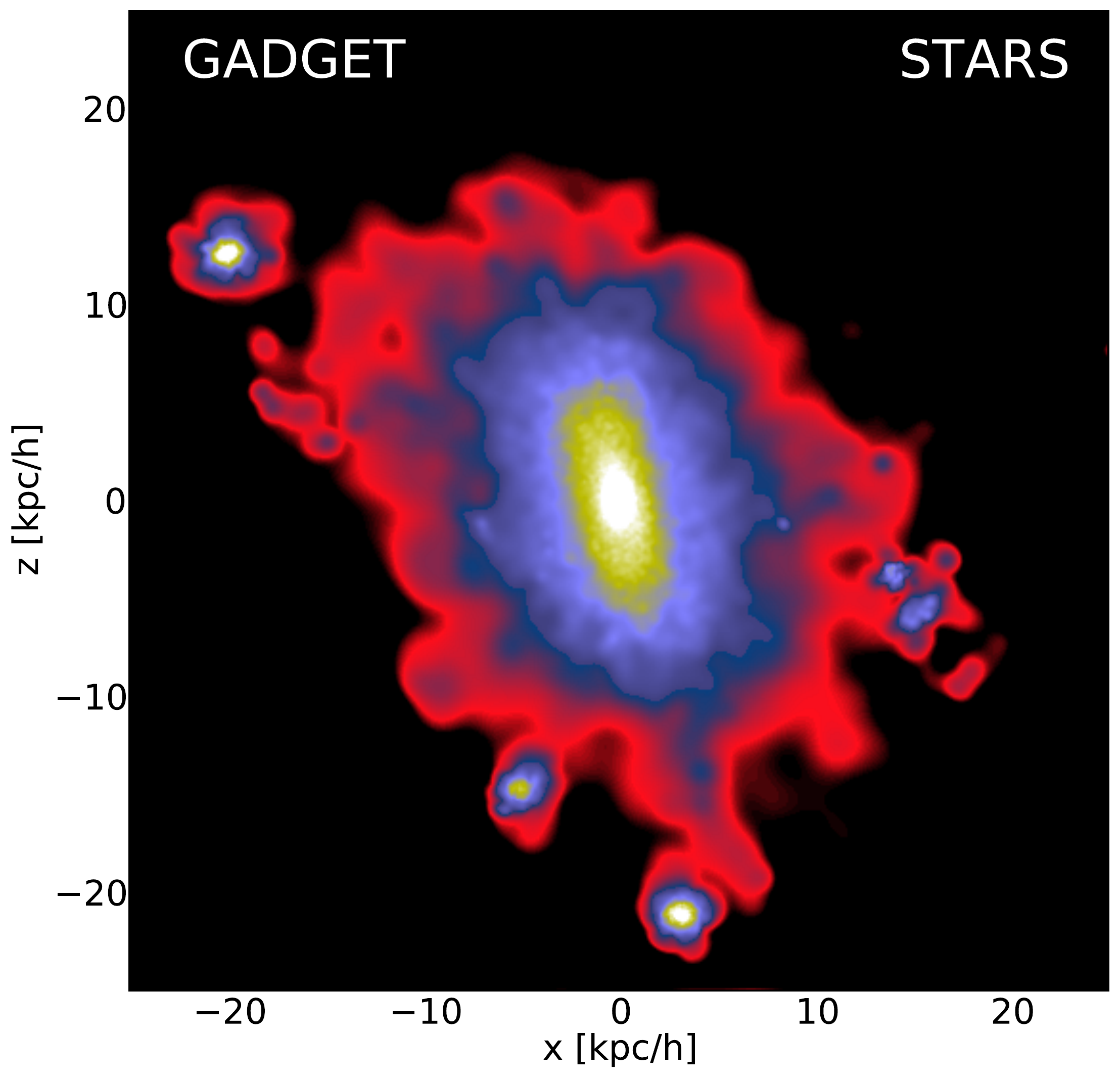}
\includegraphics[width=0.325\textwidth]{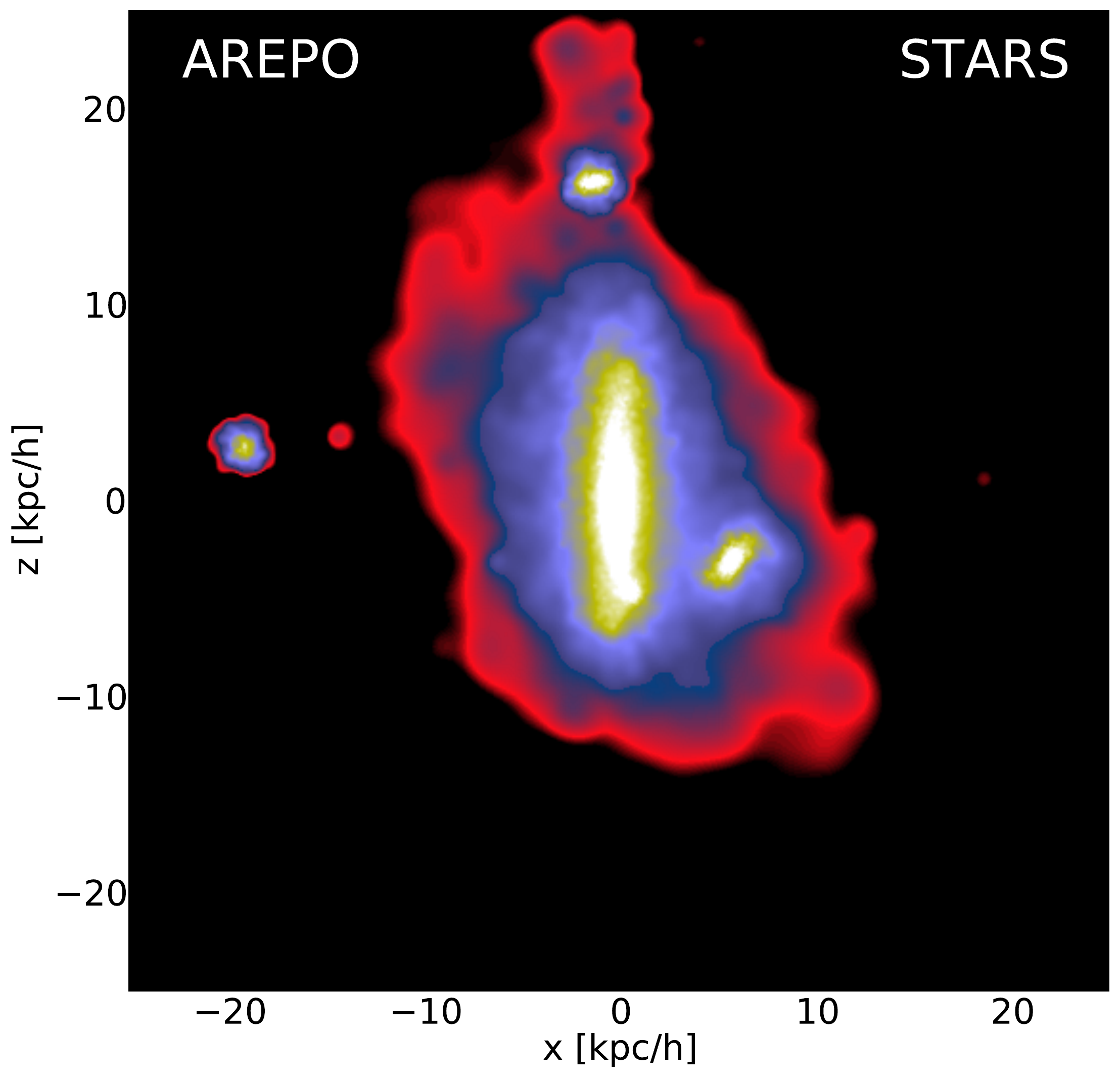}
\vskip -5pt
\caption{Example of a massive disk galaxy at redshift $z=2$ in {\small AREPO}
  and {\small GADGET} L20n512 simulations. The upper two rows of panels show
  the projected surface density of gas in a $50\,h^{-1}{\rm kpc}$ physical box
  centered on the galaxy for two different lines-of-sight. The galaxy is
  clearly more extended in {\small AREPO} and has a smoother distribution of
  dense gas. Some quantitative properties of the object are summarized in
  Table~\ref{table:disks}. The edge-on projection in {\sm AREPO} reveals a
  structure that looks more disky. The lower two rows of panels show projected
  stellar surface density in the same volume and for the same
  lines-of-sight. Clearly, also the stellar component of this large galaxy has
  a more disky morphology in {\sm AREPO} compared with {\sm GADGET}. Note that
  gaseous and stellar disks in the edge-on projection exhibit higher degree of
  alignment in the moving-mesh code.}
\label{fig:gas_examplez2}
\end{figure*}

As an illustration of differences in the sizes and shapes of galaxies,
we compare in Figure~\ref{fig:gas_examplez2} the gaseous
and stellar morphology of a typical late-type galaxy formed by our two
simulation codes.  Each projection shows the gas or stellar mass density
within a box of physical side-length $50\,h^{-1}{\rm kpc}$ centered on the
galaxy. The galaxy shown is one of the most massive galaxies in our simulated
volume at $z=2$, and the same object was selected for a discussion of its
large-scale environment in Paper I.

\begin{figure*}
\includegraphics[width=0.325\textwidth]{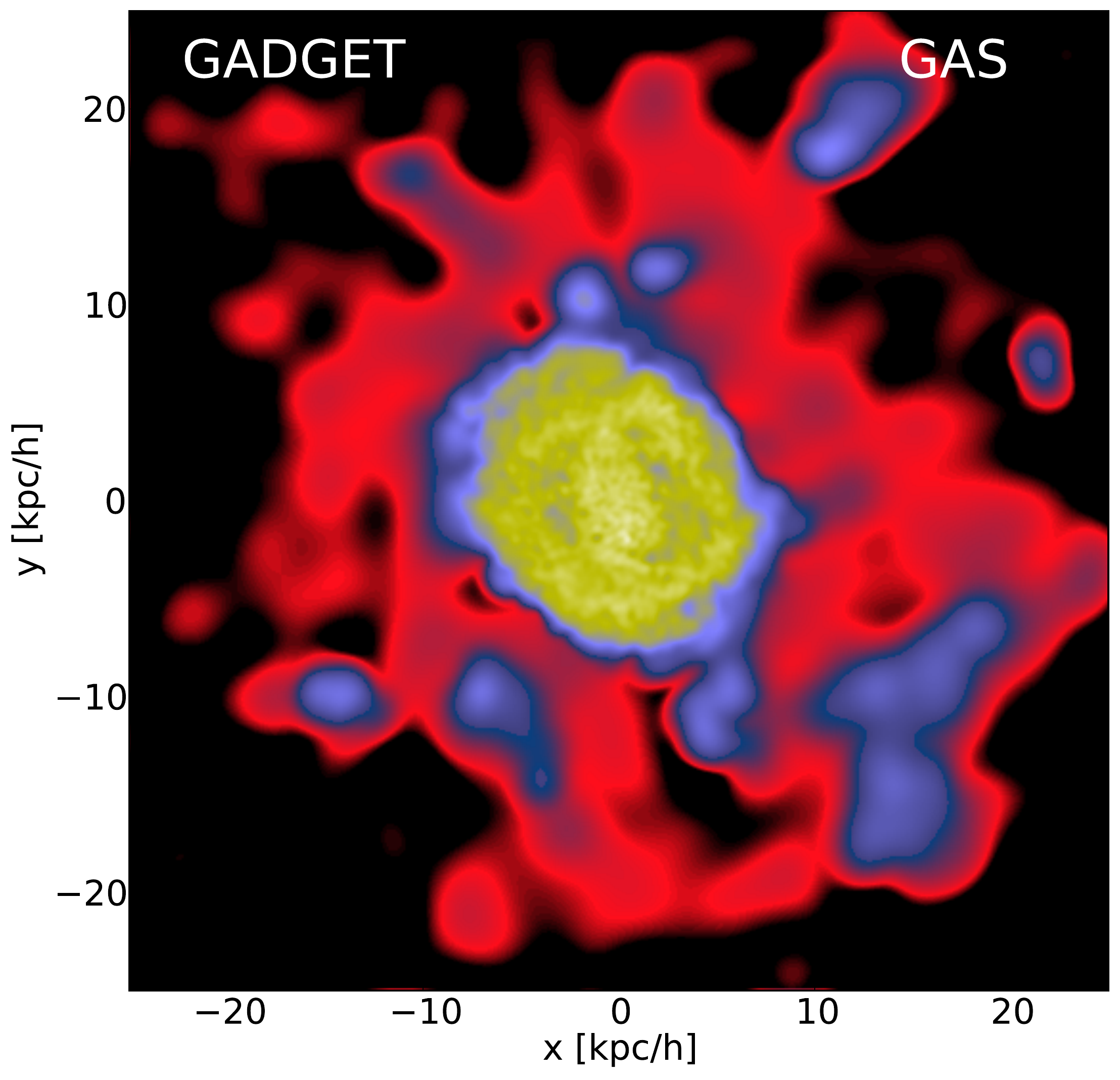}
\includegraphics[width=0.325\textwidth]{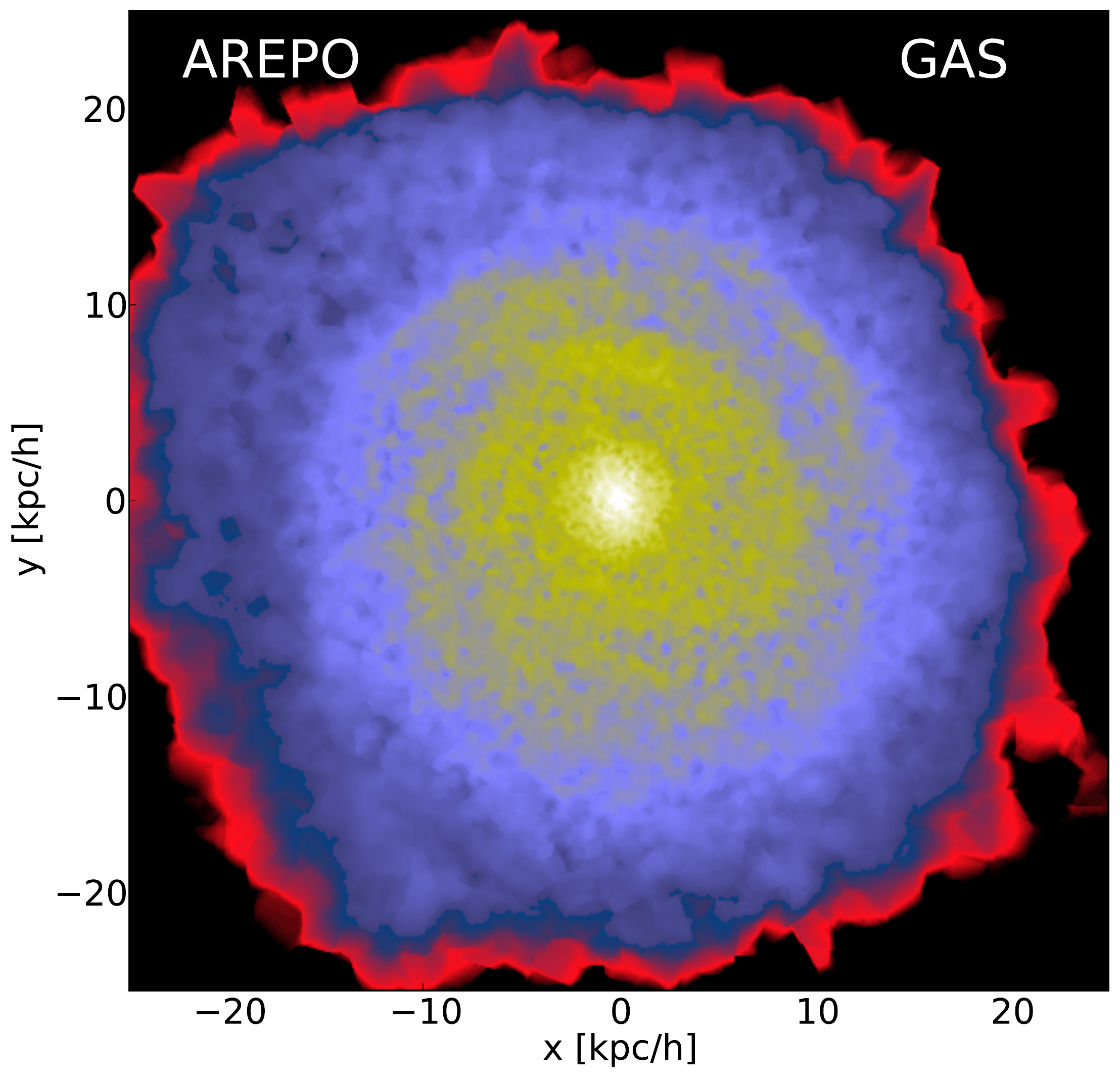}\\
\includegraphics[width=0.325\textwidth]{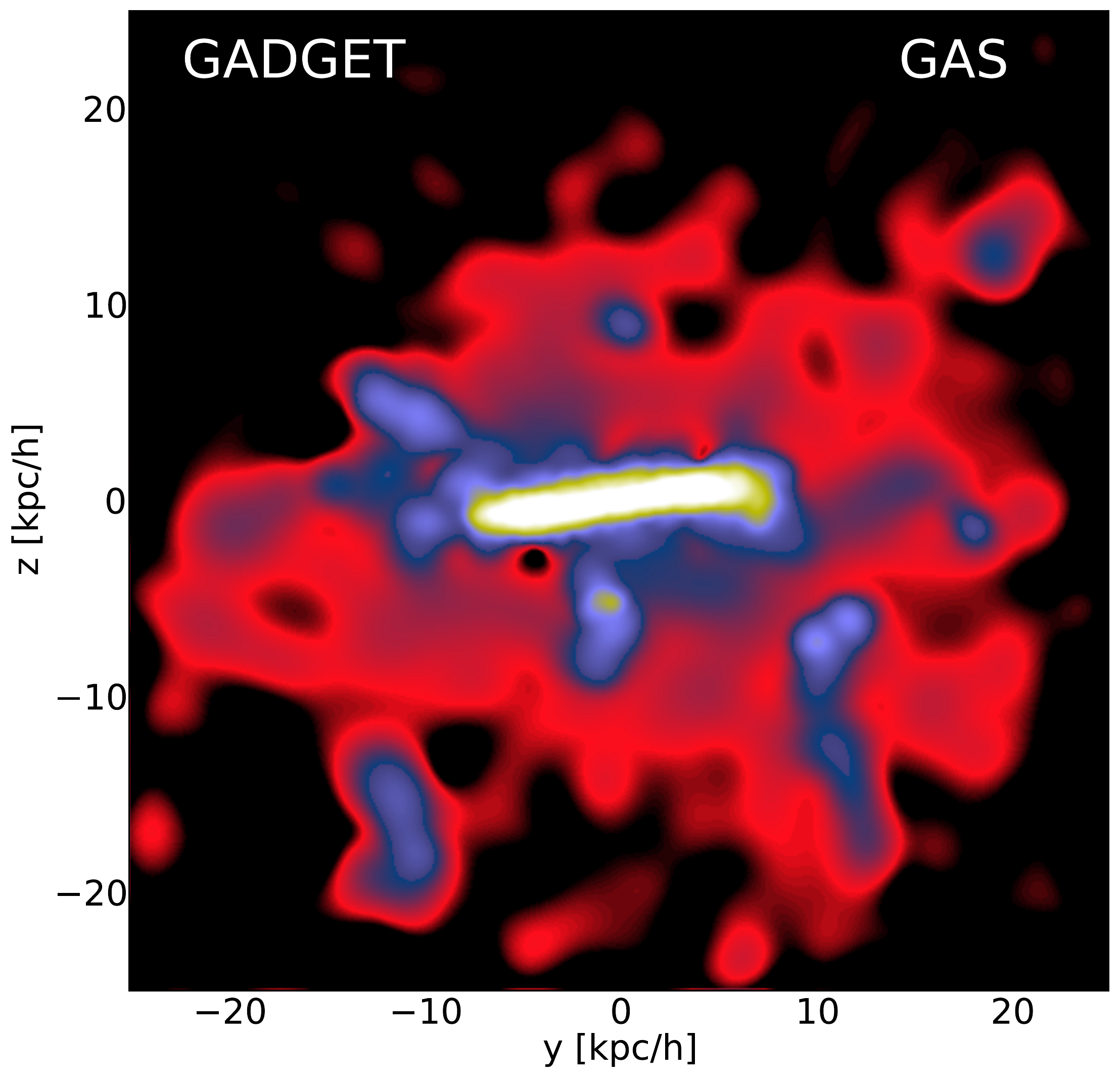}
\includegraphics[width=0.325\textwidth]{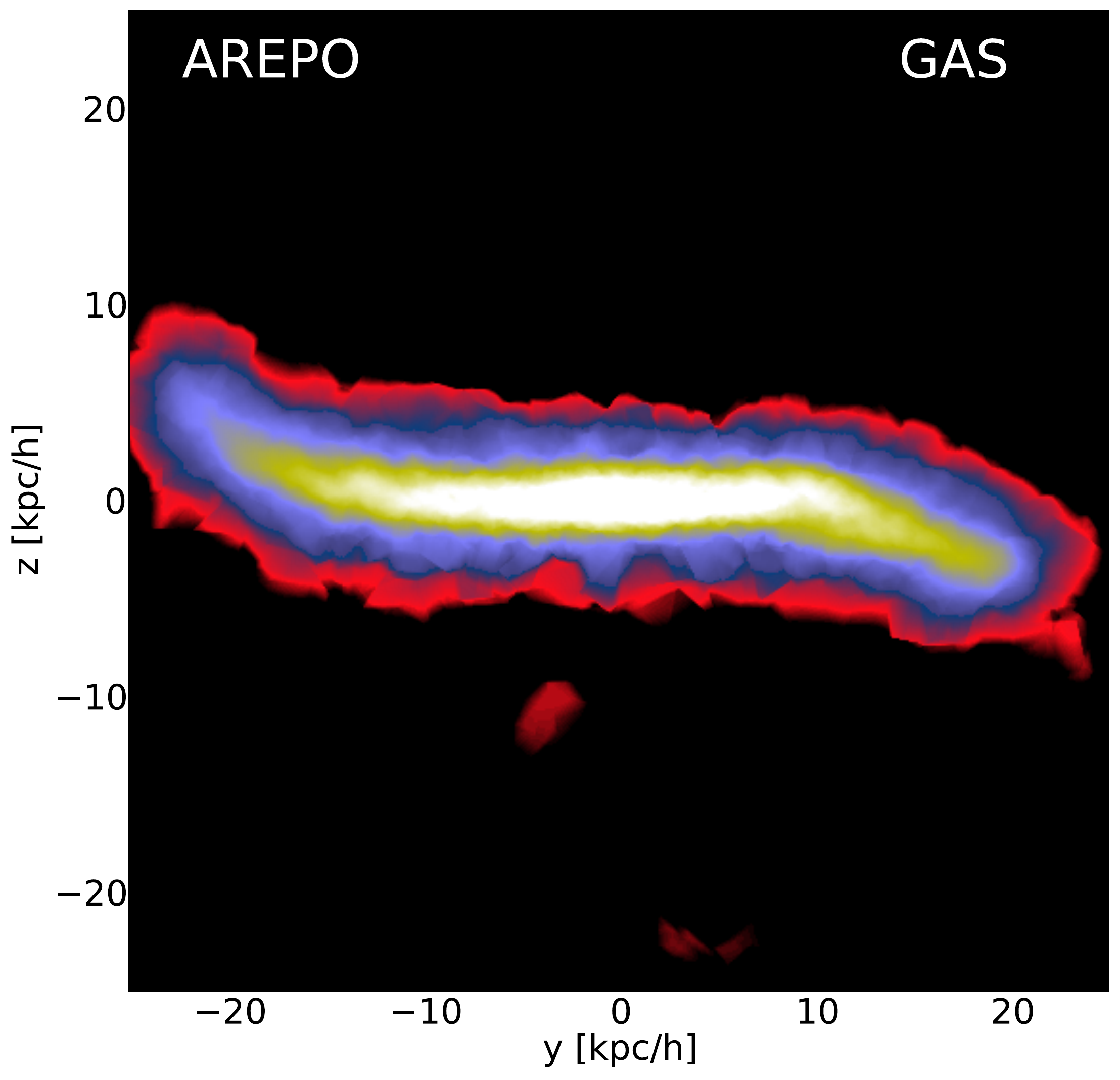}\\
\includegraphics[width=0.325\textwidth]{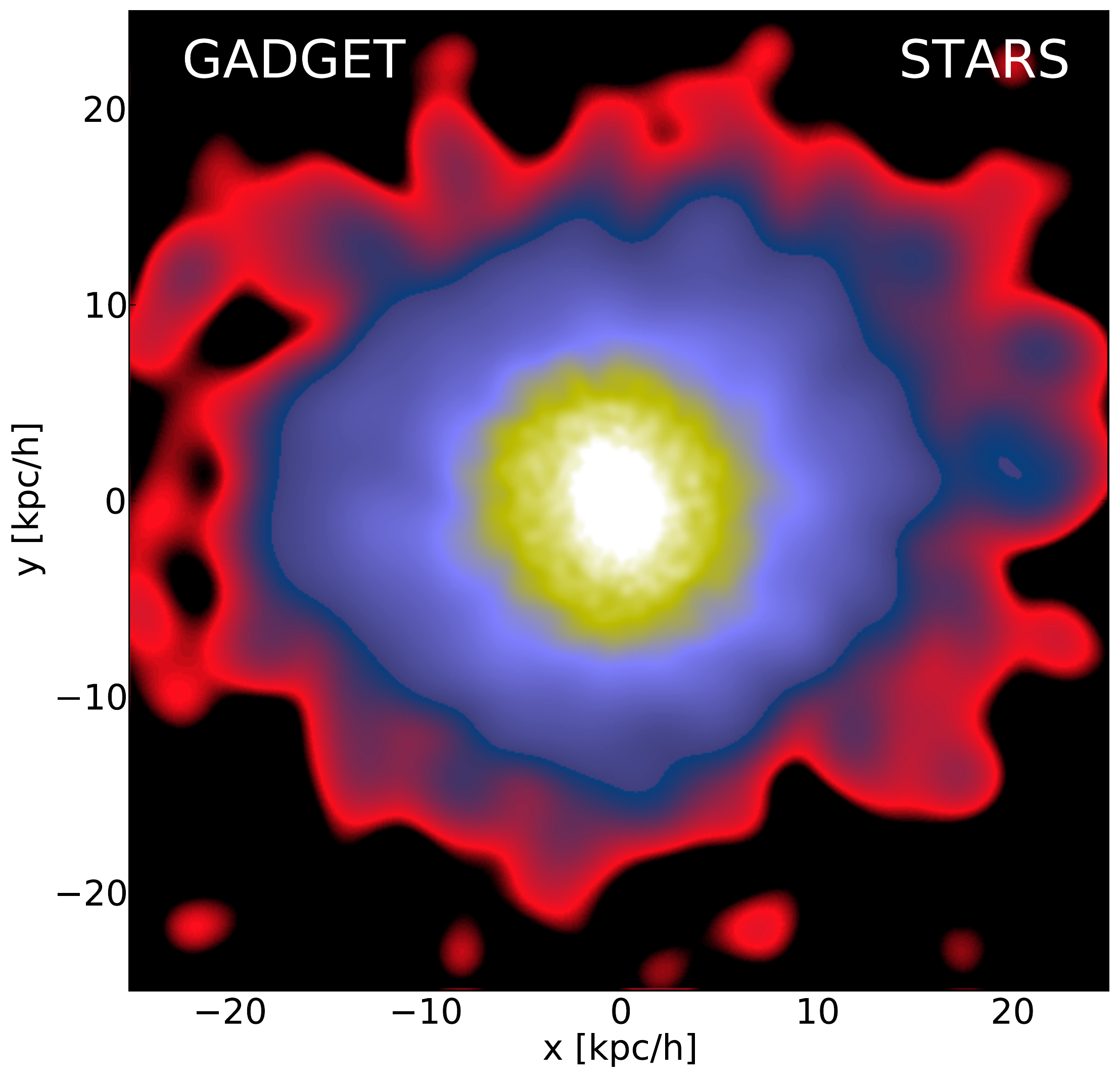}
\includegraphics[width=0.325\textwidth]{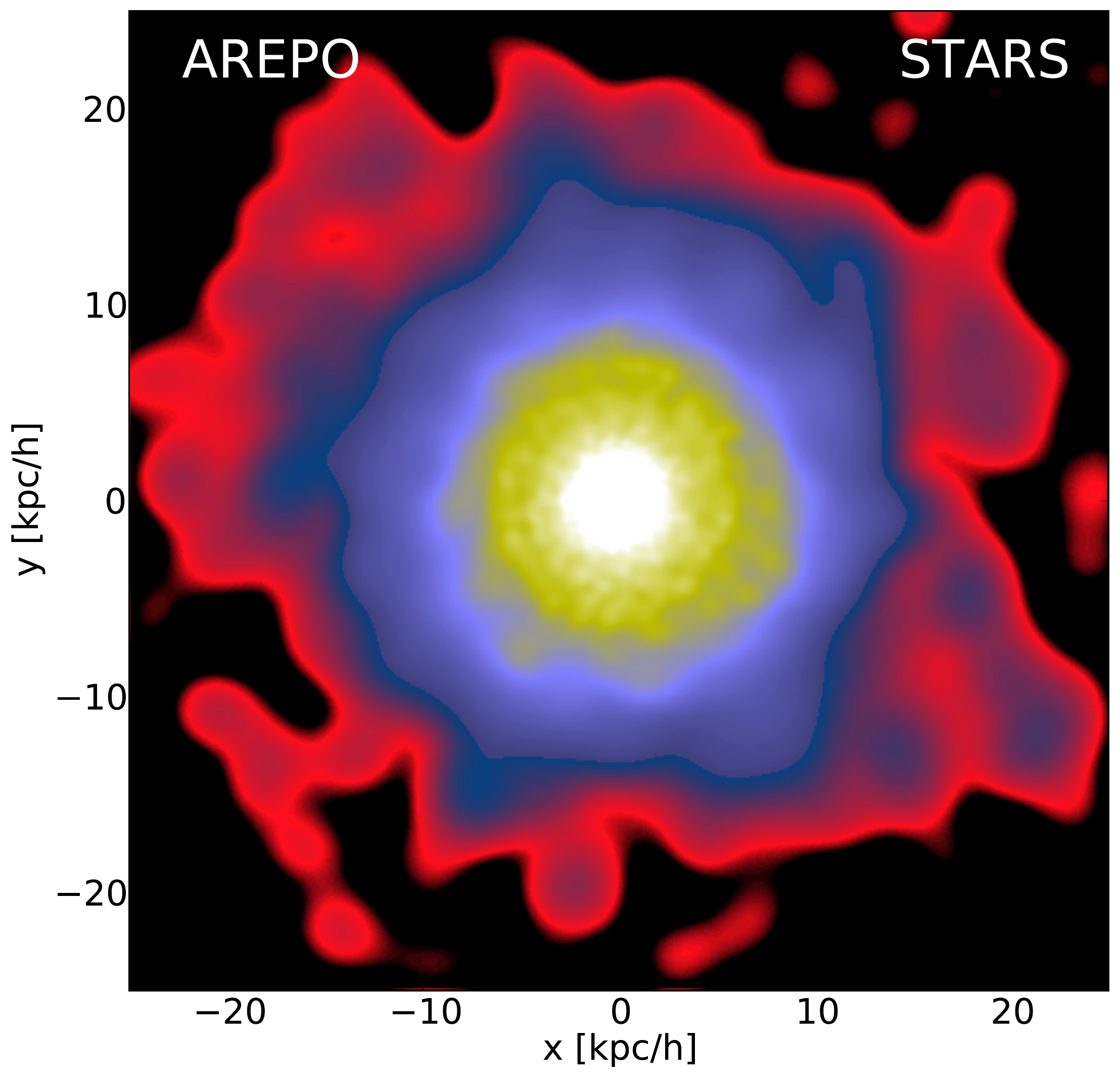}\\
\includegraphics[width=0.325\textwidth]{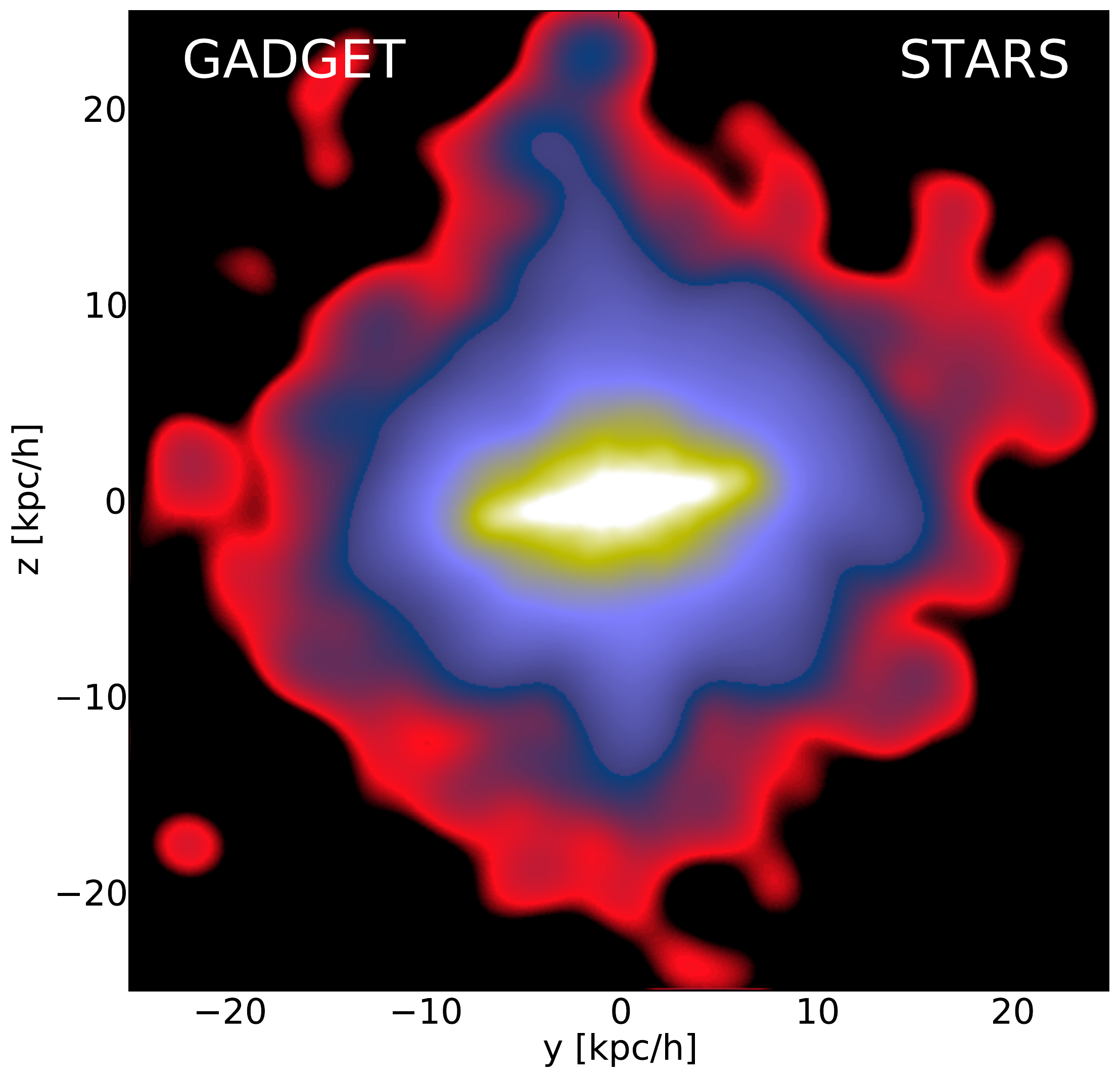}
\includegraphics[width=0.325\textwidth]{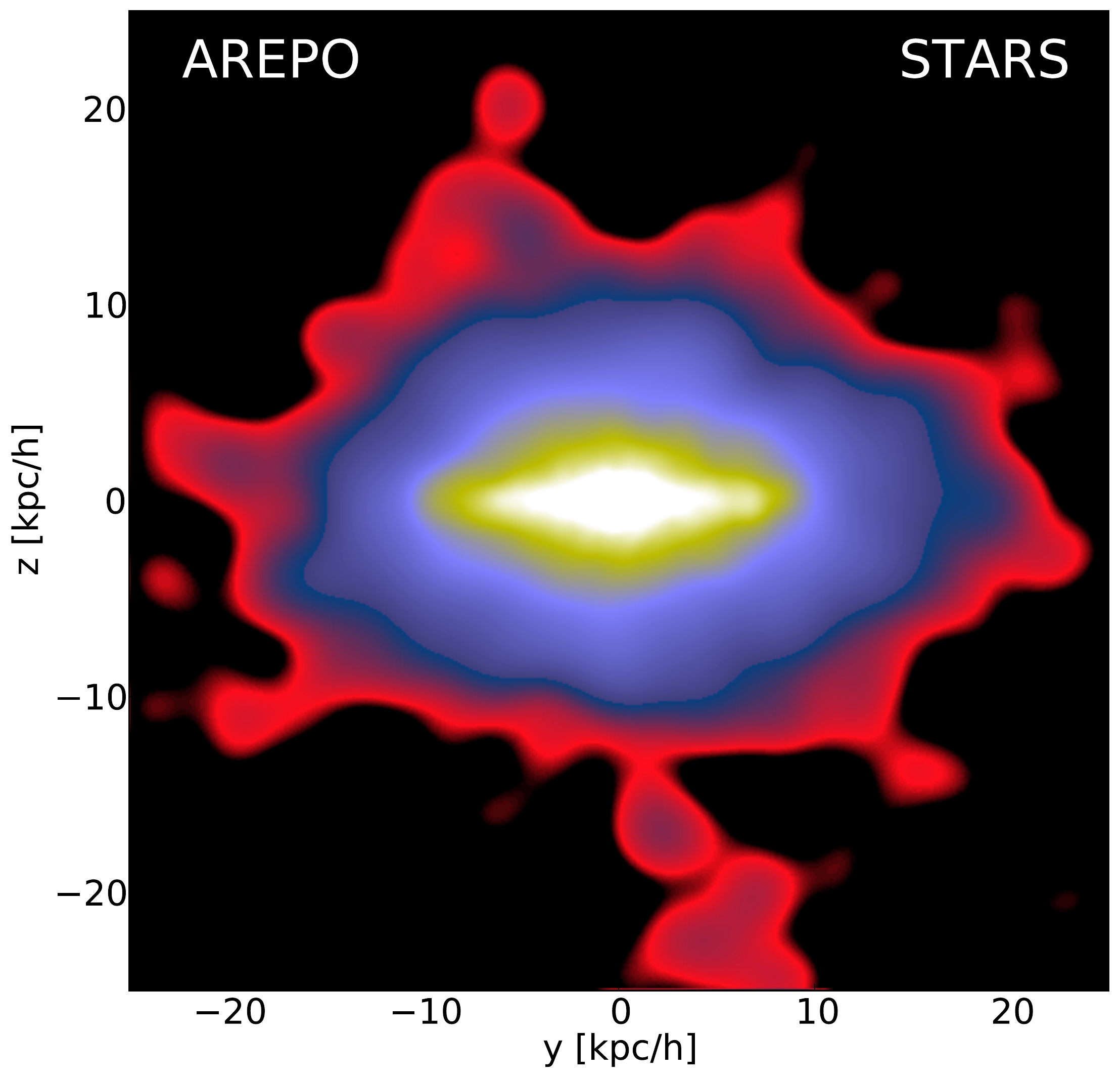}
\vskip -5pt
\caption{Example of a disk galaxy at redshift $z=0$ in {\small AREPO} and
  {\small GADGET} L20n512 simulations. The upper two rows of panels show the
  projected surface density of gas in a $50\,h^{-1}{\rm kpc}$ physical box
  centered on the galaxy for two different lines-of-sight. The galaxy is
  clearly more extended in {\small AREPO} and has a smoother distribution of
  dense gas. Some quantitative properties of the object are summarized in
  Table~\ref{table:disks}. The edge-on projection in {\sm AREPO} reveals a
  structure that looks more disky. The lower two rows of panels show projected
  stellar surface density in the same volume and for the same
  lines-of-sight. The stellar component of this large galaxy also has a more
  disky morphology in {\sm AREPO} compared with {\sm GADGET}.}
\label{fig:gas_examplez0}
\end{figure*}

\begin{table*}
\begin{tabular}{ccccccccc}

\hline Code & $M_{\rm gal}$ & $M_{\rm gas}$ & $r_{1/2\, \rm gas}$ &
 SFR &$j_{\rm gas}$ & $j_{\rm star}$&
Angle&\\ &$[10^{10}\,\msunh]$ & 
$[10^{10}\,\msunh]$ 
 &[$h^{-1}{\rm kpc}$] & $[M_{\odot}\,{\rm yr}^{-1}]$&[ $h^{-1} \rm
  km\, s^{-1} kpc$] &[$h^{-1} \rm km\, s^{-1} kpc$]&[$\rm ^o$] &\\ \hline
$z=2$ Example &&&&&&&&\\
\hline
A\_L20n512 & 12.88 & 2.03  & 5.91 & 37 & 2108 & 573 &  7\\
G\_L20n512 & 7.56 & 0.60 & 3.75 & 18 & 1150 & 352 & 18\\
\hline
$z=0$ Example &&&&&&&&\\
\hline
A\_L20n512 & 3.35 & 0.75 & 10.1 & 1.3 & 1576 & 285 &  5\\
G\_L20n512 & 2.2 & 0.24 & 5.3 & 1.4 & 817 & 218 & 10\\
\hline
\end{tabular}
\caption{Properties of the example galaxies shown in
  Figures~\ref{fig:gas_examplez2} and \ref{fig:gas_examplez0} as measured
  in our L20n512 simulations. In column (1) we list the simulation name,
  column (2) gives the galaxy baryonic mass, and column (3) the gas mass of
  our selected galaxy. In column (4) we give the effective half-mass
  radius of gaseous disk in physical units. Column (5)
  indicates the star formation rate of this galaxy, while columns (6) and (7)
  give the specific angular momentum of gas and stars. Finally, in column
  (8) we give the angle between the angular momentum vectors of galactic gas
  and stars. We note that this galaxy has a much higher gas fraction, gas
  extension and a higher star formation rate in {\small AREPO} than in {\sm
    GADGET}, which is a typical systematic difference we find for massive
  galaxies between the codes.}
\label{table:disks}
\end{table*}

\begin{figure*}
\includegraphics[width=1.\textwidth]{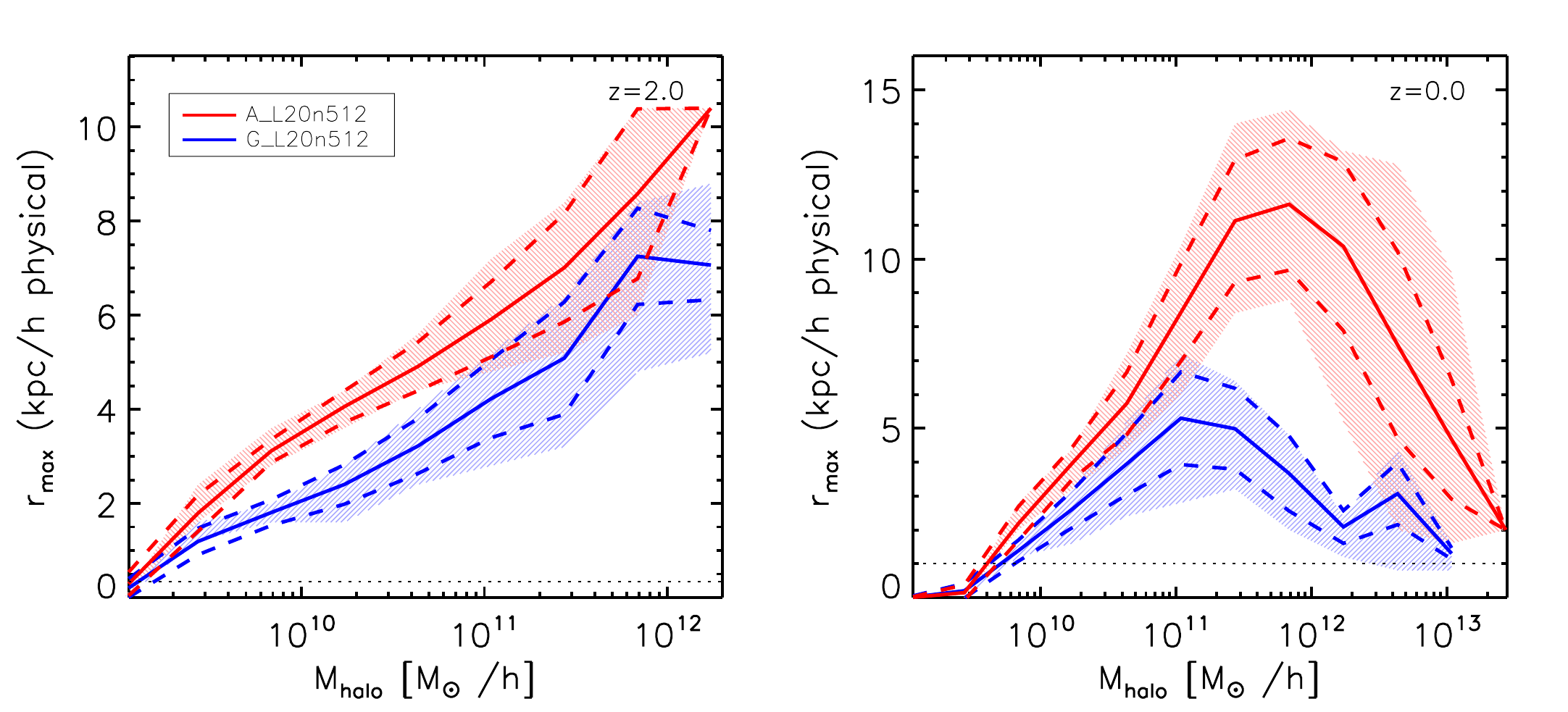}
\caption{Dependence of the size of gaseous disks of central galaxies in
  {\small GADGET} and {\small AREPO} simulations on the parent halo mass (for
  a bins size of $0.4\,{\rm dex}$ in halo mass).  The solid lines show the
  average ``edge'' radius of the dense gaseous disk, i.e.~the largest radius
  within which the azimuthally averaged column density of cold dense gas
  projected onto the disk plane is greater than $N_H=2\times10^{20} \rm
  cm^{-2}$.  Shaded regions show the 25-75\% range
  of the distribution of disk sizes at a given mass, while the dashed lines
  indicate the standard deviation around the mean.  It is apparent that the
  dense, continuous gaseous disks in {\sm AREPO} are much more extended than
  their counterparts in {\sm GADGET}.  This holds at all halo masses, both at
  $z=2$ and $z=0$, and it is most dramatic in haloes around $10^{12}\, \msunh$
  at $z=0$.}
\label{fig:galsize_rmax}
\end{figure*}

\begin{figure*}
\includegraphics[width=1.\textwidth]{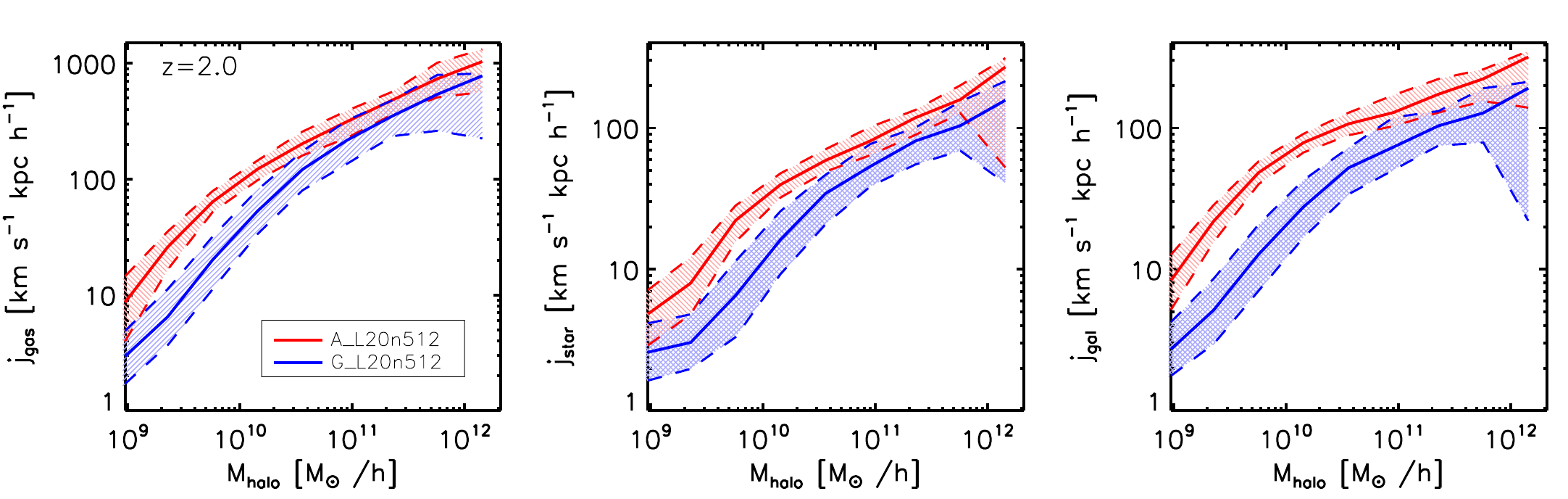}
\includegraphics[width=1.\textwidth]{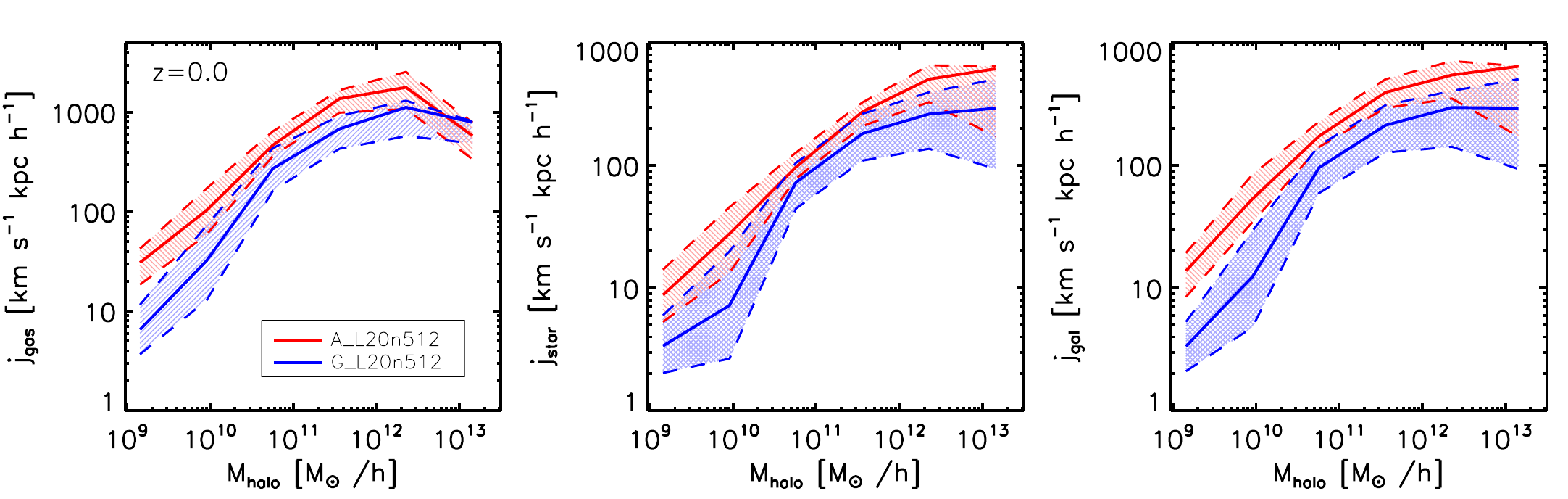}
\caption{Specific angular momentum of galactic gas (left panels), galactic
  stars (middle panel) and for the whole galaxies (right panels), at $z=2$ and $z=0$, as a function of the
  parent halo mass. The measurements demonstrate that in {\sm AREPO} (red lines) both the
  gaseous and stellar components have a higher specific angular momentum than
  in {\sm GADGET} (blue lines), indicating diskier, more rotationally
  supported baryonic configurations. Shaded regions and corresponding dashed
  lines show  25-75\% of the angular momentum distribution in a 0.4 dex mass
  bin at $z = 2$, and 0.8 dex mass bin at $z =0$. }
\label{fig:am}
\end{figure*}

The galaxy resides in a halo of mass $M_{\rm halo}\sim 9.1\times
10^{11}\,\msunh$, and its most important quantitative measurements are
summarized in Table~\ref{table:disks}.  The visual impression already
indicates that the galaxy is more massive, more extended and has a much higher
gas fraction in {\sm AREPO}, which is born out by our quantitative
measurements.  In fact, the star formation rate is a factor of $\sim 2$ higher
in {\sm AREPO} than in {\sm GADGET}, illustrating the principal difference
between the codes we discussed in the previous sections. In addition, the
gaseous disk is clearly much more extended in {\sm AREPO}, as well as being
thinner and more well-defined. While the difference in the half-mass radius is
$\le 60\%$, the net specific angular momentum of the gaseous disk is almost 2
times larger in {\sm AREPO} than for the same galaxy in {\sm GADGET}.

Similarly, the stellar disk is somewhat more extended and thinner in {\sm
  AREPO}, as seen in Figure~\ref{fig:gas_examplez2}. It also has a higher
rotational support, as indicated by the net stellar specific angular momentum
which is about 60\% higher in {\sm AREPO} compared to {\sm GADGET}. This is
consistent with the visual impression of a more disky stellar distribution in
{\sm AREPO}. It is also evident that the stellar and gaseous disks are better
aligned in {\sm AREPO}, as measured by the angle between the angular momenta
of gas and stars in the galaxy.  We note that our findings in this halo
represent typical differences in the properties of massive galaxies, and are
by no means restricted to a single system.

Figure~\ref{fig:gas_examplez0} shows another example of extended gaseous disk,
this time of a central galaxy in $\sim 2.2 \times 10^{11} \msunh$ halo at $z =
0$. The figure demonstrates that gaseous disks are even more extended at late
times and the relative difference between {\sm AREPO} and {\sm GADGET} is now
even larger. The example galaxy contains $3$ times more gas in its large disk,
most of which is non-star forming and represents a large extended gaseous
reservoir. Galactic interactions between such disks can lead to strong, extended
tidal features such as the ones visible in  Figure~\ref{fig:tidal}, while
compact gaseous disks of {\sm GADGET} galaxies are unlikely to produce them in
cosmological simulations. Differences in the stellar distribution are visible
in the bottom panels of Figure~\ref{fig:gas_examplez0} with the galaxy appearing
more disky in {\sm AREPO} as a consequence of higher specific angular momentum
of stars in our moving-mesh simulation (see Table~\ref{table:disks}).
This galaxy, however, does show a large bulge component that is similar in 
both simulations (and is a fairly typical example for this halo mass),  
suggesting the need for an efficient feedback mechanism that can remove a large 
fraction of low-angular momentum baryons \citep[e.g.][]{vandenbosch01, 
maller02, brook11}.

In order to demonstrate this point more systematically we show a measure of
the extent of gaseous disks in our highest resolution simulations in
Figure~\ref{fig:galsize_rmax}. For this purpose we project the mass of cold,
dense gas onto the gaseous disk plane of a central galaxy. Next, we find the
azimuthally averaged surface density of the cold, dense gas in $0.4\,{\rm
  kpc}$ wide rings. We define ``disk radius'' as the radius at which the
ring-averaged column density drops to a column density characteristic of
damped Lyman-${\alpha}$ absorbers, $N_H=2\times 10^{20} \rm {cm^{-2}}$. From
the results in Figure~\ref{fig:galsize_rmax} it is clear that galaxies in our
high-resolution {\sm AREPO} simulation are much more extended than galaxies in
{\sm GADGET}. At $z=2$, the difference in the average extent of gaseous disks
is typically around a factor of $\sim 1.5$ while for the most compact galaxies
it is closer to factor of $\sim 2$.  In addition, visual inspection indicates
that the disks in {\sm AREPO} are not only more extended but are also
smoother, thinner in edge-on projections and have better aligned gas at
different radii (as visible in Figure~\ref{fig:gas_examplez2}). This makes
their appearance more similar to the observed galactic disks. In
contrast, {\sm GADGET} galaxies show a more irregular gas distribution in the
disk, and numerous dense gaseous clumps in the galactic surroundings.  

At $z=0$, below a halo mass of $10^{11}\, \msunh$, {\sm AREPO} galaxies are
also a factor of $\sim 1.5$ larger, but in more massive haloes the size
difference is even more dramatic than at $z=2$. For about an order of
magnitude in halo mass around $10^{12}\, \msunh$, galaxies formed in our
moving-mesh simulations are systematically larger than in SPH by a factor of
3-5. Notice that this is a mass range of great interest as it corresponds to
the halo masses of massive late-type disk galaxies such as our own Milky Way. 

The measurement of gas disk sizes, as shown in Figure~\ref{fig:galsize_rmax},
has been specifically chosen to capture as closely as possible visual
impressions from Figure~\ref{fig:gas_examplez2} and \ref{fig:gas_examplez0}
(see also Figures 2 and 4 in Paper I). The trends of
Figure~\ref{fig:galsize_rmax} are fully consistent with a more in depth
analysis presented in \citet{torrey11}, which is based on detailed Sersic
and exponential disk profile fits. We note that stellar half-mass radii are
only barely resolved even in our highest resolution simulations, being of the
order of $1-3$ times the gravitational softening length. We can therefore not make
definitive statements about the radial extent of the stellar component, based
on a rather coarse diagnostic such as half-mass radii. Nonetheless, the full
extent of the stellar component is well resolved in our simulations and can be
used to infer at least some of the stellar disk properties, such as total
angular momentum (see below).  

To characterize the degree of orderly rotation of stellar and gaseous
components of galaxies we calculate the angular momentum of individual gas
cells/particles and stellar particles with respect to the galaxy center and
find the total angular momentum vectors for gas and stars in a galaxy. We
normalize this by the gaseous and stellar galaxy mass, respectively, before
calculating the magnitude of the net specific angular momentum. We apply the
same procedure for the galaxy as a whole. The results are shown in
Figure~\ref{fig:am} in terms of the net specific angular momenta of gas, stars
and total baryons (gas+stars) in galaxies in our highest resolution
simulations. Not too surprisingly, for the cold, dense galactic gas, the
structural differences we have observed earlier in the characteristic radii
are also reflected in the specific angular momentum of the galactic gas, which
is higher in {\sm AREPO} than in {\small GADGET} by a factor of $\sim 2$,
explaining the larger sizes of galaxies forming in our moving-mesh
simulations.

Substantial differences are also visible in the distribution of the stellar
specific angular momentum shown in Figure~\ref{fig:am}. 
Because galaxies in {\small AREPO} are diskier, their specific angular
momentum is much higher than in {\small GADGET} galaxies. This statement
applies to both $z=2$ and $z=0$, with typical angular momentum differences at
$10^{12}\, \msunh$ of the order of $1.5-2$. 

In Figure~\ref{fig:am}, we also indicate the spread of the distribution in
specific angular momentum by indicating the 25\% and 75\% percentiles of the
distribution at a given mass. The distribution of specific angular momentum at
a given mass is much narrower in {\sm AREPO} than in {\sm GADGET}.  In terms
of specific angular momentum of both gas and stars in the $10^{9}-10^{12}
\msunh$ mass range, there is very little overlap between galaxies in {\sm
  AREPO} and {\sm GADGET}. For example, galaxies within the lowest 25\% of the
distribution in {\sm AREPO} often have similar specific angular momentum as
the top 25\% of galaxies in {\sm GADGET}.

Interestingly, the differences between the net specific angular momentum for
the total baryonic content of galaxies are even larger than for the stellar
content, where {\small AREPO} galaxies have up to a factor of $\sim 2$ higher
specific angular momentum than {\sm GADGET} galaxies. This results from a
combination of higher gas content and higher net specific angular momentum in
{\sm AREPO}, and a better alignment of the gaseous and stellar components.

In our measurements, we see a trend of an increase of the gaseous disk extent
and the specific angular momentum of galaxies with halo mass and time. Both
effects are consistent with expectations of simple models of the formation of
galactic disks \citep{fall80, mo98, somerville08a}.  In fact, these trends are
qualitatively easily understood in the inside-out model of disk formation
\citep{fall80} where the galaxy sizes will approximately scale like $R_{\rm d}
\propto \lambda R_{\rm vir}$, where $\lambda$ is the dimensionless spin
parameter \citep{peebles69} and $R_{\rm d}$ is the disk scale length. The
specific angular momentum then approximately scales as $j \propto \lambda
\times R_{\rm vir} \times V_{c}$. 
Because the spin parameter is approximately independent of mass and redshift
\citep[e.g.][]{barnes87}, the sizes and specific angular momenta of galactic
gas both increase with time and mass. Both $R_{\rm vir}$ and $V_{c}$ 
increase with mass, causing the dependence of $j$ on halo mass to be 
stronger than for the disk radius, correctly approximating the
trends in our simulations. For a more detailed quantitative model,
one however has to account for the mass and redshift dependence of the halo
concentration and the angular momentum distribution within haloes. Also, one
has to postulate conservation of angular momentum of the gas during infall, an
assumption we do not have to make in our detailed hydrodynamical simulations.

\begin{figure*}
\includegraphics[width=1.\textwidth]{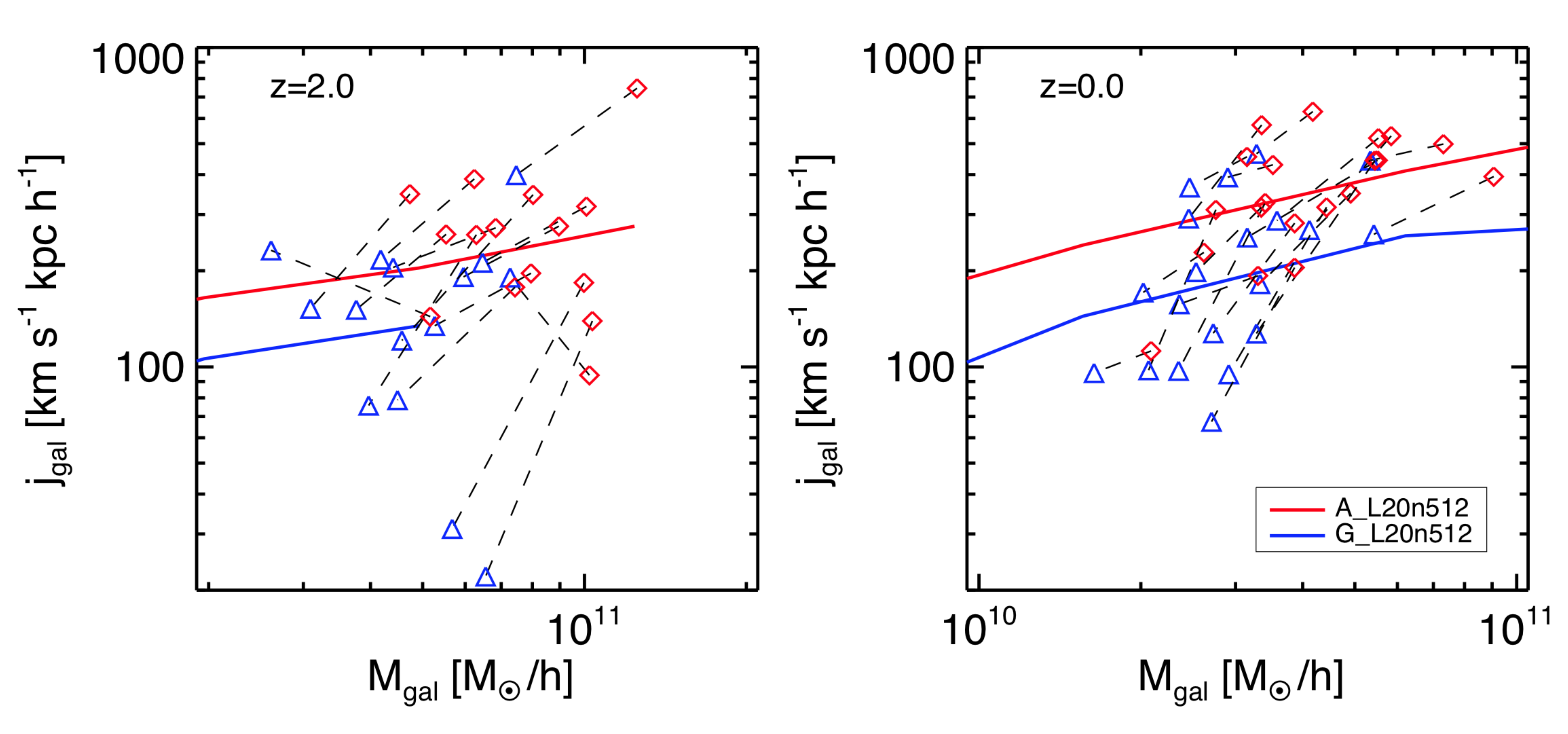}
\caption{ Specific angular momentum of galaxies (stars+gas) at
  $z=2$ and $z=0$, as a function of the galaxy mass. The measurement
  demonstrates that in {\sm AREPO} (red lines) galaxies at a given
  mass have higher specific angular momentum than in {\sm GADGET}
  (blue lines).  Thick lines show the medians in 0.4 dex mass bins at
  $z = 2$, and 0.6 dex mass bin sat $z = 0$. Symbols connected with
  thin dashed lines show a matched sample of galaxies in {\sm AREPO} (red,
  diamonds) and {\sm GADGET} (blue, triangles). The matched sample
  includes the first 20 galaxies from Figures 2 and 4 in Paper I,
  excluding galaxies undergoing major mergers, where the angular
  momentum changes rapidly. It is clear that for most of the galaxies
  in our simulations differences in specific angular momenta of
  galaxies are well above the trends expected from their mass
  differences.}
\label{fig:am_gal}
\end{figure*}

 Owing to the increasing specific angular momentum as a function
  of halo and galaxy mass, one may wonder if a large part of
  the differences in Figure~\ref{fig:am} is caused by the mass difference
  of galaxies inhabiting halos of the same mass. As we show in
  Section~\ref{sec:gal_to_halo}, in our highest resolution simulations
  the galaxy masses at a fixed halo mass are very similar between the two
  codes for $M_{halo} \lesssim 10^{11}\msun$.  This means that trends
  with halo mass also directly reflect trends with galaxy
  mass. However, one needs to be more careful at higher masses where
  there is a systematic offset in galaxy masses between {\sm AREPO}
  and {\sm GADGET} at a given halo mass.  In Figure~\ref{fig:am_gal},
  we show the specific angular momentum for galactic baryons as a
  function of galaxy mass. In addition, we show a matched sample of
  galaxies in the mass range $10^{10}-2\times 10^{11} \msun$, connected
  with dashed lines to indicate corresponding differences in specific angular
  momentum for galaxies in the same collapsed structures.  It is clear
  that the specific angular momentum of {\sm AREPO} galaxies is
  systematically larger than for {\sm GADGET} at a given galaxy
  mass. This trend is even clearer for our matched sample: for the large
  majority of galaxies, {\sm AREPO} galaxies have much higher specific
  angular momentum than galaxies of the same mass in {\sm GADGET}.  In
  other words, the baryons that make galaxies in our moving mesh
  simulations either retain more angular momentum or are forming from
  baryons with systematically higher angular momentum.

Finally, we note that by looking at measurements of disk sizes as a function
or resolution, we find that there appears to be a qualitative difference
between our simulation techniques with respect to the sensitivity of predicted
disk sizes on numerical resolution. In both codes, the sizes of the galactic
gas distributions spuriously increase with poorer resolution, but this effect
is somewhat stronger in {\sm AREPO}.  This basic trend is to be expected as
poor spatial resolution will tend to lower the central densities in galaxies
and make them puffy. In SPH, part of this trend is likely offset by the
angular momentum transport caused by the artificial viscosity in SPH, which
will tend to makes galaxies more compact, as we discussed in
Section~\ref{sec:gas_fractions}.
 
To summarize, gaseous disks of {\sm AREPO} galaxies are systematically
larger than their counterparts in {\sm GADGET}. {\sm AREPO} galaxies
show a higher degree of rotational support as measured by their
specific angular momentum, indicating that they are more disky and
less dispersion-dominated.  These differences are particularly
prominent for the range of halo masses in which large disk galaxies
are expected to form in the real Universe. We note that previous
attempts to model disk galaxy formation in cosmological simulations
with SPH often led to a too low specific angular momentum of simulated
galaxies \citep[e.g.][]{navarro00}. Higher resolution simulations 
of individual galaxies with SPH hinted that the problem might be less 
severe when the resolution is better \citep[e.g.][]{governato04, governato07}. 
However, this has not been demonstrated yet for a large population of 
galaxies.  Furthermore, we show that our moving-mesh simulations provide 
a dramatic improvement over corresponding SPH results and have much higher 
specific angular momentum at a given galaxy mass, facilitating formation of 
late-type galaxies.

\section{Discussion and conclusions}
\label{sec:discussion}

In this study, we have systematically compared results obtained for
cosmological simulations of galaxy formation with the SPH code {\small
  GADGET} against the new moving-mesh code {\small AREPO}. The
simulations started from identical initial conditions, employed the
same high-resolution gravity solver, and used a completely equivalent
modeling of the physics of radiative cooling, star formation and
associated feedback processes, the latter being incorporated into a
subresolution model for the multi-phase structure of the ISM. Hence
the outcome of our simulations yields a direct test of the impact of
hydrodynamical solvers on the predictions of cosmological galaxy
formation simulations.

In Paper I of this series, we have focused on a comparison of the
global baryonic properties predicted by the two simulation techniques,
while here we consider properties of individual
galaxies. Reassuringly, we find that many of their properties are in
fact in broad agreement, for example the galaxy mass functions are
quite similar at low masses. However, when scrutinized in detail, and
especially towards later times and in high mass systems, we find
rather significant systematic differences in the predicted galaxy
properties.

Specifically, massive haloes in {\sm AREPO} host galaxies that are
larger, diskier, more massive, and have a higher star formation rate
than their {\sm GADGET} counterparts. In the extreme end of the galaxy
mass function, {\sm AREPO} can accumulate up to a factor of $\sim 2$
more mass in central galaxies than {\sm GADGET}.  The moving-mesh
galaxies also have a higher specific angular momentum, and the
morphology of the gas surrounding the galaxies is clearly different,
being much less clumpy than in the SPH code. These clumpy
condensations of gas are largely absent in {\sm AREPO}; instead,
comparatively smooth gas disks are formed which tend to align better
with the stellar disks.

There appears to be primarily a single driver for the higher star
formation rates, lower central entropies and higher gas fractions in
massive galaxies in {\sm AREPO}.  Multiple lines of evidence show (see
also Paper~I) that the hot gaseous haloes formed in {\sm AREPO} cool
out gas more efficiently than their counterparts in {\sm GADGET},
causing in turn much higher star formation rates in the central
galaxies of these haloes. We find that these different effective
cooling rates occur despite the fact that the two codes create rather
similar gas profiles in large haloes at low redshift, which do not
suggest an a priori cause for a substantially different total cooling
emission. In fact, we find that the culprit does not lie in a
different cooling per se. It lies in a different {\em heating}. As we
have shown in Paper~I, the local dissipation rates in haloes simulated
with SPH and the moving-mesh code are quite different. Whereas more
entropy (and hence dissipative heating) is produced in the infall
regions of haloes in {\sm AREPO} compared with {\sm GADGET}, the
opposite is true for most of the halo volume inside the virial region,
and in particular in the outer parts of haloes where the cooling
radius for many of these haloes lies. This dissipative heating offsets
part of the cooling losses in SPH, reducing the strength of the
cooling flows.

There are good reasons to believe that this heating effect in SPH is
in fact spurious. First of all, as we demonstrate explicitly in Paper
III, the standard formulation of SPH that we use here tends to
suppress gas stripping out of infalling satellites, as well as any
entropy mixing with background material.  As a result, dense gaseous
clumps penetrate further into haloes in {\sm GADGET} upon infall,
heating them through ram-pressure interactions at smaller radii than
corresponding {\sm AREPO} simulations, where material is stripped and
mixed already at much larger radii.  The stripping processes
themselves, as well as the curved accretion shocks around haloes,
inject subsonic and mildly transonic turbulence into the outer parts
of haloes. However, as \cite{bauer12} show, subsonic turbulent
cascades are not represented in a physical way in SPH at the
achievable resolution (unlike in the moving-mesh code).  Instead, the
turbulence is damped efficiently close to the injection scale.

In addition, there is significant viscous damping of small-scale
subsonic noise in SPH. The latter is ultimately sourced by errors in
SPH's gradient estimate \citep{abel11}, which trigger jittering
motions of the gas particles on the scale of the SPH kernel
\citep{springel10b}. If the artificial viscosity is increased, the
amplitude of this sonic noise can be lowered, but the total
dissipation through this effect stays nearly invariant
\citep{bauer12}.  The dynamic environment around forming haloes can
efficiently regenerate the SPH noise, such that its constant damping
by the artificial viscosity constitutes a spurious heating
source. Especially in the slow cooling regime encountered for large
haloes at low redshift, this can then affect the cooling flows in
haloes. We note that this problem is neither easily cured by higher
resolution, lower artificial viscosity, nor a larger number of SPH
neighbors. What might help is a formulation of SPH that features a
more accurate gradient estimate, which seems a prerequisite to
eliminate the sub-sonic velocity noise.

The differences in cooling induce substantial differences in the SFRs
of galaxies, which are largest for halo masses around $M_{\rm halo}
\sim 10^{12}\, \msunh$ at $z=0$, i.e.~approximately at the mass scale
of our own Milky Way galaxy.  At this mass scale, we also find
substantial differences of the predicted angular momentum content of
both gas and stars, which is reflected in more extended disks galaxies
forming in our moving-mesh simulations when compared to SPH.  The
differences in the gaseous disk sizes are apparent at all halo masses,
but around $10^{12}\, \msunh$ they are particularly large and reach a
factor of $\sim 3$.  Because the angular momentum content of halo
  gas increases with time (see section \ref{sec:gal_radii}), a more
  efficient gas supply at late times will lead to a relative increase
  of the specific angular momenta and sizes of galaxies. This suggests
  that the efficient late time cooling is not only increasing galaxy
  masses but is also a significant contributor to the differences in
  galaxy sizes, especially at mass ranges where the cooling
  differences are large. However, our results in
  Figure~\ref{fig:am_gal} show that this is not the only cause.

 Also, we note that even low-mass galaxies have more extended gas
  distributions and higher gas content in the moving-mesh simulations
  when compared to SPH. These properties together with more efficient
  ram pressure stripping in moving-mesh runs (as we show in Paper
  III), will lead to a much larger mass loss from low-mass galaxies as
  they infall into more massive halos. This means that the fraction of
  the infalling material that will loose angular momentum via
  dynamical friction to dark matter halo will be much smaller in {\sm
    AREPO} leading to a relative increase of specific angular momentum
  of galactic baryons with respect to {\sm GADGET}.

The results we have presented in this paper hence clearly show that it does
matter for simulations of galaxy formation which hydrodynamical solver is
used. We find significant systematic offsets in the predicted properties of
galaxies formed in {\sm GADGET} and {\sm AREPO} at all mass scales, but the
most striking differences occur in haloes around $\sim 10^{12}\, \msunh$. Here
the more accurate hydrodynamics provided by {\sm AREPO} leads to late-type
galaxies that are large, disky, gas-rich, and have high star formation
rates. Such galaxies, especially at late times, simply appear not to form in
cosmological SPH simulations when the same physics is included.  
Our findings also suggest that {\sm AREPO} is able to significantly 
alleviate the low angular momentum problem of galaxies formed in cosmological 
hydrodynamic simulations, which has been a long-standing issue
in galaxy formation for the past two decades \citep[e.g.][]{katz92a,
navarro97a,navarro00}. Strong feedback mechanisms, however, appear to 
be needed to both regulate galaxy masses \citep[e.g.][]{keres09b} and 
remove the low angular momentum baryons that form excessive bulges 
\citep[e.g.][]{vandenbosch01}. Given the extended gaseous disks of galaxies
in {\sm AREPO} and diskier stellar morphology when compared to SPH results, 
it is likely that the strength and type of the required feedback discussed in 
previous SPH work needs to be revisited. This emphasizes that an accurate numerical 
treatment of hydrodynamics is a key factor for successful cosmological simulations that 
aim to form realistic galaxies in a complex environment.

\section*{ACKNOWLEDGMENTS}

We would like to thank the referee, Fabio Governato, for constructive comments.
The simulations in this paper were run on the Odyssey 
cluster supported by the FAS Science Division Research Computing Group at
Harvard University. DK acknowledges support from NASA through Hubble
Fellowship grant HSTHF-51276.01-A and NASA ATP NNX11AI97G.
DS acknowledges NASA Hubble Fellowship through grant HST-HF-51282.01-A.

\end{document}